\documentclass[ aps, pre, a4paper, amsfonts, amssymb, amsmath, reprint, showkeys, nofootinbib, twoside ,floatfix]{revtex4-1}
\usepackage[english]{babel}
\usepackage[utf8]{inputenc}
\usepackage[colorinlistoftodos, color=green!40, prependcaption]{todonotes}
\usepackage{amsthm}
\usepackage{mathtools}
\usepackage{physics}
\usepackage{xcolor}
\usepackage{graphicx}
\usepackage[left=23mm,right=13mm,top=35mm,columnsep=15pt]{geometry} 
\usepackage{adjustbox}
\usepackage{placeins}
\usepackage[T1]{fontenc}
\usepackage{lipsum}
\usepackage{csquotes}
\usepackage[pdftex, pdftitle={Article}, pdfauthor={Author}]{hyperref} 
\usepackage{xcolor}
\usepackage[justification=Justified,singlelinecheck=false]{caption}
\usepackage{ragged2e}
\usepackage{subcaption}
\usepackage{graphicx}
\usepackage{placeins}
\usepackage[percent]{overpic} 

\begin{document}
\title{ParABS system  induces topological modifications to spatio-temporally organize its mother and daughter
chromosomes in the replicating \emph{C. crescentus} cell}

\author{Shailesh Sathe}
    \email{shailesh.sathe@students.iiserpune.ac.in}
\author{Apratim Chatterji}
    \email{apratim@iiserpune.ac.in }
    \affiliation{Department of Physics, Indian Institute of Science Education and Research (IISER-Pune), \\ Dr. Homi Bhabha Road, Pune, India-411008.}
\date{\today} 

\begin{abstract}
We identify an entropy-mediated mechanism underlying the spatio-temporal organization of the chromosome in the bacterial cell \emph{Caulobacter crescentus}. Effective entropic interactions between replicating DNA segments emerge from their topological modifications of the daughter DNA polymers as a consequence of binding between ParA and ParB proteins. Enthalpic contributions arising from the ParABS system further influence their dynamics. In {\em C. crescentus},  chromosome replication and segregation occur concurrently prior to cell division. The origin of replication \emph{oriC}  of the mother chromosome of {\em C. crescentus} is tethered to the old cell pole. Replication begins at \emph{oriC}, generating two daughter \emph{oriCs}, one of which remains anchored at the old pole. The other \emph{oriC} is actively transported toward the new pole by the ParABS system of proteins even as replication of the rest of the chain is in progress.  How the ParABS system generates a directed force capable of transporting one of the daughter \emph{oriCs} to the opposite pole on the timescale of chromosome replication has remained unclear, despite previous efforts to elucidate the underlying mechanism. Furthermore, the two  daughter chromosomes get neatly segregated into two halves of the cell, overcoming entanglement during the time taken to complete replication.  
 In a bead-spring polymer model of the replicating \emph{C. crescentus} chromosome, we model the transient contacts between ParA and the ParB proteins. The ParB protiens are loaded onto \emph{oriC}-proximal \emph{parS} sites. This leads to the transient formation of DNA loops. These loops generate an effective entropic repulsion, which drive chromosome segregation and result in \emph{oriC} trajectories, as observed in experiments~\cite{shebelut2010caulobacter}, aided by the concentration gradient of ParA molecules near the new pole.  This minimal framework, incorporating the physics of topologically modified (ToMo) polymers, allows us to quantitatively reproduce the time-dependent spatial organization of different tagged loci and replisomes observed in multiple experiments. 
 Incorporating a minimal model of extrusion also allows us to recover the linear arrangement of the chromosome arms, as is inferred from the characteristic diagonals of Hi-C maps.




\end{abstract}


\maketitle

\section{Introduction} \label{sec:outline}
Cells contain a large number and a wide variety of macromolecules. Many of these macromolecules are spatially organized and localized in specific regions within the cell, while others remain more broadly distributed throughout the cellular volume. These macromolecules act in synchrony with each other to perform various tasks in a timely manner, which allows the cell to be {\em alive}. Understanding how synchrony, coordination, and large-scale reorganization of macromolecules emerge from their interactions within the inherently stochastic environment of the cell remains one of the central challenges--and perhaps a holy grail--of the physics of life.
DNA is one such extremely long molecule, which undergoes spatio-temporal reorganization in a timely manner \cite{Dekker2024,Woldringh2024,Badrinarayanan2015,Woldringh2015}. This enables the contact probabilities between enhancer--promoter sites on the DNA to adjust themselves appropriately using a variety of mechanisms, which in turn regulate the expression of particular genes at different points of the life cycle of the cell ~\cite{Notani2026,Jost2026,Brahmachari2018,Brahmachari2019,Hofmann2015,Marko2019,Karoui2010}. How does this tightly controlled reorganization of molecules/molecular segments happen using stochastic and active forces? 

Researchers often turn to the study of the bacterial cell because it is deemed to be much simpler than the eukaryotic cell with multiple chromosomes. Most bacterial cells have just one DNA chromosome, which is confined within the cell, i.e., it does not have the nucleus~\cite{Lioy2018,Lioy2021}. The DNA molecule associated with suitable proteins is compacted at a hierarchy of length scales driven by the action of various SMC and NAP proteins~\cite{Remus2024,Hirano2005,Sheratt2014,Sheratt2020}, and this compacted chain is referred to as the chromosome. The mother DNA undergoes replication to form two daughter DNAs. In the bacterial cell, DNA replication and segregation of daughter chromosomes to two halves of the cell (usually) occur concurrently \cite{Badrinarayanan2015,Lioy2021,reyes2012chromosome,woldringh2006structural,draper2002bacterial}. This is a critical step before cell division to ensure the presence of one chromosome in each daughter cell~\cite{Lioy2021,Woldringh2024}.

When DNA replication and segregation are in progress, the replicated DNA segments have to relocate suitably and in a timely manner and reorganize within the two halves of the cell before cell division proceeds~\cite{Cass2016,Youngren2014}. 
In the recent past, we have proposed that the replicating DNA of \emph{E. coli} is a topologically modified (ToMo) ring polymer \cite{dna1Mitra2022, mitra2022topology}, in which topological modifications are brought about by linker proteins or DNA segments. 
The linkers could either be the extruding SMC proteins or rrn operons, which co-localize in space~\cite{Gaal2016}. This then enables the partly formed daughter DNAs to relocate to their new home addresses within each half of the cell in a timely manner, harnessing entropic interactions between different segments of the ToMo polymer. Using these minimalist principles, we have reproduced experimentally observed spatio-temporal reorganization of \emph{E. coli} chromosomes in both slow~\cite{dna1Mitra2022} and fast growth~\cite{dna2Pande2024} conditions in our simulations. 

The next question is whether other bacterial cells employ similar mechanisms to reorganize the replicating mother DNA and the newly formed daughter DNA. In this work, we establish how the {\em C. crescentus} chromosome uses the ParABS system of proteins to introduce suitable topological modifications to the replicating DNA, which in turn entropically drives the directed motion of the origin of replication (\emph{oriC}) of one of the two daughter chromosomes from one pole to the other. This in turn helps in the segregation of the two daughter chromosomes into two halves of the cell, aided by extruding proteins. It has long been known that the ParABS system drives the relocation of one of the daughter \emph{oriCs} from the old pole to the new pole of the cell. However, despite evidence for bridging interactions among its constituents, the mechanism by which the ParABS system ensures the timely and directed transport of \emph{oriC} remains unclear. This paper investigates and resolves this question.

ParA binds nonspecifically to the DNA chain, while ParB binds specifically to the \emph{oriC}-proximal \emph{parS} sites along the DNA contour. In this work, we propose that transient binding between ParA and ParB effectively generates topological modifications, in the form of internal loops, in the replicating \emph{C. crescentus} chromosome. We show that the resulting entropic repulsion between different loops within the DNA and enthalpic interactions mediated by the concentration gradient of ParA are sufficient to drive the directed transport of one daughter \emph{oriC} from the old pole to the new pole. 
Consequently, the trajectories of other chromosomal loci are obtained directly from the simulations, without additional inputs, and are in quantitative agreement with experimental observations~\cite{shebelut2010caulobacter,lim2014evidence, zhang2024chromosome,Harju2025.04.03.647011}. 


It is important to emphasize that, although we invoke entropic forces arising from topological modifications to explain chromosome organization, the cell is not in equilibrium in the statistical-physics sense. Our model explicitly incorporates several non-equilibrium processes, including replication, cell elongation, ParA–ParB–mediated interactions and their modulation, ParA turnover, and loop extrusion. The relevance of entropic effects despite these non-equilibrium processes follows from a clear separation of length and time scales: chromosome reorganization occurs over large length scales ($3$--$6~\mu$m), whereas the non-equilibrium active processes occur at localized points in space ($\approx 10$nm length scale) and time. As an example, to model replication we introduce monomers at regular but well-separated intervals at two specific replication forks. At intermittent times, the simulation is a standard Langevin simulation with a fixed number of particles, a fixed length of cylinder, and well-defined interactions between monomers. Consequently, most of the polymeric chain is in a quasi-equilibrium state, as the intervals are long enough for the relaxation of a thirty-monomer polymer with excluded volume interactions, as verified separately and to be reported elsewhere. 

Recent studies~\cite{Harju2025.04.03.647011} have also reconstructed trajectories of fluorescently tagged chromosomal loci using data-driven approaches. In one such approach, time-resolved Hi-C data at six stages of the cell cycle were used as inputs for a maximum-entropy (MaxEnt) model to obtain chromosome conformations at different stages of the cell cycle. The distance between the two replicated \emph{oriC} loci, as observed by FISH, was additionally used as an input for predicting the positions of other chromosomal loci. In a separate mechanistic implementation, a time-dependent external force was applied to one of the daughter \emph{oriC} loci to mimic the action of the ParAsBS system and reproduce its directed transport from the old pole to the new pole~\cite{harju2024loop}.

In contrast, our approach is neither data-driven nor reliant on externally imposed forces. Instead, we explicitly incorporate the transient topological modifications that arise from effective interactions within the ParABS system. We show that these modifications are sufficient, within our model, to drive the timely and directed transport of one daughter \emph{oriC} from the old pole to the new pole~\cite{shebelut2010caulobacter,lim2014evidence}. Without additional inputs, our simulations also reproduce the trajectories of multiple tagged loci and the time-dependent repositioning of replisomes, in quantitative agreement with experimental observations~\cite{Harju2025.04.03.647011,zhang2024chromosome}.  Thus, a single minimalist mechanistic framework, based on effective interactions within the ParABS system that transiently modify chromosome topology and thereby modulate entropic interactions between polymer segments, accounts for six experimentally observed features of chromosome organization and dynamics. To facilitate direct comparison with experiments, we plot the simulation results using conventions used for presenting data from experiments.

We give a brief summary of \emph{C. crescentus} DNA phenomenology in the second section before describing our model and results in the third and fourth sections of this manuscript, respectively.  This brief background is for polymer physicists who are unfamiliar with \emph{C. crescentus} phenomenology. The results section has six different subsections. We conclude in the fifth section.

\section{The \emph{ C. crescentus} Chromosome} \label{sec:developenent}
\emph{Caulobacter crescentus} is a valuable bacterial model system for studying asymmetric 
cell division, as well as chromosome segregation \cite{COLLIER2019685,laub2002genes}. In the \emph{C. crescentus} cell chromosome replication is initiated only once per cell cycle, unlike that of \emph{E.  coli} (in fast growth conditions). Thus, it is easy to separate and analyze cell cycle events in synchronized populations. The \emph{C. crescentus} has a single chromosome  
with 4 million base pairs, i.e., 4000 kilobase pairs (kbp). We refer to the DNA polymer that is hierarchically compacted through the action of NAP and SMC proteins~\cite{Lioy2021,Badrinarayanan2015} as the chromosome. The chromosome is a ring polymer and is within a crescent-shaped cell.  Bacterial cells do not have a nucleus. 

The part of the bacterial cell cycle in which the mother chromosome replicates into two daughter chromosomes is the $S$-phase. The $S$-phase can vary from $60-90$ minutes depending on nutrient conditions, while the $G2$-period generally lasts $30$ minutes. The $G2$-period starts once replication is complete, i.e., at the end of the $S$-phase, and lasts until the cell divides at the end of the $G2$-period~\cite{Fang2013-qv,Lin2010-uy,Xu2022-ve}.  Replication and segregation occur in parallel, thereby one obtains two segregated daughter DNAs by the end of replication.  Replication of the chromosome starts at the \emph{oriC} loci, and thereafter, the replisome moves bidirectionally along the two arms of the ring polymer. The replisomes at the two replication forks locally unwind the mother DNA and form two daughter DNAs~\cite{zhang2024chromosome,Badrinarayanan2015}. Termination of the replication occurs when the two replisomes reach the \emph{ter}.

 \emph{C. crescentus} has the ParABS protein system, which is known to drive the 
 segregation of replicated daughter DNA to two halves of the cell. ParABS consist of three components- ParB, \emph{ parS} and ParA. \emph{parS} is a specific DNA sequence proximal to the origin of replication (\emph{oriC}).
 The DNA binding protein ParB gets loaded at the \emph{parS} sites to form ParB-{\em parS} partition complex \cite{livnyet.al.2007,jensen1999caulobacter}. Each {\em parS} site consists of $16$ bp, and there are approximately five \emph{parS} sites on the {\em C. crescentus} DNA.  The ParB dimer associates with the {\em parS} sit but then slides onto the neighboring non-specific DNA.  Once ParB binds to the {\em parS} site, it spreads into adjacent regions, covering 
 $5$ to $10$ kbp of flanking DNA near the {\em oriC}~\cite{Tran2018,Jalal2020}. 

 The ParA protein is released near the new pole, and thereafter a concentration gradient of ParA gets created from the midcell to the new pole. The ParA molecules form dimers in the solution and bind ATP. This ATP loaded 
 ParA dimers bind non-specifically onto the DNA,  i.e., ParA can bind to any point on the DNA contour \cite{leonard2005bacterial,hester2007soj}. ParA-ATP dimers can also assemble into filaments \cite{ptacin2010spindle,barilla2005bacterial,barilla2007tail} covering 
 many kbp along the DNA contour.

 When a DNA-associated ParA--ATP dimer comes into proximity with ParB, it can bind to ParB and thereby form a transient ParA--ParB complex. ParB stimulates ATP hydrolysis by the DNA-associated ParA--ATP dimer, causing ParA to detach from ParB and dissociate from the DNA into solution. ParA can subsequently bind ATP and reassociate nonspecifically with the DNA. 
 
Thus, as replication progresses, ATP-loaded ParA dimers get loaded onto the mother 
 and daughter DNAs, which can then bind to the {\em oriC}-proximal ParB to form a transient loop,
 before ParA drops off and the loop reopens. The stimulation of ATP hydrolysis is enhanced at high local concentrations of ParB  \cite{lim2014evidence,easter2002parb,leonard2005bacterial,hester2007soj,scholefield2011spo0j,barilla2007tail}. Simultaneously, one of the two copies of ParB/\emph{parS} complex (along with \emph{oriC}) is pulled toward the 
 new pole along the ParA gradient and the other remains tethered to the old pole \cite{ptacin2010spindle,schofield2010cell,shebelut2010caulobacter}

The average length of a newly born daughter cell is 2.4 $\mu$m to 2.5 $\mu$m, with an average cell width of $0.7 \, \mu$m to $0.8\, \mu$m \cite{wright2015intergenerational}.  Estimates for the size of a $10$-kBP segment of the chromosome, corresponding to compacted DNA associated with proteins, give a characteristic length (or bead diameter) of $\approx 88$~nm~\cite{messelink2021learning,Harju2025.04.03.647011}. Assuming self-similar packing of DNA at the $4$-kBP and $10$-kBP scales, a $4$-kbp segment is estimated to have a characteristic length of $\approx 65$~nm. The volume fraction occupied by DNA-polymer inside the cell is estimated to be $\sim 0.1$--$0.2$~\cite{Joyeux2023,Jost2026}. We therefore consider $5$
(or~$10$)~kbp of compacted DNA as a coarse-grained polymer segment with characteristic length scale of approximately $70$~($100$)~nm. Previous studies of eukaryotic chromosomes have reported a radius of gyration ($R_g$) of approximately $62$~nm to $68$~nm for a $5$-kbp DNA segment~\cite{kadam2023predicting}.

Two additional features of the \emph{C. crescentus} cell are relevant to our model and are incorporated into the chromosome simulations. First, in \emph{C. crescentus}, the PopZ protein is localized at both cell poles. The {\em oriC} proximal \emph{parS}/ParB protein complex is tethered to PopZ at one pole prior to the initiation of DNA replication \cite{bowman2010caulobacter}. Following replication,  the \emph{oriC} segment of one daughter remains tethered to the old pole, while the other is initially free and becomes tethered to the new pole upon reaching it. Second, the ParA concentration exhibits an approximately exponential gradient along the long axis of the cell, with the highest concentration near the new pole and a substantially lower concentration toward the cell center~\cite{lim2014evidence,shebelut2010caulobacter}.

 The average doubling time of {\em C. crescentus} is $146 \pm 5$ min in M2G minimal medium~\cite{campos2014constant}, whereas in nutrient-rich PYE medium it is  $96 \pm 1$ min~\cite{hottes2004transcriptional}. Although the doubling time in M2G medium varies from $120 \pm1.8$ to $157 \pm 5.4$ min depending on the \emph{C. crescentus} strains \cite{boutte2008genetic}.  The division of the mother cell leads to the formation of two phenotypically different daughter cells; one of the daughters is a swarmer cell, and the other is a stalked cell~\cite{jensen2002dynamic,ryan2003temporal,degnen1972chromosome}.  A new round of replication starts soon after the new stalked cell is born. However, the swarmer cells have to first be converted into a stalked cell before the replication process can begin.
\section{Model} \label{sec:develop}

In this section, we provide an overview of our modeling framework, focusing on the minimal ingredients needed to understand the results. Details of the simulation protocol and parameter values are provided in the Supplementary Information (SI--I). We model the chromosome as a bead--spring ring polymer, with each bead representing approximately $4$ kbp of DNA (together with its associated proteins), such that the entire chromosome is represented by $1000$ beads. We construct a model of a replicating chromosome that incorporates the actions of proteins involved in specific cellular processes. Below, we describe the essential features of the model, which incorporates the following active processes in an effective manner: (i) elongation of the cell; (ii) replication of the mother chromosome into two daughter chromosomes; (iii) release of topological constraints associated with topoisomerase-IV activity; (iv) binding of ParA to the monomers on chromosome, while maintaining ParA concentration gradient with its maximum near the new pole; (v) transient association of ParA and ParB, resulting in intermittent topological modifications of the chromosome;  (vi) tethering of the \emph{oriC} loci to the poles of the cylindrical cell. 

We describe each of these processes and interactions separately below, in an itemized form.
Each of these processes is incorporated in a coarse-grained, effective manner rather than explicitly modeling the detailed molecular interactions, biochemical reactions, and associated energy consumption that underlie these processes in the biological cell. Instead, we represent their relevant physical consequences through effective interactions and dynamical rules. This allows us to retain the essential nonequilibrium effects of the molecular machinery while avoiding the complexity associated with explicitly resolving its molecular details. We refer to the replicating ring-polymer model incorporating all the features described above as {\em Model-1}. 

However, the contact maps obtained from simulations using {\em Model-1} do not exhibit the characteristic cross-diagonal features observed in the experimental Hi-C map of \emph{C. crescentus} ~\cite{Harju2025.04.03.647011,le2013high}. 
The cross diagonals indicate a parallel arrangement of the left and right chromosome arms along the longitudinal axis of the cell. To obtain a coarse-grained contact map consistent with the experimental Hi-C map, we additionally incorporate chromosome extrusion into the model. Extrusion begins near the {\em oriC}. At the end of the extrusion process, it introduces topological modifications through additional cross-links (CLs) between specific monomers. The locations of these cross-links in our $1000$-monomer bead--spring model are shown schematically in Fig.~\ref{fig:fullreplicationscheme}(a). The positions of the two cross-links are motivated by our previous work~\cite{mitra2022topology}, in which we considered a simpler, non-replicating chromosome model without the ParABS system. We refer to the model incorporating these additional features as {\em Model-2}. We emphasize {\em Model-1} considers a ring polymer without the topological modifications shown in Fig.~\ref{fig:fullreplicationscheme}(a).

We model the \emph{C. crescentus} cell as a cylinder within which the chromosome is confined. The length of the
cylinder is doubled over the course of the simulation. This is implemented in the simulation to mimic the elongation of the cell, which doubles its length without increasing its width even as replication proceeds inside it. 
We do not model cell division. We now systematically describe the active and diffusive processes included in the model. 

\begin{enumerate}
\item We modeled the mother chromosome as a coarse-grained bead--spring ring polymer comprising $N=1000$ monomers confined within a cylinder of aspect ratio $3.5$. The cylindrical confinement mimics the spatial constraints imposed by the cell boundary. For simplicity, we neglect the spherical end caps, the nucleoid, and the crescent-shaped geometry of the cell. 


The mean distance between adjacent beads along the chain contour is $a$. We set $a$ to be the unit of length in the simulations, the thermal energy $k_BT$ as the unit of energy, and the mass $m$ of the bead as the unit of mass. Thus, we report all other lengths, energies, and masses in units of $a$, $k_BT$, and $m$ respectively. The unit of time is then fixed at  $\tau= \sqrt{m a^2/k_BT}=1$, since we 
use $a=1$, $m=1$ and $k_BT=1$ for our simulations. The friction coefficient, i.e., $\gamma$ of the Langevin simulation is chosen to be $\gamma = m/\tau$. Each spherical bead of diameter $\sigma=0.8a$ represents a DNA segment of $4$ kilobase pairs (kbp). We use Langevin dynamics to update the coordinates of monomers, using  the implementation of HOOMD-Blue~\cite{ANDERSON2020109363}, with a time step of $dt=0.01 \tau$. 

\begin{figure}[htbp]
\centering
    \includegraphics[width=0.73\linewidth]{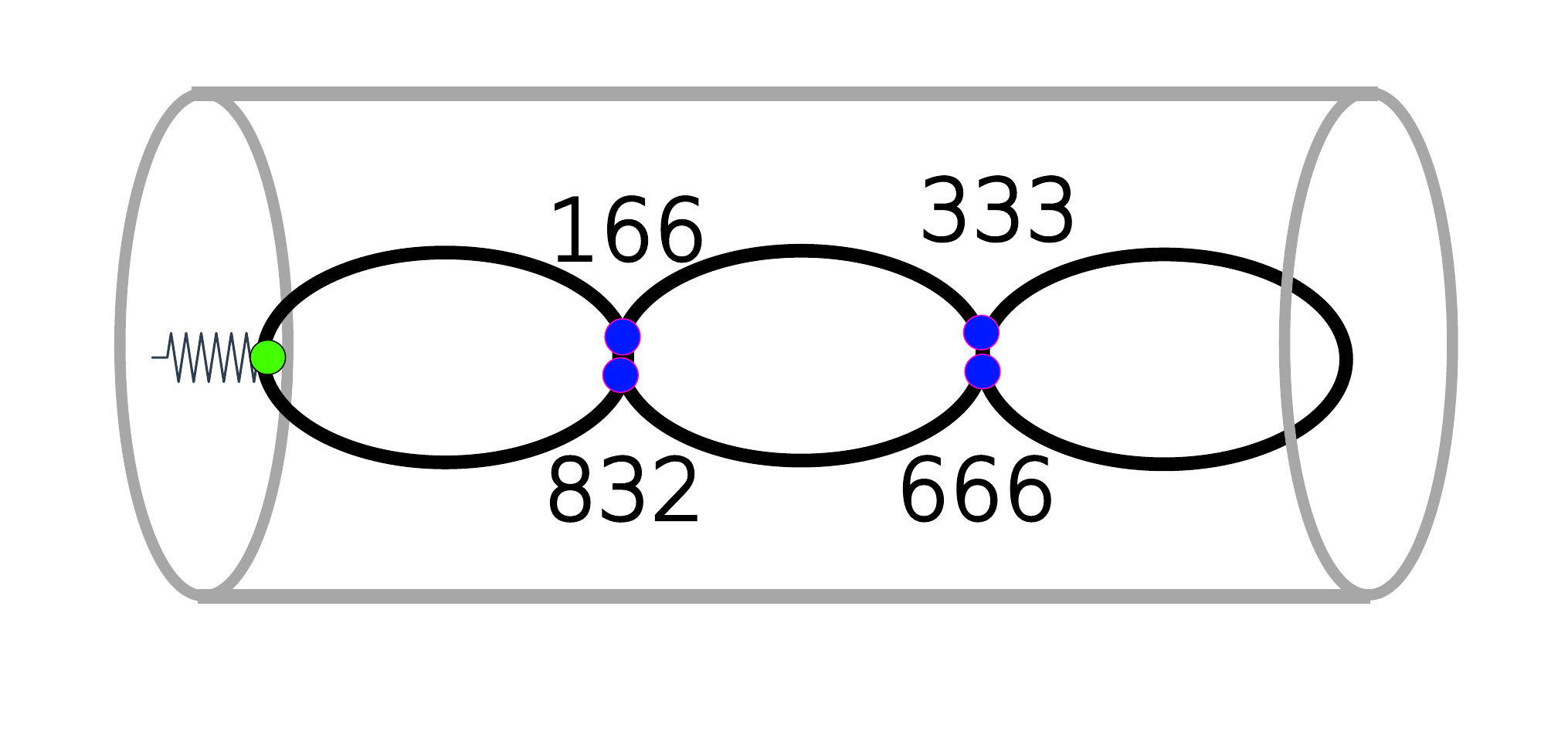}  
    \put(-190,78){\large\textbf{(a)}}\\
    \includegraphics[width=0.75\linewidth]{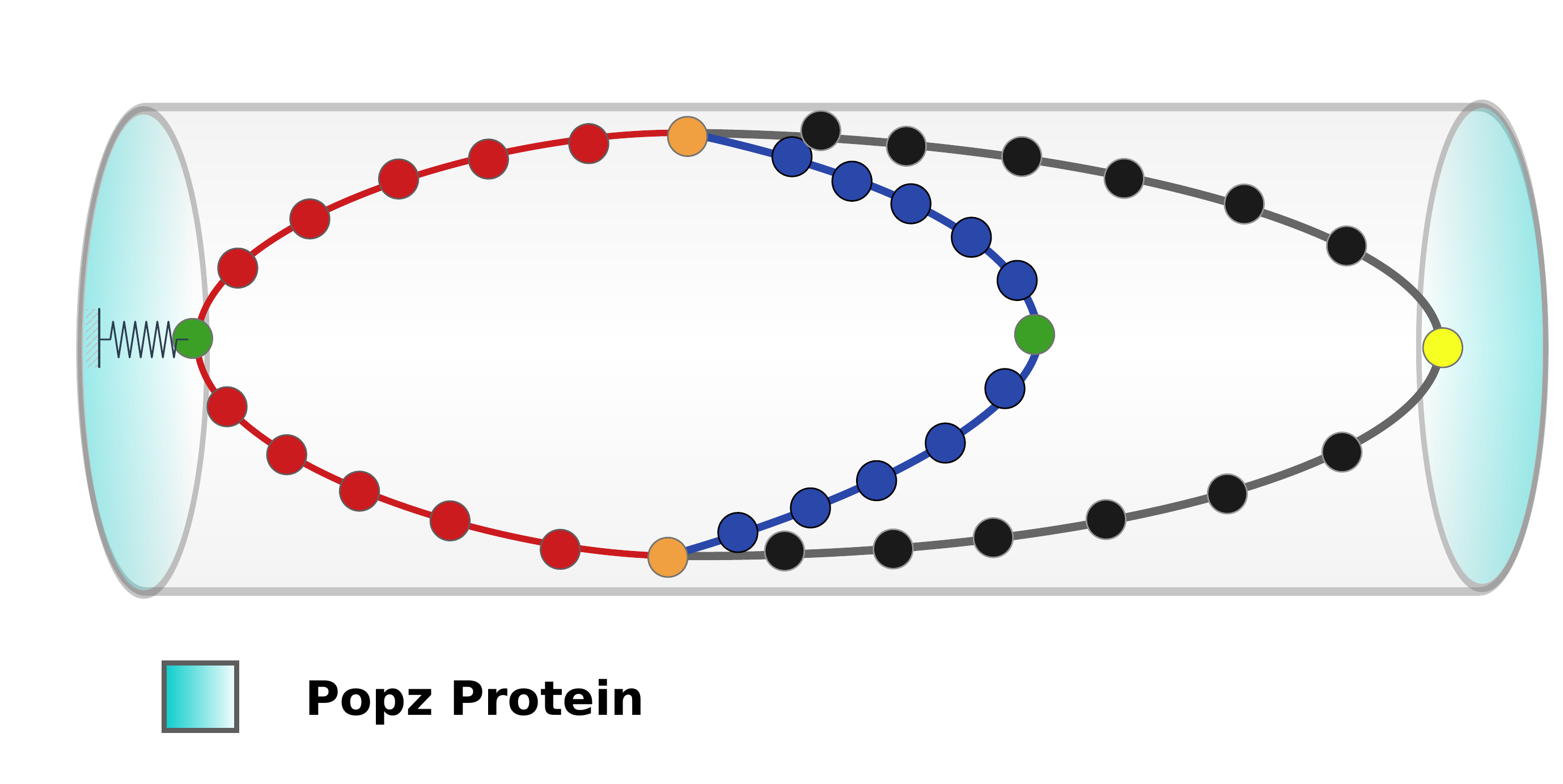}
    \put(-190,78){\large\textbf{(b)}}\\
    \includegraphics[width=0.9\linewidth]{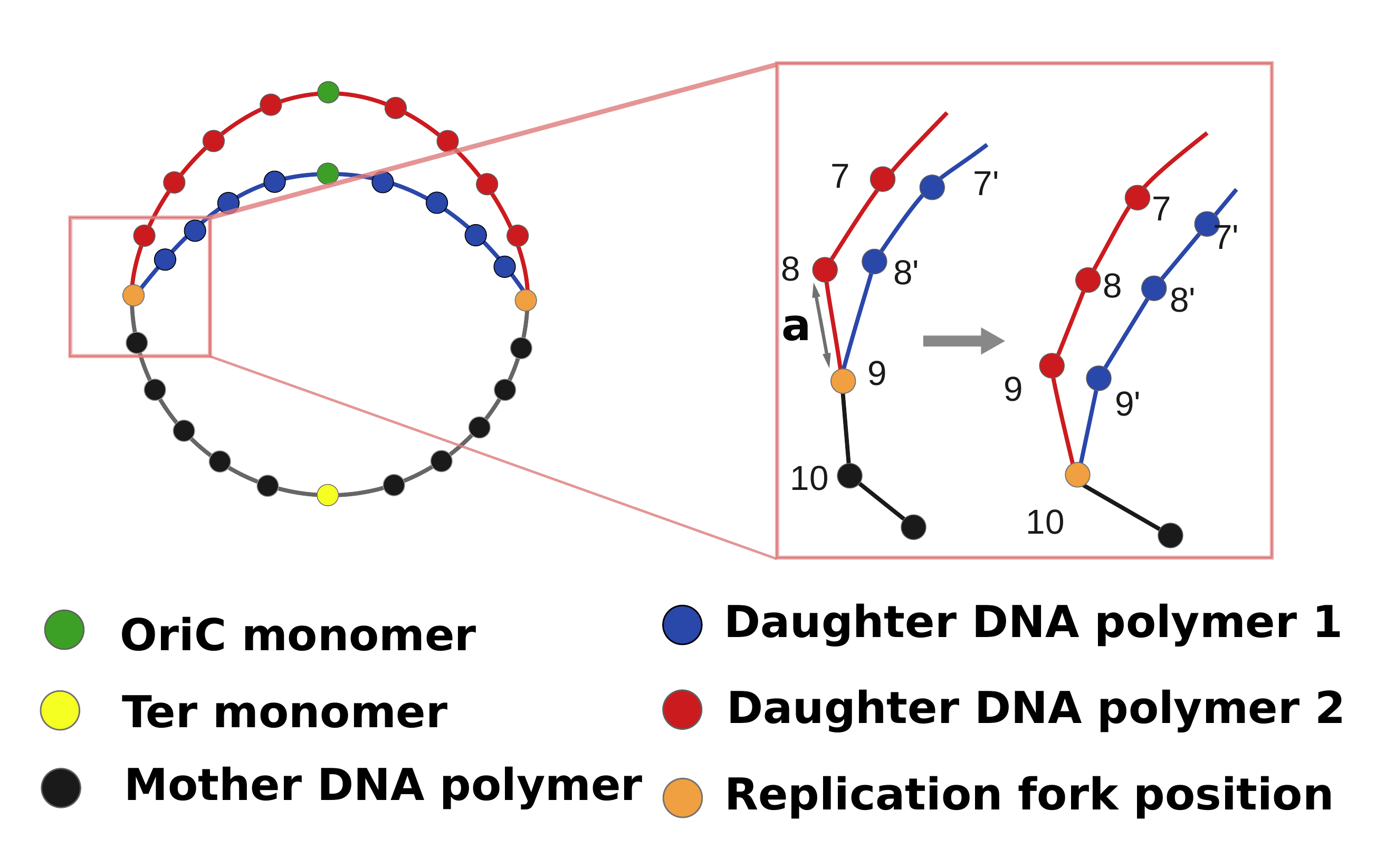}
    \put(-210,120){\large\textbf{(c)}}
    \caption{\justifying 
    \textbf{(a)}  For {\em Model-2}, we schematically show the polymer topology used for the {\em C. crescentus} chromosome at the initialization stage, i.e., at the onset of replication. The topology contains cross-links between specific monomer pairs, which are modeled using harmonic springs. At the end of replication, the two daughter chromosomes have the same topology. The procedure used to introduce the cross-links in the daughter chromosomes is described in the text.
    \textbf{(b)} We show schematically a partially replicated mother chromosome with daughters DNA-polymer-$1$ and DNA-polymer-$2$, hereafter referred to as DNA-1 and DNA-2.  Replication starts from the \emph{oriC}, which is tethered to the old pole using a spring mimicking the effect of the PopZ protein complex, whereas the sister \emph{oriC} gets transported toward the new pole as replication progresses and gets tethered there. The partially replicated mother DNA polymer is shown by black beads, and the two daughter-DNA polymers are shown by blue and red beads, respectively.  \textbf{(c)} A schematic of the method by which we model replication of a bead-spring mother DNA polymer. Monomers are added at the two points on the chain contour marked as the Replication Forks (RFs)  to mimic replication in our simulations. On the right, we show a zoom of one of the RFs at the $9-$th monomer of the mother DNA attached to two daughter DNAs.   Monomer-$9'$ is added as the RF moves to the $10-$th monomer of the mother DNA. Refer to the text and SI--I, subsections A-C, for details.}
    \label{fig:fullreplicationscheme}
\end{figure}

To incorporate the action of protein topoisomerase IV, we employed soft repulsive interaction potentials between the beads of the coarse-grained polymers if the 
distance between the two monomers is less than $\sigma$. The use of a soft potential permits the release of topological constraints by allowing polymer segments to cross one another while maintaining excluded volume repulsion. It also provides an effective description of DNA segments decorated with NAP (nucleoid-associated proteins) and SMC proteins, where transient partial interpenetration and overlap of beads (DNA segments) can occur. Refer to SI--I, subsection-A, for the details of the potentials and parameters.

\item We designate the monomer with index $1$ as the \emph{oriC}, corresponding to origin of replication, and the monomer $N/2$ is designated as the \emph{ter}, corresponding to the replication termination site. We modeled replication as a process in which a polymer chain (mother DNA) containing $N$ monomers gives rise to two polymer chains (daughter DNAs), each consisting of $N$ monomers, by the end of the simulation. Refer to Fig.~\ref{fig:fullreplicationscheme}(b) and  Fig.~\ref{fig:fullreplicationscheme}(c) for a schematic of the replication process.

 The two replication forks (RFs) start out from the \emph{oriC} position of the mother DNA polymer and propagate along the two arms of the mother DNA polymer to meet at the \ emph {ter} monomer at the completion of replication. The monomers of the mother DNA polymer are numbered $1$ to $1000$, with the $500$-th monomer is labelled as the  {\em ter} monomer. After the mother DNA (of $1000$ monomers) has been completely replicated, we have two $1000$-monomer chains, each with its own \emph{oriC} and \emph{ter}. The two \emph{ter} monomers remain connected by a spring till the end of the simulations. Further details are in SI--I, subsection-B.

\item To mimic the effect of the PopZ protein complex, which anchors the \emph{oriC} region at a cell pole, we tether one of the \emph{oriC} monomers to one end of the cylindrical confinement using a harmonic spring, as shown schematically in Fig.~\ref{fig:fullreplicationscheme}(b). We refer to this as the old pole, and the other pole at the opposite end of the cylinder is named as the new pole. Once the second replicated \emph{oriC} segregates to the new pole, we tether it to the new pole, mimicking the effect of the PopZ protein complex at the new pole. Further details can be found in SI--I, subsection-C.

\item Simultaneously, the length of the confining cylinder was increased exponentially such that it doubled by the end of the replication process. Following \cite{Harju2025.04.03.647011}, we have estimated  the diameter of one bead to be around $\sigma = 0.8a \equiv 65nm$.  The cylinder radius was kept fixed at $8a =0.65 \mu m$, throughout the simulation.  The length $L_\mathrm{cyl}$ of the cylinder was increased from  $28a \equiv 2.3 \mu$m to $56a \equiv 4.6 \mu m$ of the course of the simulation.  Refer to SI--I, subsection-D, for other details. 

\begin{figure}[htbp]
    \centering
    \includegraphics[width=0.9\linewidth]{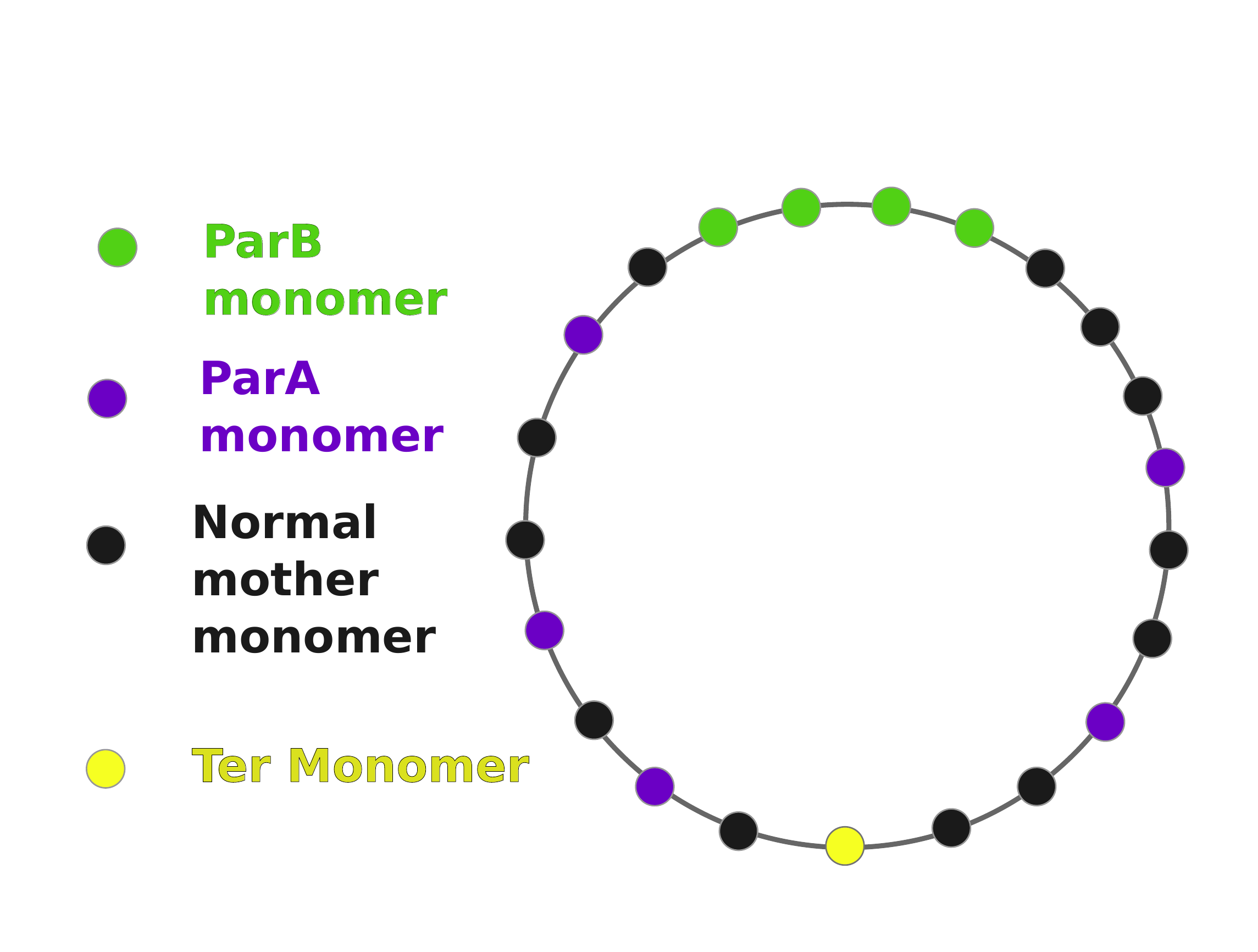}
    \caption{\justifying The schematic shows the locations of the coarse-grained monomers representing ParB/\emph{parS} complexes and the monomers associated with ParA proteins along the DNA polymer. Each bead represents 4kbp of  DNA segments. ParA proteins are loaded onto randomly selected monomers along the chain contour, subject to a prescribed density distribution along the cylinder axis (see the main text and SI--I for details). Monomers associated with ParA proteins are hereafter referred to as \emph{ParA monomers}. Four monomers, including the \emph{oriC} monomer, represent monomers with the ParB/\emph{parS} complex and are referred to as \emph{ParB monomers}. When a ParA monomer and a ParB monomer come into close spatial proximity through diffusive motion, they interact via an attractive potential that persists thereafter for a maximum duration of $ n_P$ iterations, i.e., $100\tau$. Since the times of association of ParA and ParB proteins are not unambiguously known, we have also provided data where the value of $n_P$ is varied systematically. The attraction is switched off thereafter, as it is observed that ParA proteins get deactivated and fall off ParB/\emph{parS} sites. A random site on the DNA contour is identified as the new ParA site. 
    }
    \label{fig:ParABschem}
\end{figure}

\item In addition, we incorporate the effects of the ParABS protein system. Specifically, two monomers, corresponding to $\approx 8$~kbp of DNA and located proximal to and on either side of \emph{oriC}, are designated as \emph{parS} sites. These four monomers are endowed with specific interaction properties to represent ParB proteins loaded onto the \emph{parS} sites. For convenience, we refer to these monomers as ParB monomers. The ParB monomers attract a second class of monomers, which we refer to as ParA monomers. 

The ParA monomers represent DNA segments loaded with ParA proteins; see Fig.~\ref{fig:ParABschem} for a schematic representation. Upon replication of the \emph{oriC}-proximal region of the mother DNA, the corresponding \emph{parS} sites are replicated, resulting in four ParB monomers on each daughter DNA polymer. The total number of ParA monomers, $n_A$, is kept fixed throughout a simulation, even as replication progressively doubles the number of DNA monomers. We perform independent simulations for different values of $n_A$ to investigate its influence on chromosome segregation.  Unless stated otherwise, we use $n_A=100$ for the results presented below. Further details are provided in SI--I, subsection-E.

\item Since ParA binds nonspecifically to DNA, ParA monomers are assigned randomly along the polymer contour according to an exponentially decaying distribution, with the ParA density decreasing with distance from the new pole. Thus, DNA segments closer to the new pole have a higher density of ParA monomers, representing the spatial distribution of ParA-loaded DNA. Further details are in SI--I, subsection-E.

\begin{figure}[b]
    \centering
    \includegraphics[width=0.9\linewidth]{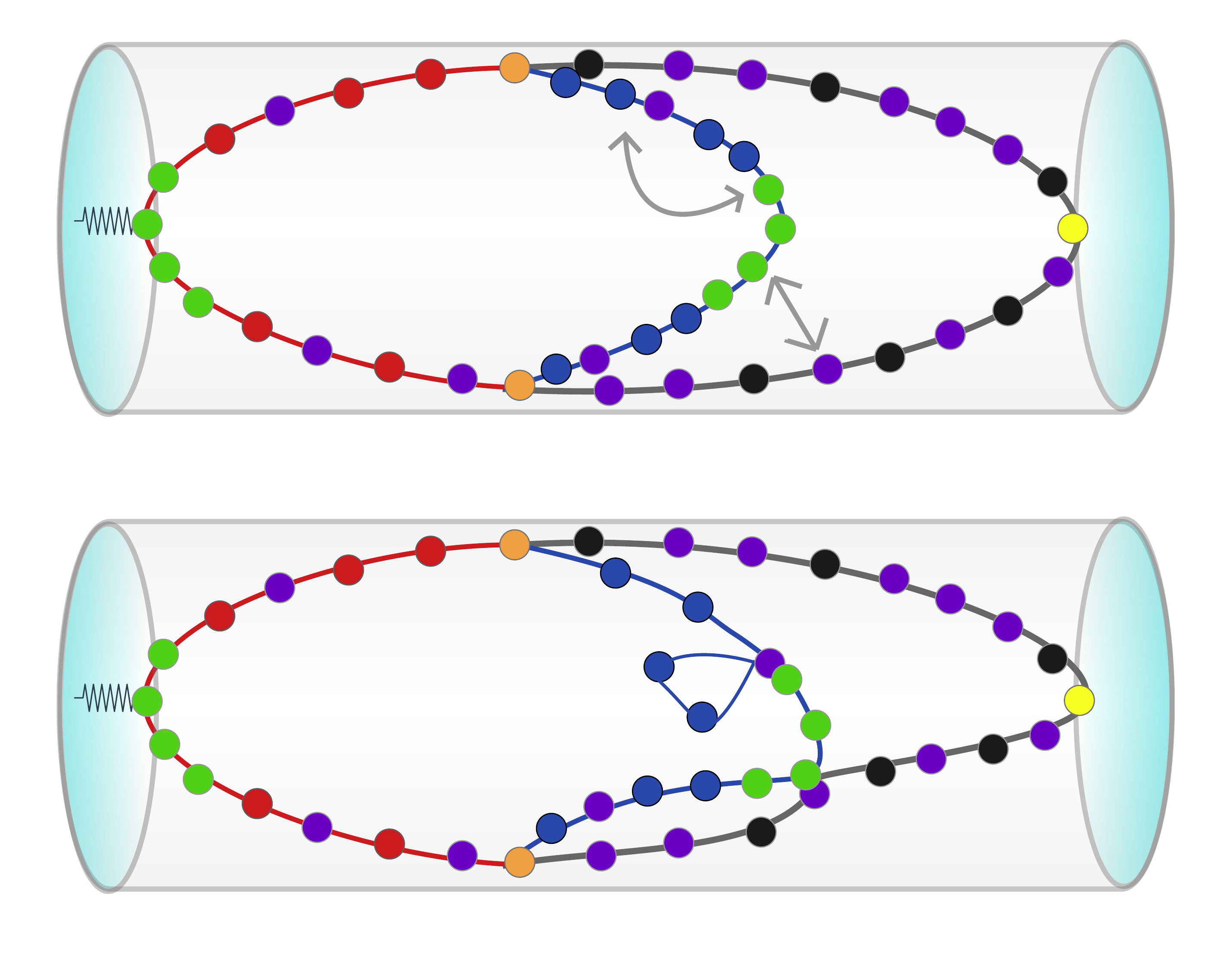}
    \caption{Schematics to demonstrate consequences of attraction between ParA and ParB 
    monomers when they are spatially proximal, even as replication is in progress in the cylinder.  Internal loops are formed as long as ParA and ParB monomers 
    remain conjoined due to transient attraction introduced between them.
    The color schemes for denoting the monomers are consistent with that marked in Figs.~\ref{fig:fullreplicationscheme} and Figs.~\ref{fig:ParABschem}. Since the {\em oriC} of DNA-1 is tethered to the new pole, loops are more likely to be on DNA-2,  as the monomers of DNA-2 diffusively move away from the old pole. The formation of transient loops increases the entropic repulsion between DNA-1 and DNA-2, and facilitates the transport of {\em oriC} of DNA-2 towards the new pole.}
    \label{fig:ParABschem1}
\end{figure}

\item ParA and ParB monomers can come into proximity through the diffusive motion of the DNA polymers. When the separation between a ParA and a ParB monomer becomes smaller than $r_c^P = 1.5a$ (unless stated otherwise), they interact through a short-ranged attractive potential that represents the effective binding of ParA to ParB monomers. The interaction persists for a stochastic duration (details below), representing the transient nature of ParA--ParB binding and its subsequent dissociation. The binding of a ParA monomer to a ParB monomer effectively generates a loop in the DNA polymer. Since the ParB monomers are located proximal to \emph{oriC}, the resulting transient loops are formed in the vicinity of the \emph{oriC} region. A schematic representation of this transient loop formation mediated by the ParABS system is shown in Fig.~\ref{fig:ParABschem1}. Further details are in SI--I, subsection-E.

\item After a fixed number of iterations, $n_P$, where  $n_{P}=10000 \equiv 100\tau$ (unless stated otherwise), we check for the  existing number of pairs of ParA--ParB monomers. Thereafter, we switch off the attractive interaction between the corresponding ParA-–ParB pairs, such that the  ParA monomer(s) becomes an ordinary monomer of the chain with only EV interaction with other monomers. This mimics the stochastic dissociation of ParA from ParB following ATP hydrolysis and the consequent loss of ParA activity. We randomly select another monomer to be a new ParA monomer, while maintaining the concentration gradient of ParA. The duration of ParA--ParB binding is stochastic and is determined by the time at which the distance between the interacting pair first becomes less than $r_c^P = 1.5a$.

We first obtained results using the modeling aspects described above.
This model captures several features of chromosome organization.
 We refer to this model as \emph{Model-1}. The mother and daughter chromosomes have ring-polymer topologies, except for the transient loop formation induced by the ParABS system. 

\item In \emph{Model-2}, we additionally incorporate loop extrusion mediated by SMC proteins to generate specific topological modifications of the daughter chromosomes during replication. In our coarse-grained model of an extruder protein, we link two monomers which are at equal distance along the chain contour from the {\em oriC}, and situated on the left arm and right arm of the newly replicated daughter DNA polymer(s), respectively. 
The extruder essentially holds two monomers on either side of {\em oriC} 
from the two arms like a handcuff. We introduce two symmetric (two-sided) extruders, each of which pulls monomers on both arms to create a loop(s) which grows bigger as the extruders move away from the {\em oriC} at a speed which is half that of the speed of RFs.  The two extruders stop moving only when they reach specific monomers, such that the topology shown in Fig.~\ref{fig:fullreplicationscheme}(a) is attained by the daughter DNA-1  and DNA-2.

Simulations with {\em Model-2} retain the directed transport of {\em oriC} of DNA-2 to new pole along the long axis of the cell obtained with {\em Model-1}, while also reproducing key features of the experimentally measured Hi-C contact maps. Before the onset of replication, the mother chromosome is assumed to adopt the same specific topology as shown in Fig.~\ref{fig:fullreplicationscheme}(a).  At the onset of replication, these cross-links on the mother DNA polymer are removed by switching off the corresponding spring potentials between the specific monomer pairs. The cross-links are subsequently reintroduced on the daughter DNA polymers through the extrusion process during replication. Refer to SI--I subsection-F for details of how we implement loop extrusion in our {\em Model-2}.


\end{enumerate}

Extrusion is the dynamical mechanism that establishes/maintains the topology during replication, whereas the topology itself is what produces the double-diagonal  structure in the contact map as seen in Hi-C maps. Despite its simplicity, this model reproduced the localization of chromosomal loci observed in FISH experiments~\cite{viollier2004rapid} as can be verified 
from  Fig.~S$4$(a) of SI--I, and consistent with our previous work with non-replicating polymers 
\cite{mitra2022topology}.

In our simulations, we also compare the results obtained if the modeling of the effects of the ParABS system or the extrusion process is switched off. This helps us identify the roles of the various processes. Although we work with a bead-spring polymer model, incorporation of the effects listed above renders our simulation to be very different from a standard equilibrium polymer simulation.
The reasons why we can invoke entropy despite modeling a nonequilibrium 
system such as the living \emph{C. crescentus} cell is discussed at the end of the  Introduction section. We now describe the results obtained from our simulations.

\section{Results} \label{Outcomes}
Before the production runs for the trajectories of specific loci, we equilibrate the mother chromosome within the cylindrical cell geometry for $10^6\tau$ using Langevin dynamics, with monomer-$1$ (\emph{oriC}) tethered to one pole of the cylinder, designated as the old pole. From the final equilibrated configuration, we generate $120$ independent initial configurations by continuing the dynamics for $400\tau$ with independent random-number seeds. The resulting configurations therefore differ in their molecular conformations. For each of these $120$ configurations, $n_A$ monomers are then selected as ParA monomers according to the procedure described above. Because the ParA monomers are selected independently in each realization, the resulting configurations provide distinct initial conditions for the subsequent nonequilibrium dynamics.

The production run is initiated from each of these $120$ configurations by starting chromosome replication at \emph{oriC}. During replication, the cylinder length is gradually increased to mimic cell growth. The positions of all monomers are recorded every $50\tau$ for subsequent statistical analysis of the dynamics of the DNA polymers.
Thus, we perform $120$ independent non-equilibrium runs using independent initial configurations of the mother chromosome. We deem a run as `successful', if one of the daughter {\em oriC}  successfully reaches the new pole of the cylinder from the old pole.  We use the resulting trajectories from `successful' runs to calculate statistical averages. We then directly compare the trajectories of multiple chromosomal loci obtained from the simulations with experimentally measured trajectories of the corresponding fluorescently tagged loci.

We performed simulations of chromosome replication and obtained trajectories of \emph{oriC}, the replisome, and several other fluorescently tagged loci using both \emph{Model-1} (without loop extrusion
and corresponding topological modifications) and \emph{Model-2}. The two models yield quantitatively similar locus trajectories, both of which are in good agreement with experimental observations. These results demonstrate that ParABS-mediated, entropy-driven transport of \emph{oriC} is sufficient to reproduce the experimentally observed dynamics of chromosomal loci. The directed transport of one of the \emph{oriC} loci from the old pole toward the new pole emerges naturally from the model and does not require the application of an external force. Thus, pole-to-pole segregation of \emph{oriC} is therefore an emergent consequence of chromosome dynamics coupled to ParABS-mediated interactions.

A distinction between the two models emerges, however, when chromosome organization is examined through contact maps. Comparison with experimentally measured Hi-C contact maps shows that \emph{Model-2} yields substantially better agreement with experiment than \emph{Model-1}. Thus, while loop extrusion is not required to reproduce the observed locus trajectories, it plays a crucial role in establishing the large-scale chromosome organization captured by Hi-C measurements.

The \emph{oriC} trajectories obtained using Model-2, together with their comparison to experimental data, are presented in the main manuscript. For completeness, the corresponding results from \emph{Model-1} are provided in SI-II, enabling a direct comparison between the predictions of the two models. 

\subsection{\emph{oriC} Trajectory.}

\begin{figure}[htbp]
    \includegraphics[width=\linewidth]{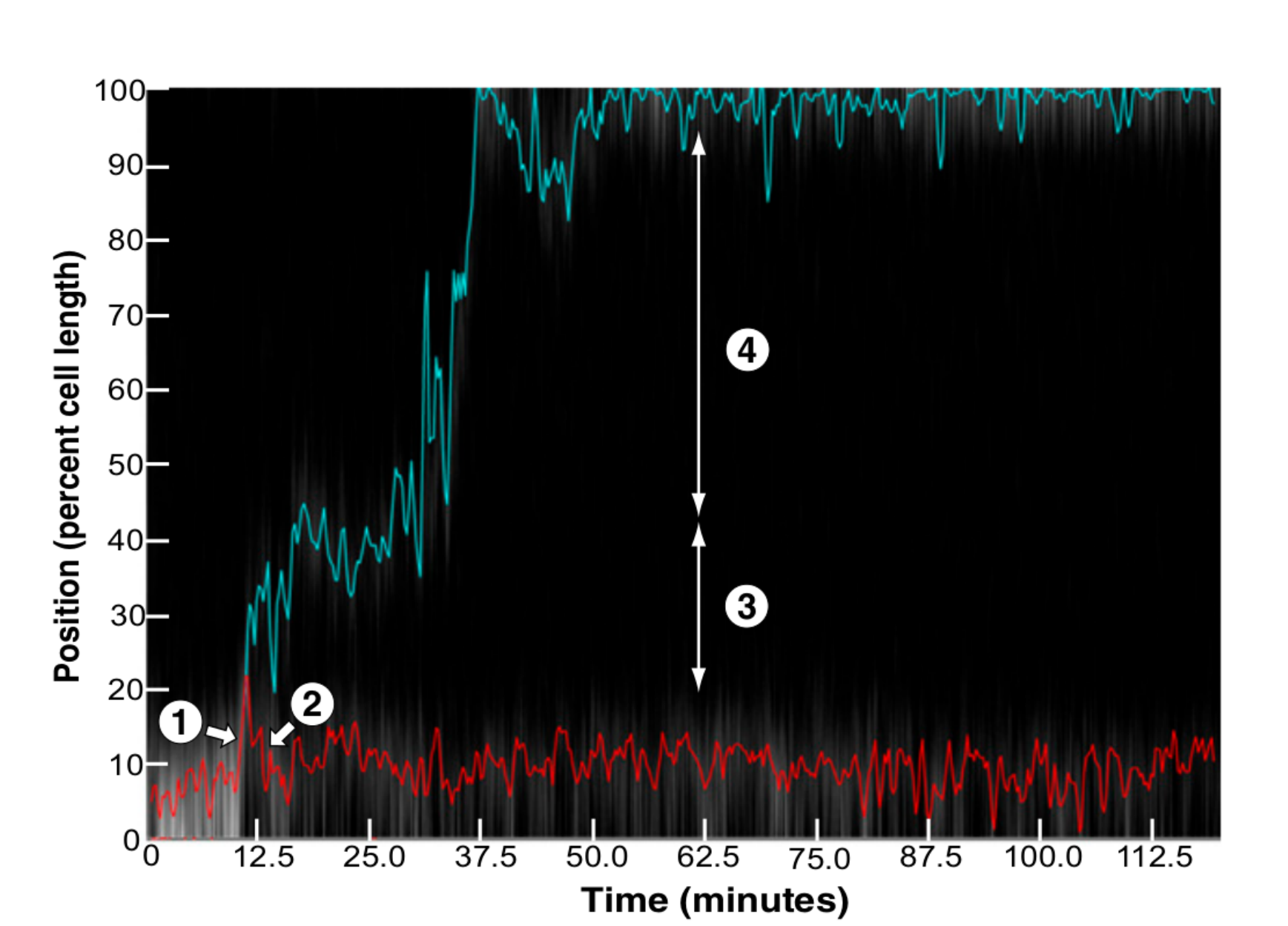}
    \put(-205,142){\color{white}\large\textbf{(a)}}
    \hfill
    \includegraphics[width=\linewidth]{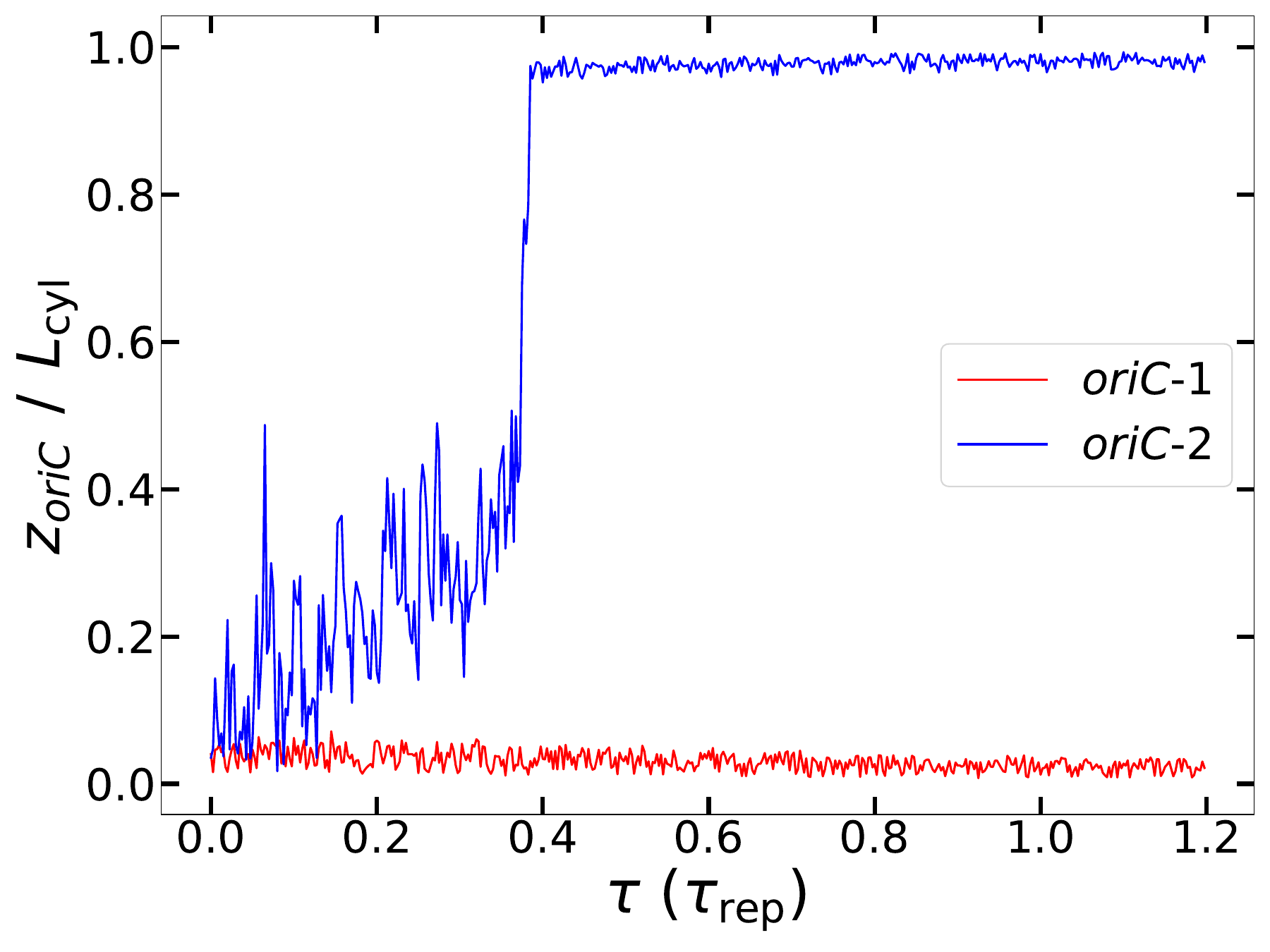}
    \put(-205,160){\large\textbf{(b)}}
    \caption{ {\bf oriC trajectory from old to new pole of the cell.} \textbf{(a)} Experimentally measured trajectories of GFP-tagged ParB/\emph{parS} complexes in a replicating \emph{C.\ crescentus} cell, reproduced from Ref.~\cite{shebelut2010caulobacter} with due permission. 
    The fluorescent ParB/{\em parS} foci serve as proxies for the positions of the {\em oriC} loci. The red and blue curves denote the positions of the two ParB/\emph{parS} complexes along the cell's long axis, normalized by the instantaneous cell length. The time taken for replication (the S-phase) can be taken 
    to be $80$ minutes to facilitate comparison with the trajectory from simulations shown in (b). The data reveal distinct stages of {\em oriC} segregation: (1) release from the old pole, (2) polar retraction, (3) early translocation (slow), and (4) late translocation (fast). \textbf{(b)} Corresponding \emph{oriC} trajectories obtained from simulations. The red and blue curves show the normalized axial positions, $\mathrm{z}_{\emph{oriC}}/L_{\mathrm{cyl}}$, of the two \emph{oriC} loci as functions of time. The simulated trajectories reproduce the characteristic stages of {\em oriC} transport observed experimentally. Note that the width of the fluctuations of $\mathrm{z}$ positions of {\em oriC}-1 decreases as the value of $L_\mathrm{cyl}$ increases with time. By convention, the origin of the figure is taken to be the position of the old pole.
}
\label{fig:Oritraj_sim_exp}
\end{figure}

We begin by comparing the dynamics of \emph{oriC} in a single cell, for which experimental trajectories have been reported by several research groups~\cite{shebelut2010caulobacter,lim2014evidence}. The experimentally measured {\em oriC} trajectory from Ref.~\cite{shebelut2010caulobacter} is reproduced in Fig.~\ref{fig:Oritraj_sim_exp}(a), while the corresponding trajectory obtained from our simulations is shown in Fig.~\ref{fig:Oritraj_sim_exp}(b). This comparison enables a direct assessment of the extent to which our minimal model captures the experimentally observed {\em oriC} dynamics.

In Fig.~\ref{fig:Oritraj_sim_exp}(a), the red and blue curves denote the positions of the two {\em oriC}-proximal ParB/\emph{parS} foci as functions of time measured from the onset of replication, which is defined as $t=0$. Distances are measured relative to the old pole. One of the loci (red curve), hereafter referred to as {\em oriC-1}, remains localized near the old pole throughout the cell cycle. The second locus (blue curve),  becomes experimentally resolvable only after  $\approx 12$ min, as the fluorescent signals from the two loci overlap at earlier times. The time at which the two foci can first be distinguished is referred to as the splitting time. This marks the transition from a single fluorescent focus to two separate foci.

Following the splitting event, the second {\em oriC}, hereafter denoted \emph{oriC-2}, undergoes directed motion toward the opposite pole, although its trajectory exhibits a brief plateau extending to approximately $30$ min. Under the growth conditions considered here, the replication time of {\em C.\ crescentus} typically lies between $70$ and $90$ min and depends only weakly on whether cells are grown in minimal (M2G) or nutrient-rich (PYE) medium \cite{li2008quantitative}. For definiteness, we adopt a replication time of $80$ min, corresponding to the S-phase of the cell cycle. 

The time required for \emph{oriC-2} to reach the new pole is approximately $37.5$ min when measured from the onset of replication and approximately $25$ min when measured from the splitting event. These times correspond to about $47\%$ and $31\%$ of the replication time, refer Fig.~\ref{fig:Oritraj_sim_exp}(a).

\begin{figure}[htbp]
    \centering
        \includegraphics[width=\linewidth]{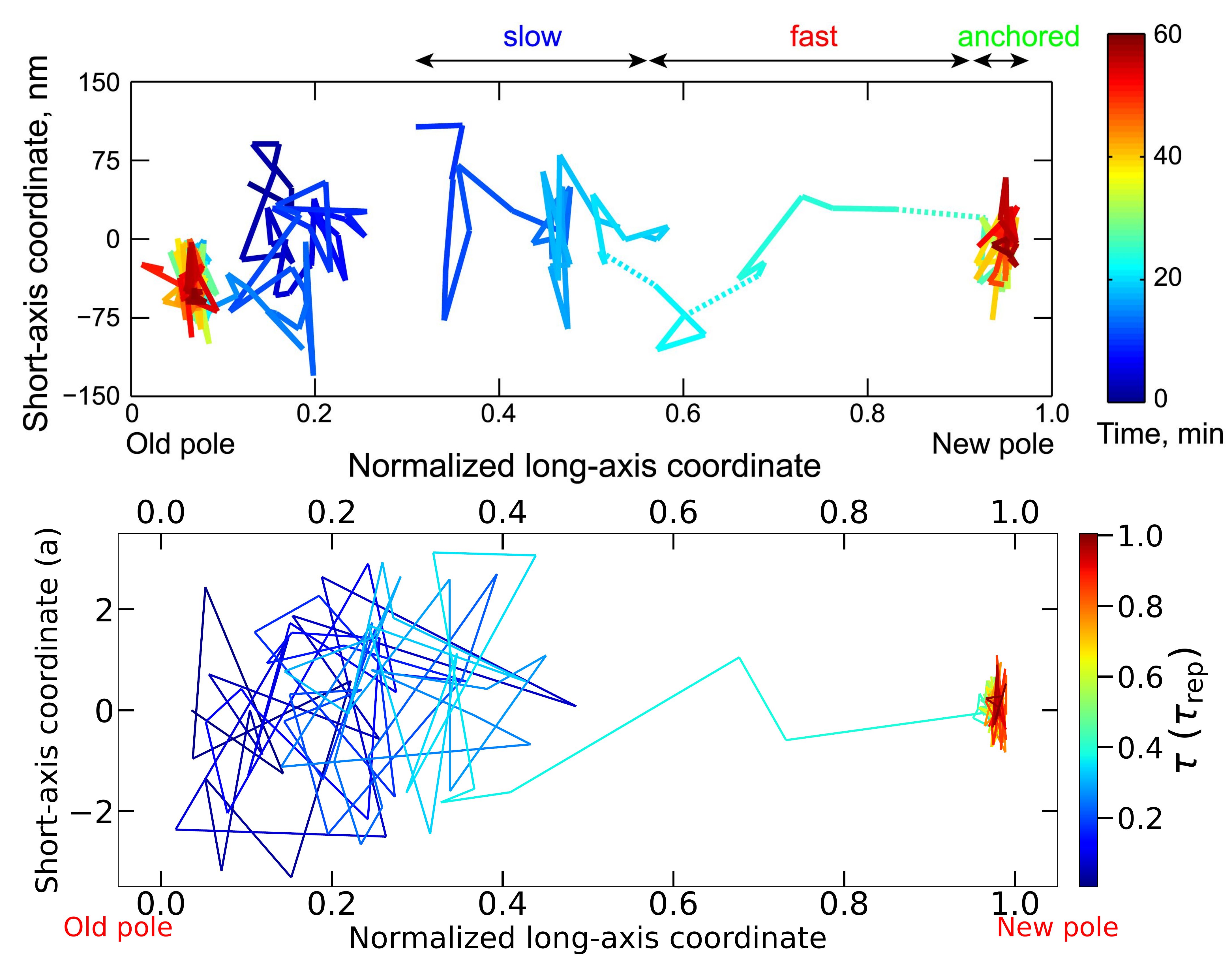}
        \put(-70,160){\large\textbf{(a)}}
        \put(-70,70){\large\textbf{(b)}}
    \hfill
    \caption{ {\bf oriC trajectory from old to new pole of the cell.} \textbf{(a)} Experimental trajectory of ParB as seen inside the {\em C. crescentus} cell. A part of the original figure (suitably cropped) has been reproduced  from 
    Ref. \cite{lim2014evidence}, under the license  \href{https://creativecommons.org/licenses/by/3.0/}{(CC BY 3.0)} © 2014 Lim et al.  The authors have plotted the positions of the observed fluorescently tagged \emph{oriC} loci as seen in the $2-d$ plane.
    A second trajectory appears when the two oriCs can be resolved into two. The different stages of segregation of segregation have been observed as also in Fig.~\ref{fig:Oritraj_sim_exp}(a). \textbf{(b)}: We also have plotted $2-d$ trajectory of \emph{oriC} that we have observed in our simulations. We have only shown the trajectory of the {\em oriC} of  Daughter DNA-2, i.e., the one which is not anchored at 
    the old pole and hence is able to get transported to the new pole. In our simulations, we have the positions of the two {\em oriC}s as soon as the replication begins.} 
\label{fig:2d_image_sim_exp}
\end{figure}

Since the cell length increases during the cell cycle, it is convenient to express the position of \emph{oriC-2} relative to the old pole as a fraction of the instantaneous cell length, $L_\mathrm{cyl}$,
as is shown in Fig.~\ref{fig:Oritraj_sim_exp}(a). We adopt the same convention when presenting the simulation results. However, the simulation time axis is expressed in units of the replication time rather than in physical units because a reliable mapping between simulation time and experimental time is difficult to establish unambiguously, refer to the Appendix for a discussion.

We discuss the quantitative agreements between the time scales obtained from the experiments and simulations (suitably normalized by the replication time $\tau_{rep}$), as shown in Figs.~\ref{fig:Oritraj_sim_exp}(a) and \ref{fig:Oritraj_sim_exp}(b), respectively. The experimental data for the early stages of replication corresponding to the time window between approximately $12$ and $30$ min show that the rate of separation of the two {\em oriC} loci is relatively slow. In this period, the position of \emph{oriC-2} fluctuates around $0.3$--$0.45$ of the cell length. Thereafter, {\em oriC} reaches the new pole within $45\%$ of the time taken to complete replication, $\tau_{rep}$. The same features can be seen in the trajectory obtained from  simulations, shown in  
Fig.~\ref{fig:Oritraj_sim_exp}(b).

The same behavior is evident in Fig.~\ref{fig:2d_image_sim_exp}(a), reproduced from Ref.~\cite{lim2014evidence}, where the trajectory from a single cell is visualized in an alternative representation. In this figure, the horizontal and vertical axes correspond to the spatial dimensions of the cell, and the trajectory of the {\em oriC}-proximal ParB/\emph{parS} focus is shown within this reference frame. The temporal evolution is encoded by the color of the trajectory, with the corresponding times indicated by the color bar.

\begin{figure}
    \centering
    \includegraphics[width=0.9\linewidth]{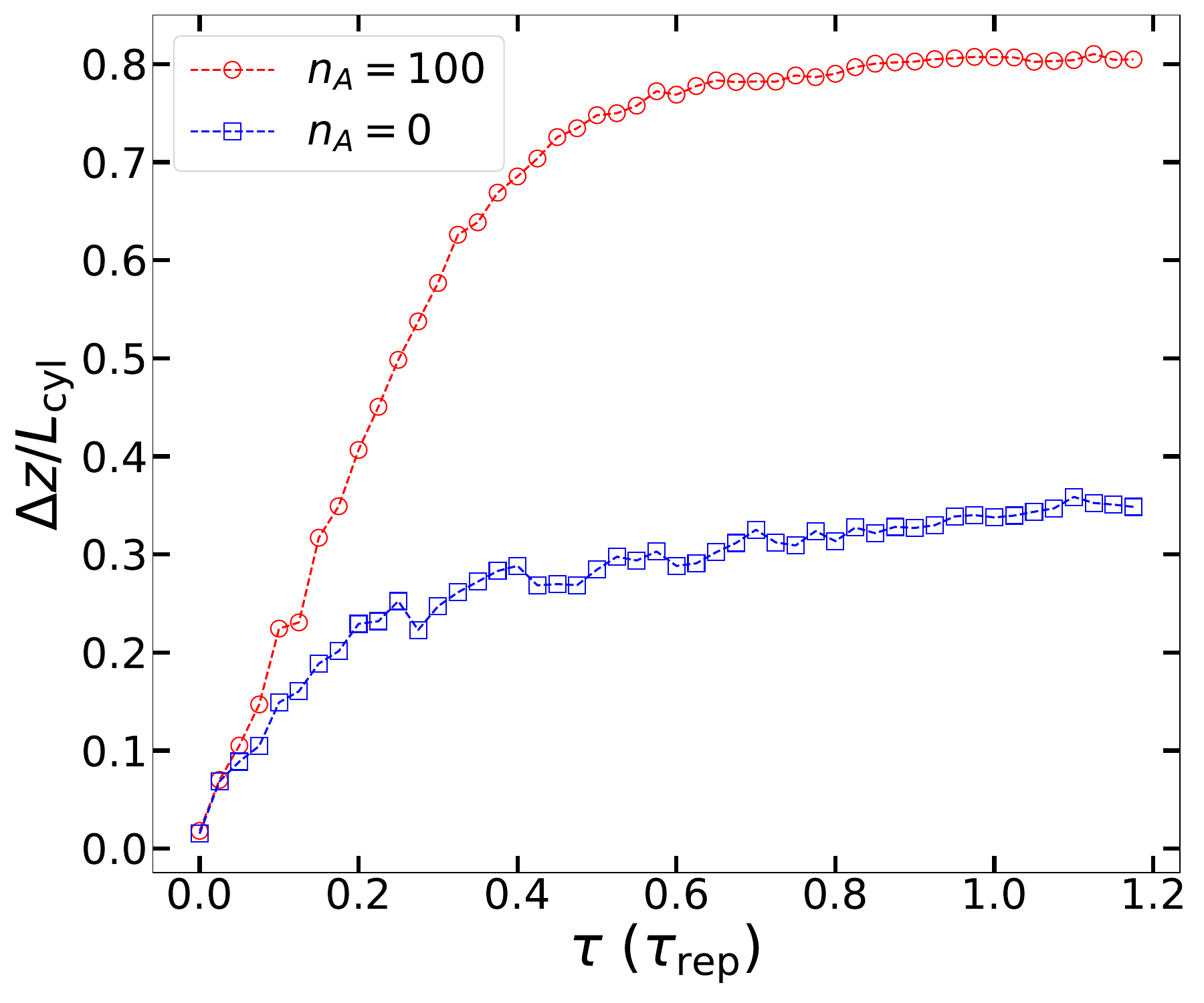}
    \put(-180,135){\large\textbf{(a)}}\\
    \includegraphics[width=\linewidth]{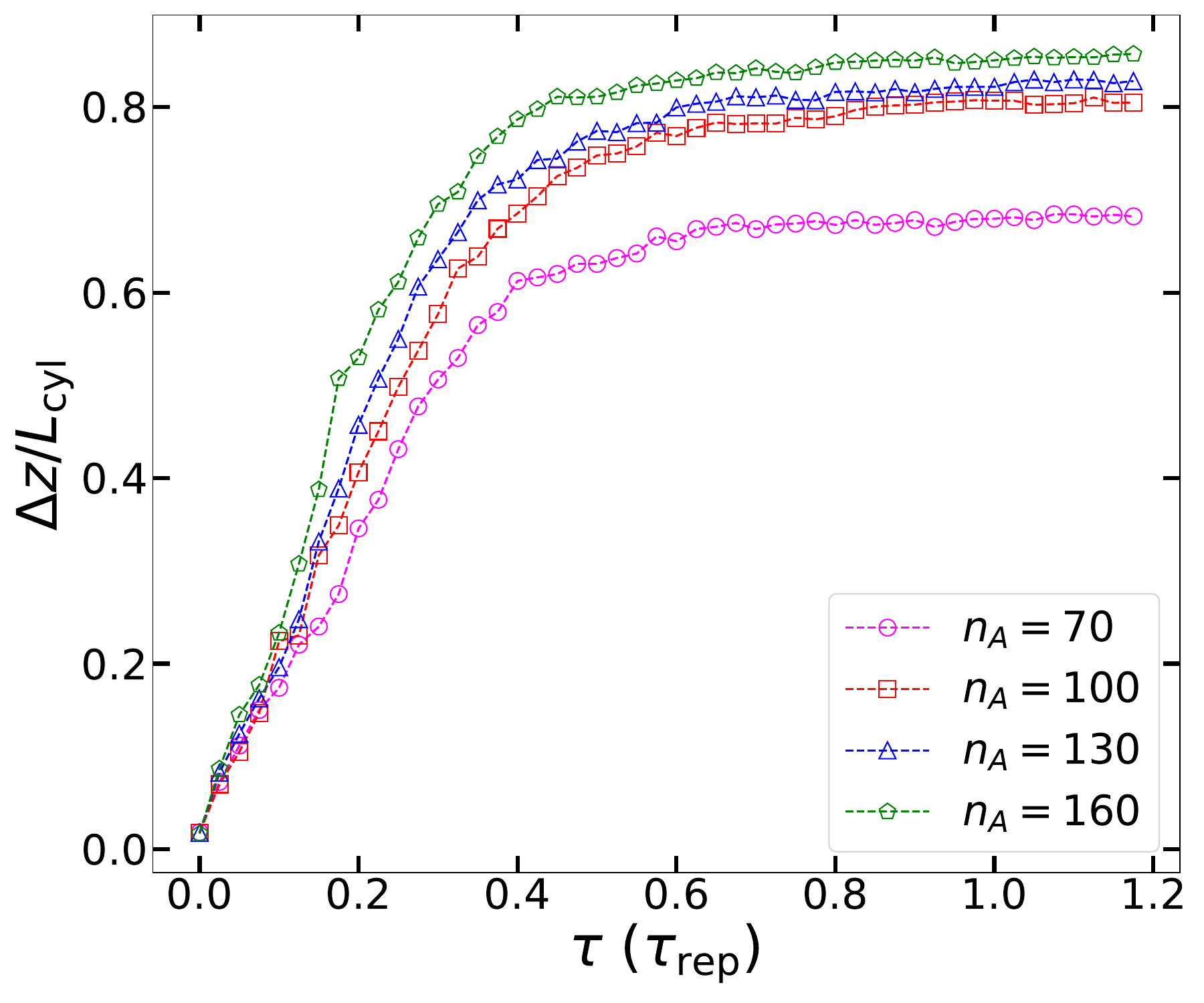}
    \put(-180,150){\large\textbf{(b)}}

    \caption{ {\bf Segregation dynamics for different numbers of  ParA monomers.} \textbf{(a)} Average distance between two replicated \emph{oriCs} ($\Delta \mathrm{z})$ normalized with instantaneous length ($L_{\mathrm{cyl}}$) with 100 number of ParA monomers (Blue) and without ParA (Orange) over time. These distances are averaged over $120$ simulation runs, independent 
    of whether they show `successful' {\em oriC} translocation or not. As shown later, the number 
    of `successful' runs increases with an increase in $n_A$. \textbf{(b)} Average distance between two replicated \emph{oriCs} ($\Delta \mathrm{z})$ normalized with instantaneous length ($L_\mathrm{cyl}$) with different number of ParA with time. These distances are averaged over $120$ runs.  We add monomers every  $20$ thousand iterations, whereas $r_c^P= 1.5a$ and $n_P=10$ thousand iterations for the data presented in this plot.}
    
    \label{fig:withwithoutParA&diffParA}
\end{figure}

In Fig.~\ref{fig:2d_image_sim_exp}(a), one observes that one of the \emph{oriC}, {\em viz.} \emph{oriC}-1 loci remains localized between approximately $0$ and $0.2L_\mathrm{cyl}$ for the entire cell cycle. It is 
germane to note that the trajectory of the other {\em oriC}, {\em viz.}, \emph{oriC}-2 appears at a position $\approx 0.25 L_\mathrm{cyl}$. This reflects the consequences of the experimental protocol rather than the actual onset of motion of \emph{oriC}-2. Specifically, the plotted positions correspond not to the {\em oriC} loci themselves, but to GFP-tagged ParB/\emph{parS} complexes located proximal to {\em oriC}. The trajectory of the distal ParB/\emph{parS} focus is recorded only after the second fluorescent spot becomes distinguishable from the spot associated with \emph{oriC}-1. Consequently, the trajectory spanning the initial stage of  the cell cycle 
(confined to positions below approximately $0.25\ell$), corresponds to the ParB/\emph{parS} complex associated with \emph{oriC}-1, which remains tethered to the old pole for much of the S phase.

In Fig.~\ref{fig:2d_image_sim_exp}(b), we show a representative trajectory obtained from our simulations. A comparison with Fig.~\ref{fig:2d_image_sim_exp}(a) demonstrates the close qualitative agreement between the simulated and experimentally observed {\em oriC} dynamics. In the experimental data, the two trajectories can be resolved only after the splitting event. In contrast, the simulations provide direct access to the positions of both \emph{oriC} loci throughout the cell cycle. For clarity, we therefore plot only the trajectory of \emph{oriC}-2
in Fig.~\ref{fig:2d_image_sim_exp}(b), as we keep \emph{oriC}-1 tethered to the old pole. Following the rapid translocation phase [region 4 in Fig.~\ref{fig:Oritraj_sim_exp}(a)], \emph{oriC}-2 becomes localized near the new pole, consistent with its anchoring by the PopZ protein complex. We model this by suitable attraction between \emph{oriC}-2 and a bead at the new pole. This is observed 
in both Fig.~\ref{fig:Oritraj_sim_exp}(b) and in  Fig.~\ref{fig:2d_image_sim_exp}(b).

Thereby, the simulated trajectory demonstrates that ParA--ParB interactions generate directed translocation of \emph{oriC}-2 toward the new pole. Importantly, this poleward transport emerges naturally from the interaction dynamics and does not require the application of an external driving force.

To elucidate the role of ParABS-mediated interactions arising from topological modifications in the directed transport of \emph{oriC}, we compare the separation between \emph{oriC}-1 and \emph{oriC}-2 during chromosome replication for two systems: \emph{Model-2}, which incorporates transient attractive ParA--ParB interactions, and a corresponding second system in which these interactions are absent. Figure~\ref{fig:withwithoutParA&diffParA}(a) shows the mean separation between the two {\em oriC} loci as a function of time for both cases. 
In the second system, the mother DNA polymer replicates into two daughter rings without the intermediate topological modifications induced by the ParABS proteins. The trajectories shown in Fig.~\ref{fig:withwithoutParA&diffParA}(a) are averaged over $120$ independent simulations and are therefore less noisy than the single-cell trajectories shown earlier.

 For simulations with $Model-2$, we observe that the  mean  separation 
 $\Delta \mathrm{z}$ between the two {oriC}s (suitably normalized by $L_{\mathrm{cyl}}$) increases to values close to 
 $\approx 1$, whereas in the absence of ParABS-mediated interactions, the corresponding value remains $ < 0.5$. Thus, in the absence of ParABS interactions,  the mean separation between the two {\em oriC} loci remains below half the cylinder length throughout the replication cycle. We have checked  that the two daughter DNA  polymers did not segregate in the system without ParA--ParB interactions in many of the runs. In the few runs, where the two daughter DNA polymers did segregate (within the time scale of simulations), the segregation times are much longer, and moreover the {\em oriC}-2 trajectory appears different from the experimentally observed trajectory. As shown in Fig.~S$8$ of SI--III, the \emph{oriC}-2 moves gradually towards the new pole without exhibiting the rapid translocation observed in the previous figures. 
 
In contrast, when ParABS interactions are included, the two {\em oriC} loci undergo progressive segregation, with their mean separation reaches approximately $0.8L_\mathrm{cyl}$ by time $0.4\tau_{rep}$. These results demonstrate that ParABS-mediated interactions are essential for the efficient pole-to-pole transport and segregation of \emph{oriC}. The underlying entropic mechanism by which smaller internal loops facilitate the rapid segregation of two polymers has been investigated previously by us, using two nonreplicating, topologically modified polymers initially prepared in an overlapped configuration in a cylinder of fixed length~\cite{doshi2026kineticssegregationtopologicallymodifiedring}. Thus our previous study did not incorporate the non-equilibrium active processes relevant for the chromosome dynamics of \emph{C. crescentus} cells.

In Fig.~\ref{fig:withwithoutParA&diffParA}(b), we have shown the distance  $\Delta \mathrm{z}$ between 
the \emph{oriCs} of the two daughter DNA polymers (normalized by the instantaneous cell length $L_\mathrm{cyl}$), as a function of time as we varied the number of ParA monomers, $n_A$, in the system. All other parameters were kept unchanged. In fact, we observe from the data presented in  Fig.~\ref{fig:withwithoutParA&diffParA}(b), that an increase in the number of  ParA monomers leads to increased values of $\Delta \mathrm{z}$ as well as a faster rate of separation between the {\em oriCs}. The reason is that increasing the number of ParA monomers ($n_A$) will lead to  an increase in the number of loops formed due to ParA--ParB interactions. This will lead to greater entropic repulsion between daughter DNA-1 and DNA-2, leading to faster translocation of \emph{oriC}-2 to the new pole. In the next subsection, we describe control simulations which further establish this, and discuss the mechanism of translocation. 


Note that a majority of the $120$ independent runs showed `successful' translocation
of \emph{oriC} to the new pole in the system where the special interaction potential between ParA and ParB 
monomers is present. In certain runs, \emph{oriC} does not reach the new pole over the length of the run, we refer to these as `unsuccessful' runs.  The data shown in  Fig.~\ref{fig:withwithoutParA&diffParA}(a) and Fig.~\ref{fig:withwithoutParA&diffParA}(b) is averaged over \emph{oriC} positions in both `unsuccessful' and `successful' 
runs. Table:\ref{table:diff_ParA_info} gives the percentage of successful runs obtained with different numbers of  ParA monomers considered in the system. 


\subsection{Is the {\em oriC} transport entropy driven? }
A natural question is the relative importance of entropic and enthalpic interactions in driving the segregation of daughter chromosomes and the pole-to-pole transport of \emph{oriC} on DNA-2. Since our model includes transient attractive interactions between the {\em oriC}-proximal ParB monomers and ParA monomers, one may ask whether the observed transport is primarily energetically driven. In particular, because the ParA monomers are distributed with a density gradient that increases towards the new pole, could the attraction between ParA and ParB simply drag the \emph{oriC} of DNA-2 toward the new pole, with the remainder of the chromosome following passively?
Experimentally, however, ParA--ParB interactions occur over nanometer length scales and are unlikely to generate directed transport over the micrometer length scales of the bacterial cell.

To address such questions, we perform a series of control simulations that isolate the roles of enthalpic attraction and topology-induced entropic forces. These simulations demonstrate that the large-scale reorganization of daughter DNA-1 and DNA-2 cannot be achieved by the local ParA--ParB attraction alone. Instead, segregation requires the long-range entropic repulsion generated by topology-modifying processes, whose loops can span a substantial fraction of the cell length. The different sets of control simulations are itemized below.

$\bullet$ In the first set of control simulations, we switch off the repulsive excluded volume (EV) interactions between the monomers such that the DNA polymers become ideal chains. The rest of the model, including chain connectivity by springs, replication processes, the confining effect of the cylinder, and the (transient) attractive interactions between ParA and ParB proteins are kept unchanged. Only the soft repulsive potential between monomers is switched off. Thus, the monomers are oblivious to the presence of other monomers (of the same polymer or other polymer) except through chain connectivity and attraction forces between ParA and ParB monomers. 

In such a scenario, we observe that  all the monomers aggregate at the old pole due to clustering of the ParA and ParB monomers. We have shown 
the distribution of monomers from each polymer along the cylinder length in Fig.~S$9$ of SI--IV, and we see that there is a peak in the density of monomers are the old pole.

\begin{table}
\centering
\begin{tabular}{|c|c|c|c|}
 \hline
    $n_A$ &  Success (\%) &  $\tau_{passage} (\tau_{rep})$ &  Std. Dev. \\
\hline
          0 &      20.0 &          0.76 &     0.19 \\
         70 &      64.2 &          0.36 &     0.16 \\
        100 &      81.7 &          0.34 &     0.15 \\
        130 &      82.5 &          0.29 &     0.12 \\
        160 &      86.7 &          0.26 &     0.12 \\
 \hline
\end{tabular}
\caption{Success rates for transport of the \emph{oriC} monomer from the old pole to the new pole as a function of the number ($n_A$) of ParA monomers, calculated over $120$ independent runs. The interaction range is fixed at $r_c^P=1.5a$, and $n_P=10^4$. The mean passage time, $\tau_{\mathrm{passage}}$, normalized by the replication time $\tau_{\mathrm{rep}}=10^7$ iterations ($=10^5\tau$), is calculated over successful runs. 
 }
\label{table:diff_ParA_info}
\end{table}

$\bullet$ In the second set of control simulations, the monomers within the same DNA polymer interact  
by EV interaction, but the EV interaction between the monomers from the two different daughter DNA polymers are set to zero.  All other interactions remain unchanged as before. In this case, we observe that the monomers of each polymer are spread throughout the cylinder, i.e., the polymers do not segregate, refer to Fig.~S$10$(a) of SI--IV. But the \emph{oriC}'s separate out due to the attraction between ParB and ParA, aided by the high density of ParA near the new pole, as shown in Fig.~S$10$(b) of SI--IV. However, the translocation of the {\em oriC} of DNA-2 is accelerated as compared to what is observed experimentally. This is because the translocation is not hindered by the EV interactions between monomers of the mother DNA and daughter DNA-2.

The directed motion of a ParB monomer toward the new pole in a ParA density gradient arises from a bias in its probability of rebinding to ParA on a separate DNA segment. Random displacements toward the new pole are more likely to bring the ParB monomer into proximity with a ParA monomer and thereby form a transient cross-link. Once bound, the ParB monomer remains relatively localized until the transient ParA--ParB interaction is released. Upon release, the ParB monomer again undergoes diffusive motion in either direction, but displacements toward the new pole have a higher probability of resulting in rebinding because of the higher local ParA density. Thus, repeated cycles of transient binding and diffusive motion produce a net bias in the ParB trajectory toward the new pole. Monomers bound by cross-links (transient or otherwise) also exhibit slower diffusion; this aspect is being investigated separately and will be reported elsewhere.

\begin{table}
\centering
\begin{tabular}{|c|c|c|c|}
 \hline
                                        $n_{P}$ &  Success (\%) &  $\tau_{passage}$ ($\tau_{rep}$) &  Std. Dev. \\
\hline
5000 &      45.0 &          0.48 &     0.24 \\
      10000 &      62.5 &          0.34 &     0.16 \\
      20000 &      69.2 &          0.40 &     0.18 \\
      40000 &      77.5 &          0.42 &     0.20 \\
 \hline
\end{tabular}
\caption{ {\bf Segregation statistics:} Success rate for directed transport of one {\em oriC} from the old pole to the new pole, calculated from $120$ independent simulation runs for different values of $n_P$ iterations. The replication rate is fixed by introducing two monomers, one at each RF, every $20,000$ iterations. The interaction range is fixed at $r_c^P=0.8a$ and  $n_A=100$ ParA monomers. For runs exhibiting successful transport, we report the mean passage time, $\tau_{passage}$, normalized by the replication time, $\tau_{rep}=10^7$ iterations ($10^5\tau$), where $\tau_{passage}$ is the time required for {\em oriC} to reach the new pole. 
}
\label{tab:n_P_variation}
\end{table}

$\bullet$ To  investigate the consequences of the decrease in EV interactions, we also  do simulations where  the diameter $\sigma$ of the monomers is reduced to $0.3a$ to reduce repulsive EV interactions. Simultaneously, the strength of the repulsive interaction between ParA-ParB also gets reduced to $A_{rep} =5 k_BT$, and furthermore, the strength of the attractive potential between ParA and ParB monomers is also reduced by a factor of $ 4$ to $ A _ {att} = -5 k_BT$ (refer Eqns. $14, 15$ of SI--I, subsection E). Note that the range of interaction potential $r_c^P$ between ParA and ParB remains unchanged, refer Fig.~S$3$(b) of SI--I for a plot of the revised potential. 
If we keep the strength of the attractive interaction unchanged, all the ParA and ParB monomers cluster near the old pole, as the effects of ParA--ParB attraction dominates over entropic repulsion. Details are in the SI--IV. From the data presented in Fig.~S$11$(a) and Fig.~S$11$(b) of SI--IV, we still observe the translocation of the {\em oriC} of DNA-2  to the new pole, although the time to reach the new pole is significantly increased, we also observe the segregation of the two polymers. We have provided the distribution of the density of monomers of daughter DNA-1 and DNA-2 polymers along the long axis after replication is complete.

$\bullet$   We also investigate the effect of varying $n_P$, keeping the range $r_c^P$ of the interaction potential between ParA and ParB fixed at $0.8a$: refer Fig.~S3(c) of SI--I for a plot of the potential and values of parameters of ParA--ParB interaction, when $r_c^P$ is reduced from $1.5a$ to $0.8a$. The range of the repulsive potential remains unchanged at $0.8a$. The parameter $n_P$ controls the duration over which the polymers experience the entropic interactions arising from transient loops formed by ParA--ParB binding. Table~\ref{tab:n_P_variation} shows that the number of `successful' simulation runs increases with increasing $n_P$. In contrast, the normalized passage time, $\tau_{passage}/\tau_{rep}$, depends non-monotonically  on $n_P$.

\begin{figure*}[htbp]
    \centering
    \includegraphics[width=0.8\linewidth]{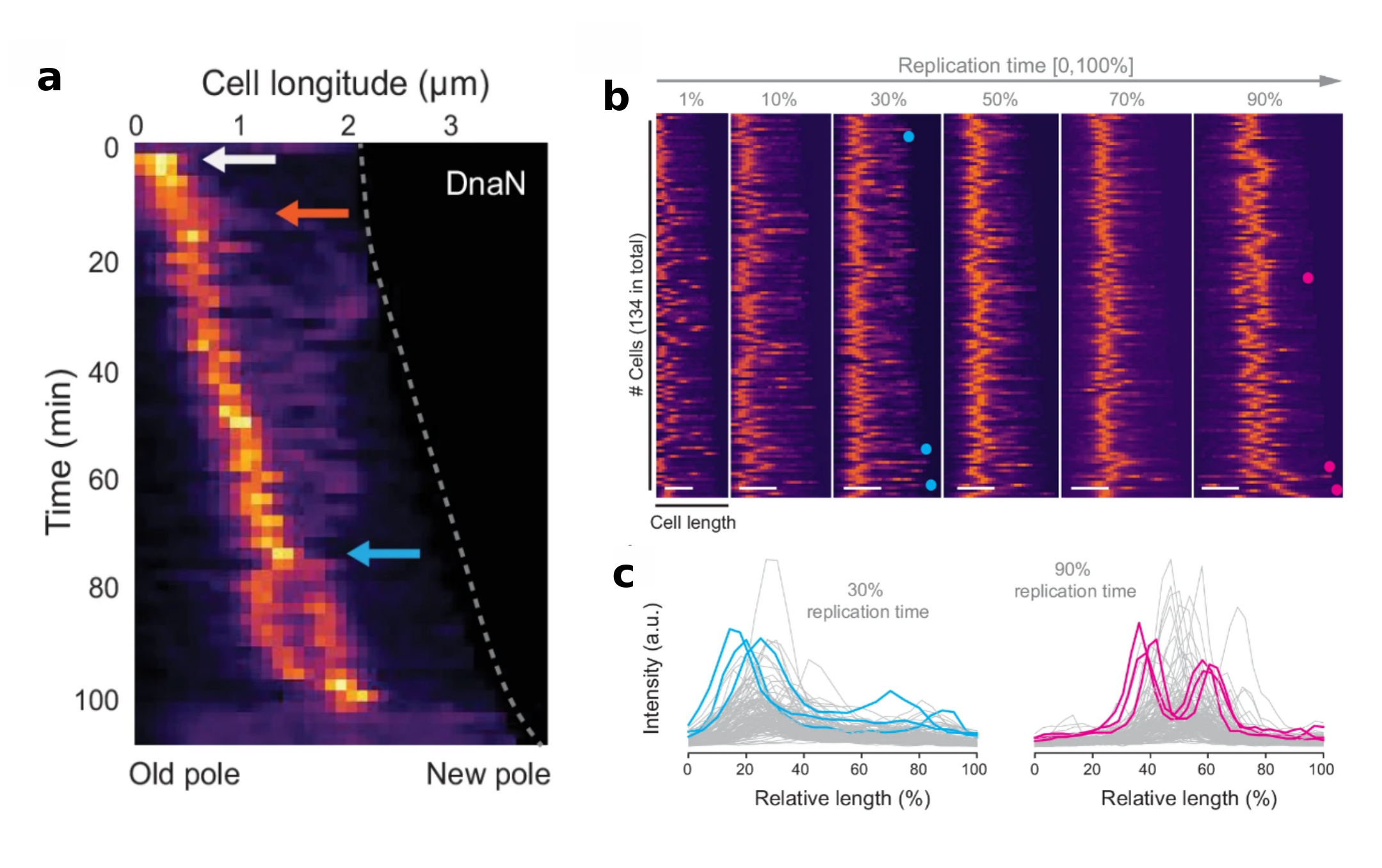}
    \caption{ {\bf Experimental data reproduced from Ref.~\cite{zhang2024chromosome}}. Spatial localization of fluorescently tagged replisomes during the cell cycle. (a)  Kymograph of time-lapse imaging of one of the representative cells. Replication initiation, appearance of early-splitting and late-splitting events are indicated by white, orange and blue arrows, respectively. The kymograph shows that in later stages of replication (around 70\% to 90\% of replication time), splitting of the replication fork, i.e., two fluorescent foci, is observed, while in early stages they see a single fluorescent focus is observed.  (b)  Demograph of normalized DnaN-sfGFP fluorescence at different time points of normalized replication duration. 134 individual cells aligned along the cell length at different stages of replication (1\%, 10\%, 30\%, 50\%, 70\%, and 90\% replication time). Cyan and magenta markers highlight representative cells at 30\% and 90\% replication time, respectively. (c) Fluorescence intensity profiles of the highlighted cells (colored curves) superimposed on the profiles of all cells (gray curves), illustrating the transition from a predominantly single replisome focus early in replication (30\%) to two separated foci near the late stage of replication (90\%). Panels (b)--(d) are cropped and rearranged from Fig.~1 of {\em Chromosome organization shapes replisome dynamics in Caulobacter crescentus}, Ref.~\cite{zhang2024chromosome}, and reproduced under the \href{https://creativecommons.org/licenses/by/4.0/}{Creative Commons Attribution 4.0 International (CC BY 4.0) license}. © 2024 The Author(s). {\em The  figure caption has been reproduced verbatim.}}
    \label{fig:repdist_experimental}
\end{figure*}

We attribute this non-monotonic behavior to a competition between two consequences of changing $n_P$.
On one hand, lower values of $n_P$ allow the transient loops to reposition themselves 
along the DNA contour as the ParB monomers move towards the new pole
with higher ParA concentration. This likely optimizes the entropic forces driving \emph{oriC} 
transport. But the loops themselves exist for a relatively short duration, which implies that the enhanced entropic repulsion acts for short times. Since the ParA--ParB interactions are switched off every $n_P$ iterations, the maximum lifetime of a loop is $n_P$ iterations. New loops can subsequently form only when other ParA and ParB monomers diffuse within the interaction range of each other.

On the other hand, increasing $n_P$ prolongs the lifetime of these loops once they are formed, thereby ensuring higher entropic repulsion between the two daughter DNA polymers. But the readjustment of the position of the loops along the chain contour is then delayed. As a consequence, the segment with ParB monomers is unable to diffuse around quickly, as it has to also drag the DNA segment that has ParA. Thereby,  the directed transport of \emph{oriC} to the new pole slows down. But note that the success rates of segregation are increased with an increase in $n_P$.

\begin{figure*}[htbp]
\centering

    \includegraphics[width=0.32\linewidth]{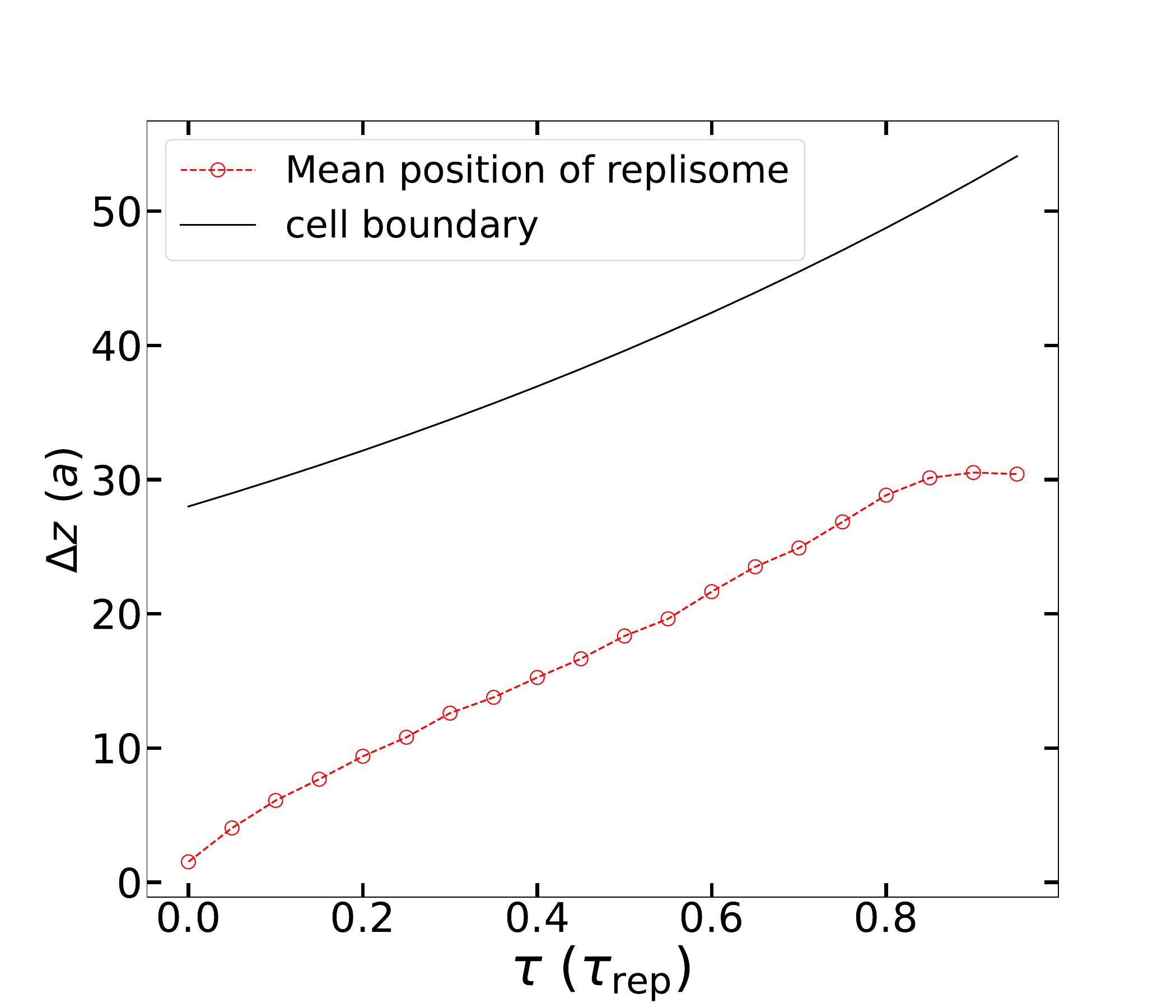}
    \put(-130,90){\large\textbf{(a)}}
    \includegraphics[width=0.32\linewidth]{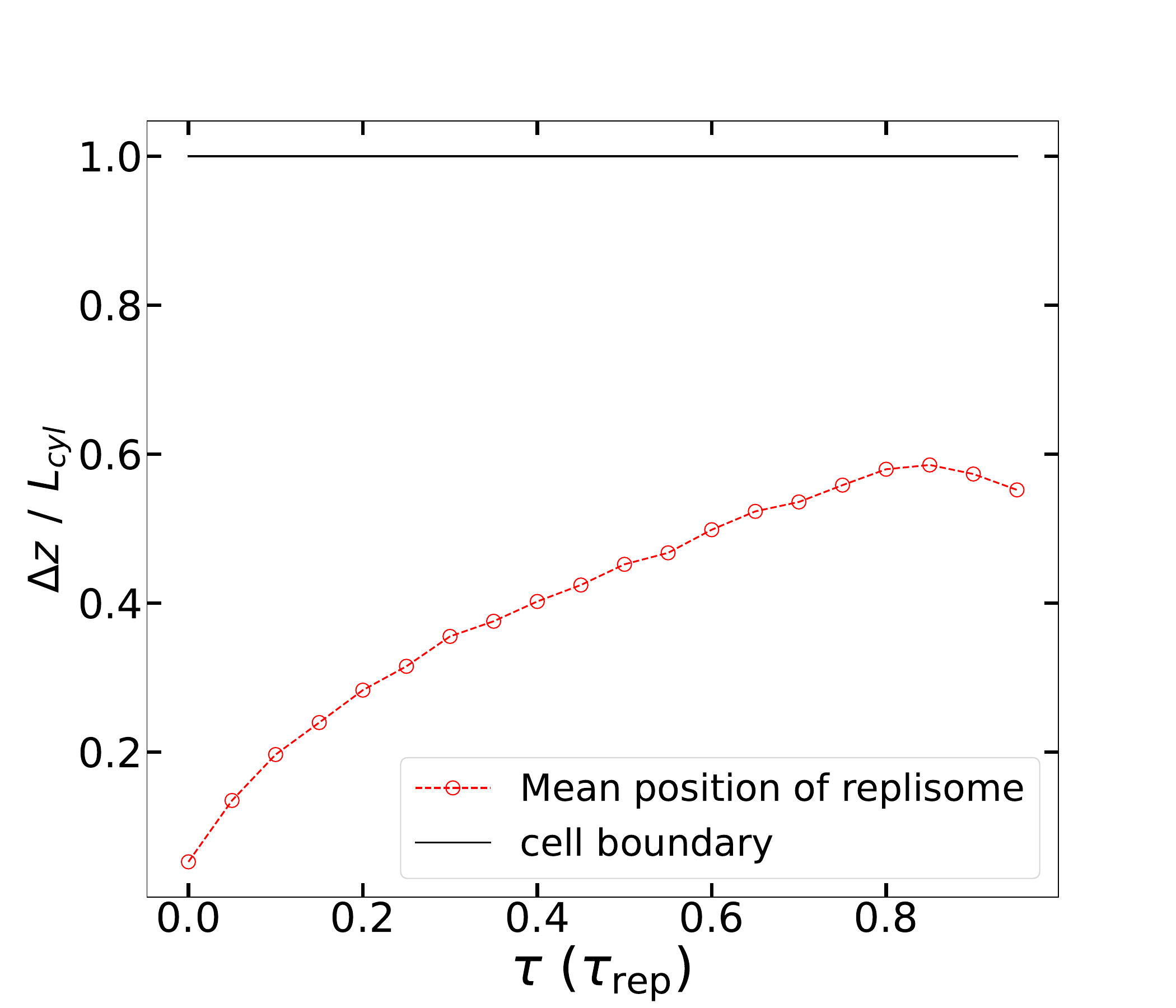}
    \put(-130,90){\large\textbf{(b)}}
\hfill
    \includegraphics[width=0.32\linewidth]{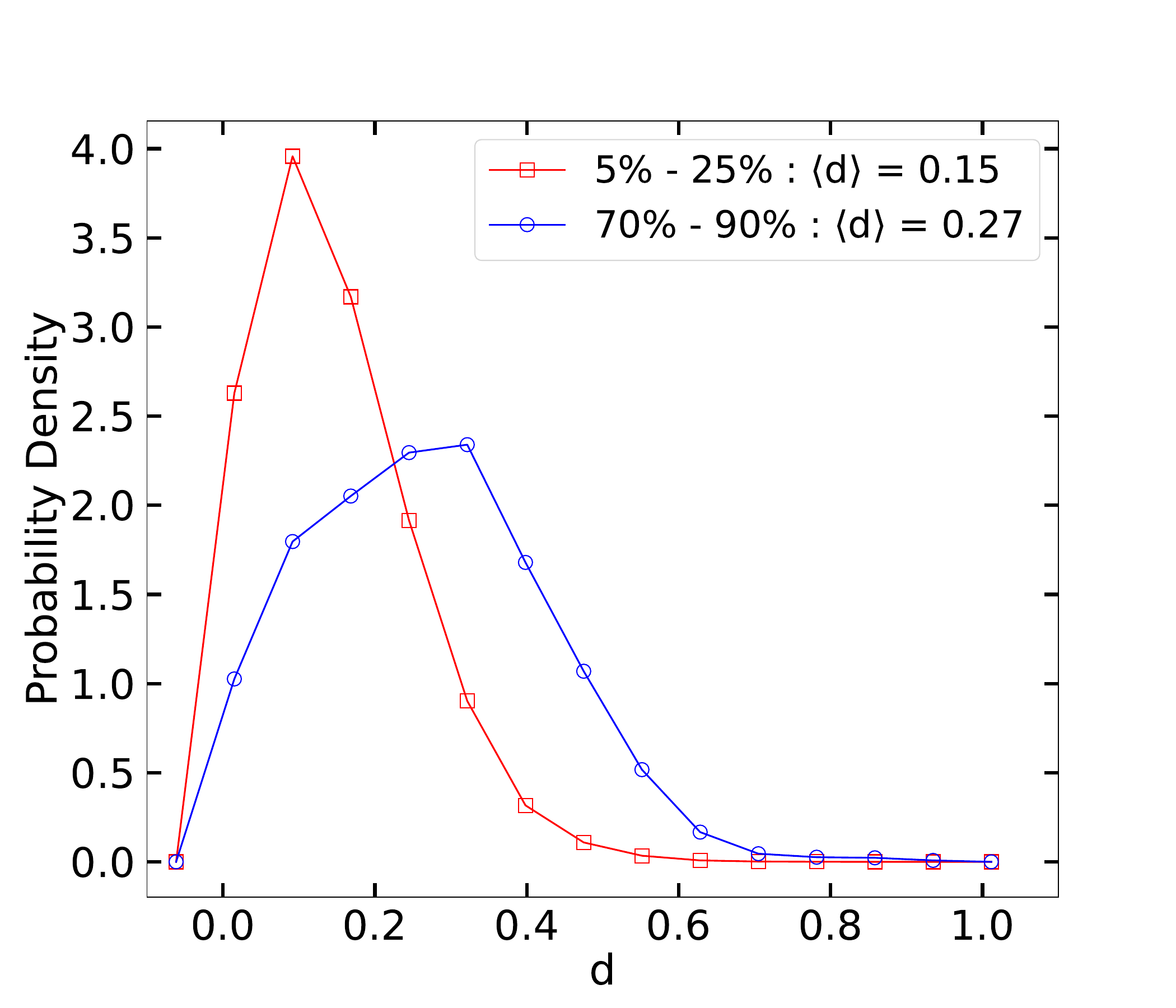}
\vspace{0.2cm}
    \includegraphics[width=0.98\linewidth]{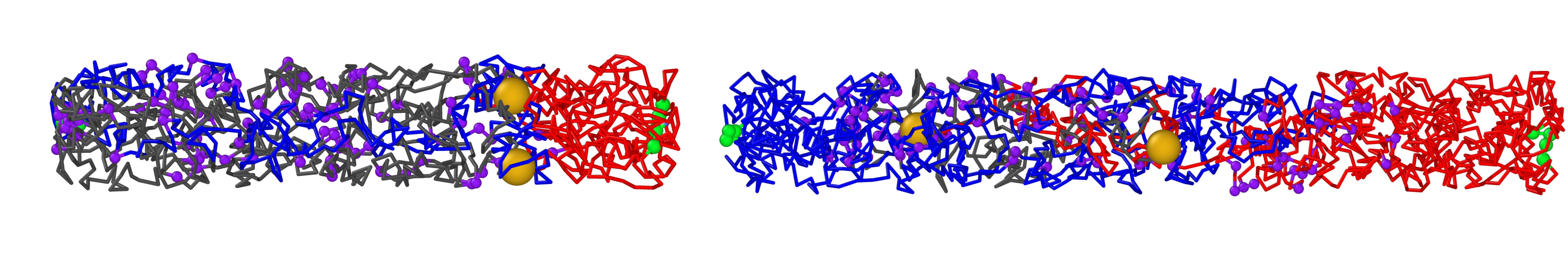}
    \put(-475,70){\large\textbf{(d)}}
    \put(-260,66){\large\textbf{(e)}}
\caption{
  {\bf Reorganization of replication forks (RFs) in simulations; RFs represent the replisome.} The simulations are performed with $100$ ParA monomers, $r_c^P= 1.5a$ 
  and $n_P =10^4$ iterations. \textbf{(a)} The plot shows the longitudinal mean position $\Delta \mathrm{z}$ of the replisomes (RFs), averaged over their instantaneous positions, as a function of time normalized by the replication time. The distance $\Delta \mathrm{z}$ is measured from the old pole of the cell. Data is averaged over the `successful' runs which showed successful {\em oriC} transport to the new pole out of $120$ independent runs. 
\textbf{(b)}  The plot shows the average long axis position $\mathrm{z}$ (normalized with instantaneous length of cylinder $L_{\mathrm{cyl}}$) of the replication fork vs time in units of replication time.
\textbf{(c)} Probability density distribution of long axis distance ($d$) between two RFs at ($5\%$ to $25\%$) of replication time (Red) and (70\%-90\%) of replication time (Blue). The long-axis distance, $d$, is normalized by the instantaneous cylinder length, $L_{\mathrm{cyl}}$.
\textbf{(d)} Representative simulation snapshot showing the replicating chromosome configuration at $30\%$ replication (early stage of replication). The old pole is at the right end of the 
cylinder. Monomers belonging to the two daughter chromosomes are shown in red (DNA-1) and blue 
(DNA-2), while the replisome is shown in orange and the mother 
chromosome in grey. Purple and green 
monomers represent the ParA and ParB monomers, respectively. 
\textbf{(e)} Representative simulation snapshot showing the replicating chromosome configuration at $80\%$ replication (late stage of replication). The color scheme remains unchanged from the snapshot shown in (d). 
}
\label{fig:main_repdist_sim}
\end{figure*}


\subsection{Replisome (Replication fork) trajectories:}

The replisome is the aggregate of protein machinery that sits on DNA at the replication fork and creates two copies of daughter DNA from the mother DNA. Experiments have shown that the two replisomes, one on each arm of the chromosome, remain co-localized, showing a single fluorescent spot in the initial stages of replication, refer Fig.~\ref{fig:repdist_experimental}(a). In the later stages of replication (i.e., when around  $70\%$  to $90\%$ of replication is complete), the single fluorescent spot splits into two, refer Fig.~\ref{fig:repdist_experimental}(a), which is reproduced from \cite{zhang2024chromosome}. The splitting happens because the two replisomes get separated spatially and thereby can be resolved.

Time-resolved experimental measurements of replisome dynamics from $134$ cells are shown in Fig.~\ref{fig:repdist_experimental}(b). In Fig.~\ref{fig:repdist_experimental}(b), the data from the individual cells are stacked vertically. Each row shows the replisome position from a single cell at different stages of the cell cycle. Thus, moving horizontally along a given row traces the positions of the replisome(s) as replication progresses for a single cell. Because the cell elongates during the replication cycle, the width  of each panel increases as the replication proceeds from $1\%$ to $90\%$ of  $\tau_{rep}$.

Two salient features emerge from these data. First, the fluorescent replisome focus shifts from the old-pole-proximal region toward mid-cell as replication proceeds. Second, the initially single fluorescent focus broadens and subsequently resolves into two distinct foci, corresponding to the separation of the two replisomes. This behavior is quantified in Fig.~\ref{fig:repdist_experimental}(c), which compares the spatial distributions of replisome positions at $30\%$ and $90\%$ of the replication cycle. At $30\%$ of the replication time, the distribution exhibits a single peak, whereas by $90\%$ it evolves into a bimodal distribution, reflecting the spatial separation of the two replisomes.

We would like to analyze whether these observations about the distribution of replisome positions are reproduced in our simulations as an emergent consequence of the entropic interactions mediated 
by our effective model of the ParABS system. Although we do not explicitly model the replisome in our simulations,
we know that the replisome is located at the replication fork (RF).  We can track the positions of 
the RFs as a function of time in our simulations. This is because we know the position of all the monomers, including the one where the RF(s) is stationed at a particular stage of replication. 
In Fig.~\ref{fig:main_repdist_sim}(a), we track the positions of the RFs by plotting the distance $\mathrm{z}$ of the RF monomers from the old pole, enabling a direct comparison with the experimental data shown in Fig.~\ref{fig:repdist_experimental}.

We also plot the position of the new pole, {\it i.e.}, the opposite end of the cell, as a function of time to account for cell elongation. Distances from the old pole are reported in units of $a$ in 
Fig.~\ref{fig:main_repdist_sim}a; the corresponding data in micrometers are provided in Fig.~S$12$ of SI--V.
 To express distances in $\mu$m, as described in the Methods, we use an estimated bead diameter of $0.8a \approx 65$~nm~\cite{messelink2021learning}. We use the conversion factor $a \approx 81$ nm.


Figure~\ref{fig:main_repdist_sim}(a) shows that the RFs relocate from the old pole toward midcell as the two daughter chromosomes segregate and the {\em ter} locus occupies the cell center. By the end of replication, the RFs are located in close proximity to the \emph{ter} locus. The segregation of daughter DNA and the relocation of the \emph{ter} loci and the replisomes (RFs)  is an emergent phenomenon as a consequence of the entropic forces of segregation. We also present the mean of the instantaneous positions of the two RFs normalized by the instantaneous cell length, $L_\mathrm{cyl}$ in Fig.~\ref{fig:main_repdist_sim}(b). Since the position is plotted in units of instantaneous $L_\mathrm{cyl}$, $L_\mathrm{cyl}=1$ is also indicated by the black line near the top of the subfigure. This again shows relocation of the RFs from the old pole to the middle of the cell.

Thereafter, since Fig.~\ref{fig:main_repdist_sim}(c) shows the distribution of 
distance between the two replisomes at the early stages and late stages of replication, we 
also plot the distribution of the distance $d$ between two RFs. The quantity $d$ is 
normalized by the instantaneous $L_\mathrm{cyl}$ of the cylinder. Fig.~\ref{fig:main_repdist_sim}(c) 
shows the distribution calculated from our $120$ independent runs for the 
early stages of replication (i.e. in the stage when $5\%$ to $25\%$ of replication is complete) 
and the late stages of replication (i.e. in the stage when $70\%$ to $90\%$  of replication is complete). From the plot, we can see that in the early stages the two RFs are co-localized, as the 
distribution is peaked at very low values of $d/L_\mathrm{cyl}$. 
At late stages of replication, the peak of the distribution of the distance between 
two RFs shifts away from zero to $d/L_\mathrm{cyl} \approx 0.22$. This is 
consistent with the experimental data shown in Fig.~\ref{fig:repdist_experimental}(c). 

 We now propose a plausible mechanism for the relocalization and separation of the RFs during 
 the later stages of replication, {\em viz.}, after approximately $75\%$ of the chromosome has been replicated. Entropic interactions play a central role in determining replisome localization in {\em C.~crescentus}. Following the replication of {\em oriC}, one of the daughter {\em oriC} loci rapidly relocates to the new pole (left end of the cylinder) through the action of the ParABS system, as shown in Fig.~\ref{fig:withwithoutParA&diffParA}a. During this process, it traverses the region occupied by the unreplicated mother chromosome (gray polymer in the snapshot of Fig.~\ref{fig:main_repdist_sim}(d)).

This directed transport stretches the corresponding daughter chromosome (blue polymer), while the RFs remain near the old pole (right end), where replication was initiated and where the second daughter chromosome (red polymer) is localized. In Fig.~\ref{fig:main_repdist_sim}(d), the replisomes are shown as orange spheres and the {\em oriC}-proximal ParB monomers as green spheres.
As replication proceeds, entropic segregation drives the two nearly replicated daughter chromosomes into opposite halves of the cylindrical cell, while they remain connected through the unreplicated {\em ter} region of the mother chromosome. We propose that the resulting entropic repulsion between the daughter chromosomes, together with their interactions with the remaining unreplicated mother chromosome, drives the probabilistic separation of the two RFs. A representative configuration illustrating this is shown in Fig.~\ref{fig:main_repdist_sim}(e).


\subsection{Loci localization}
Experimental measurements show that the trajectory of specific fluorescently tagged 
loci on the {\em C. crescentus} chromosome show distinct trajectories as 
the mother chromosome gets replicated to form daughter chromosomes  \cite{Harju2025.04.03.647011}. Each locus from the daughter chromosomes occupies certain home positions along the long axis of the cell at the end of the replication process \cite{viollier2004rapid}.
However, experimental measurements of loci trajectories by fluorescence microscopy are subject to several unavoidable challenges that complicate their quantitative interpretation. We briefly summarize these issues.

 \begin{figure*}
\centering
    \begin{subfigure}[b]{0.22\linewidth}
        \centering
        \includegraphics[width=\linewidth]{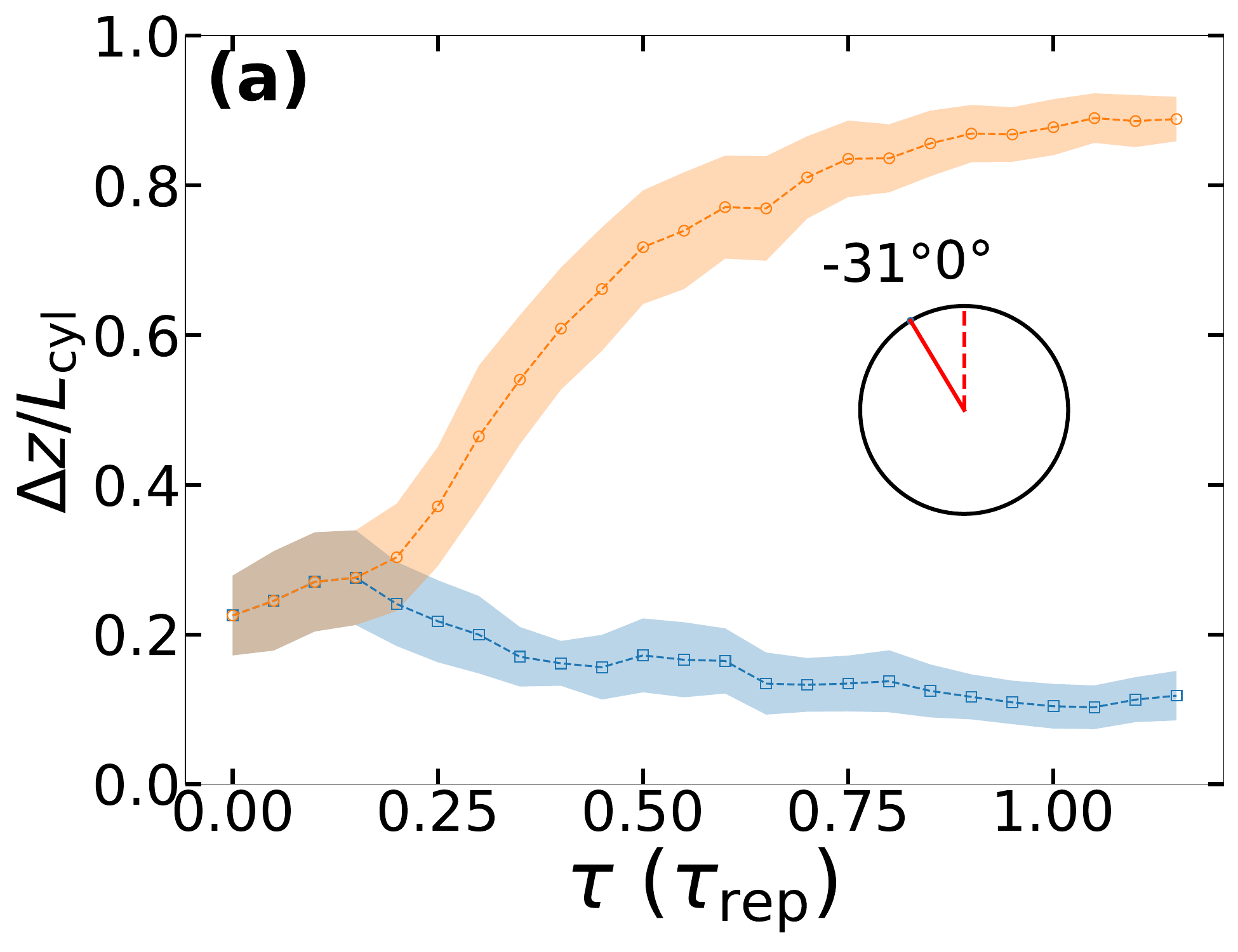}
    \end{subfigure}
    \hfill
    \begin{subfigure}[b]{0.22\linewidth}
        \centering
        \includegraphics[width=\linewidth]{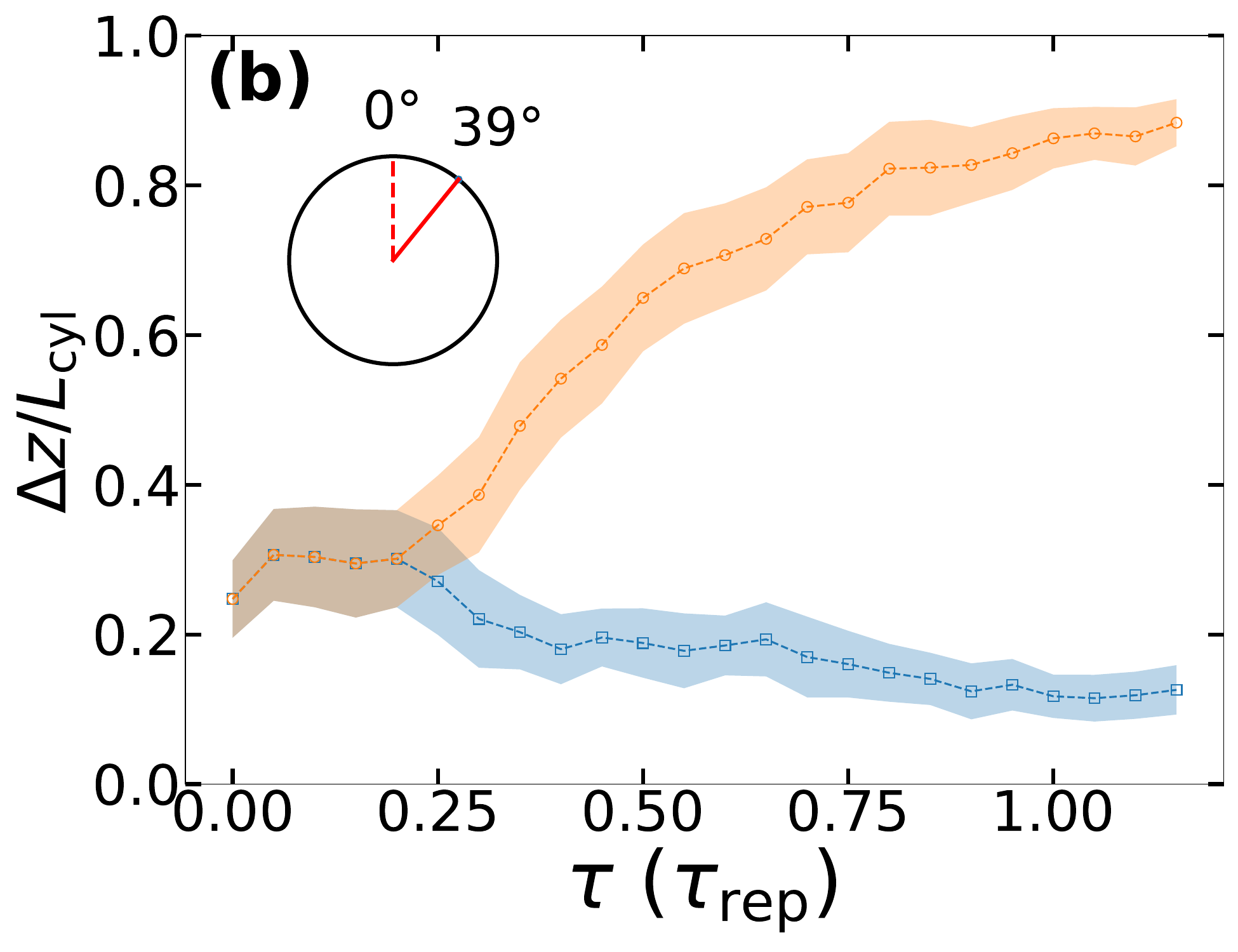}
    \end{subfigure}
    \hfill
    \begin{subfigure}[b]{0.22\linewidth}
        \centering
        \includegraphics[width=\linewidth]{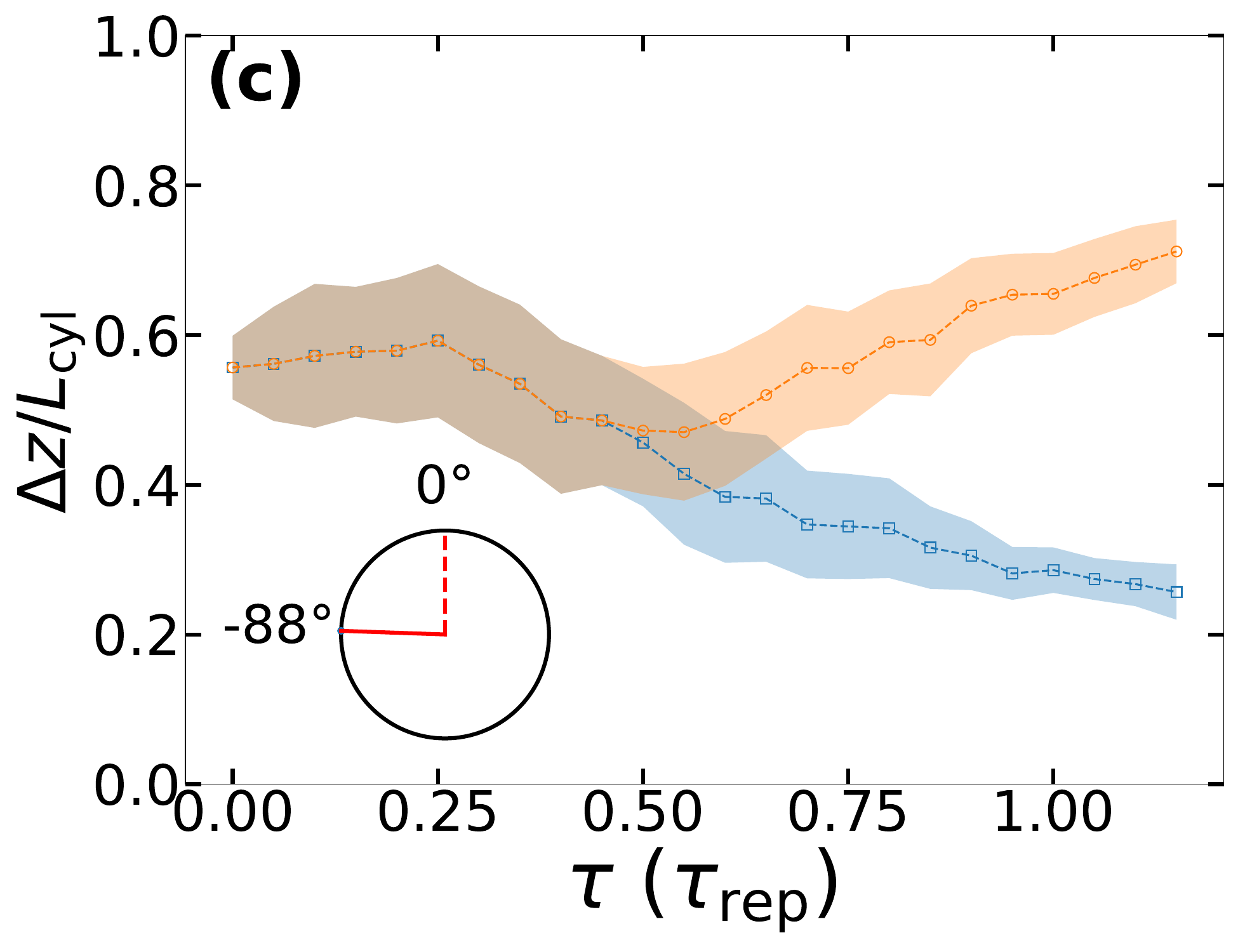}
    \end{subfigure}
    \hfill
    \begin{subfigure}[b]{0.22\linewidth}
        \centering
        \includegraphics[width=\linewidth]{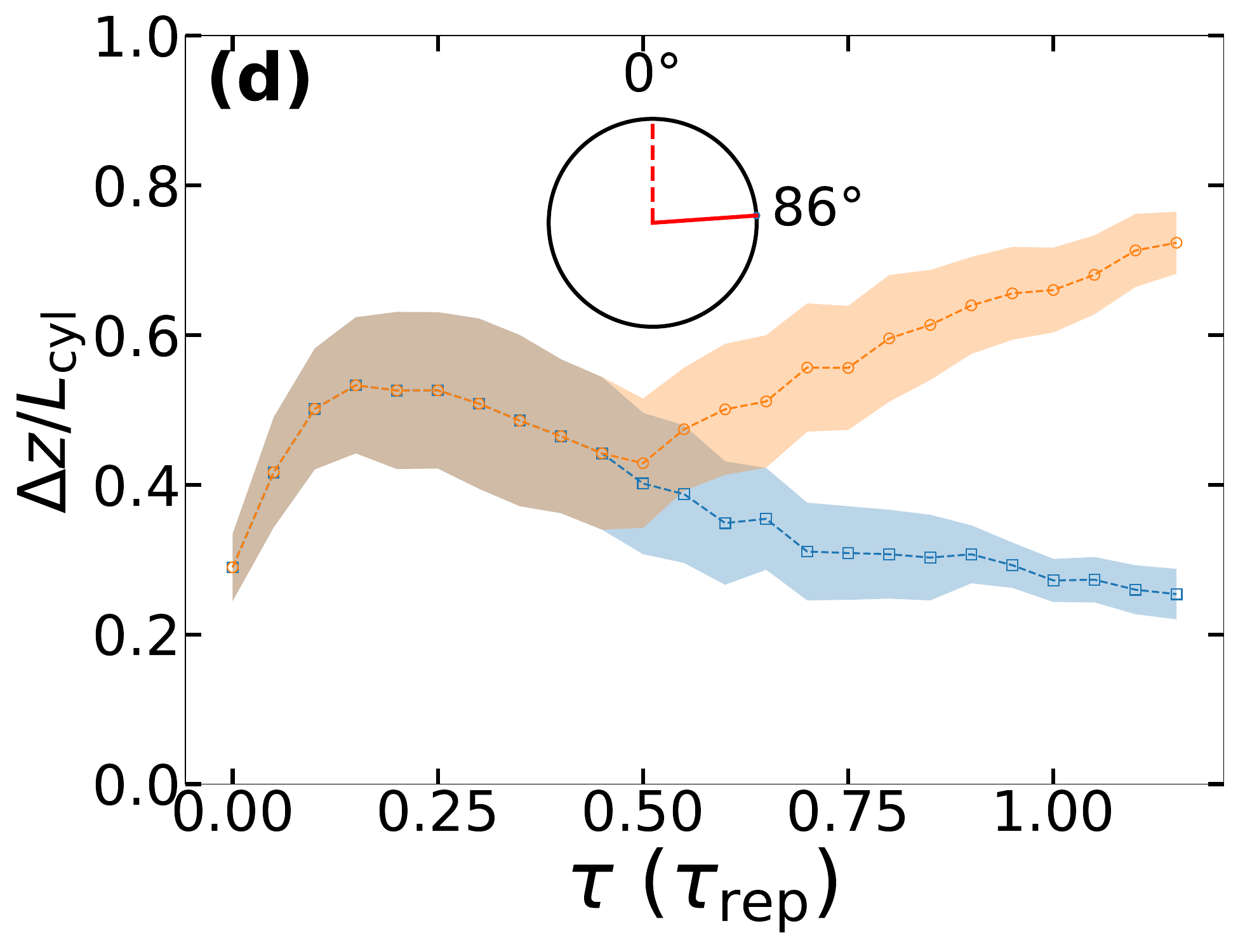}
    \end{subfigure}
    \hfill
    \begin{subfigure}[b]{0.22\linewidth}
        \centering
        \includegraphics[width=\linewidth]{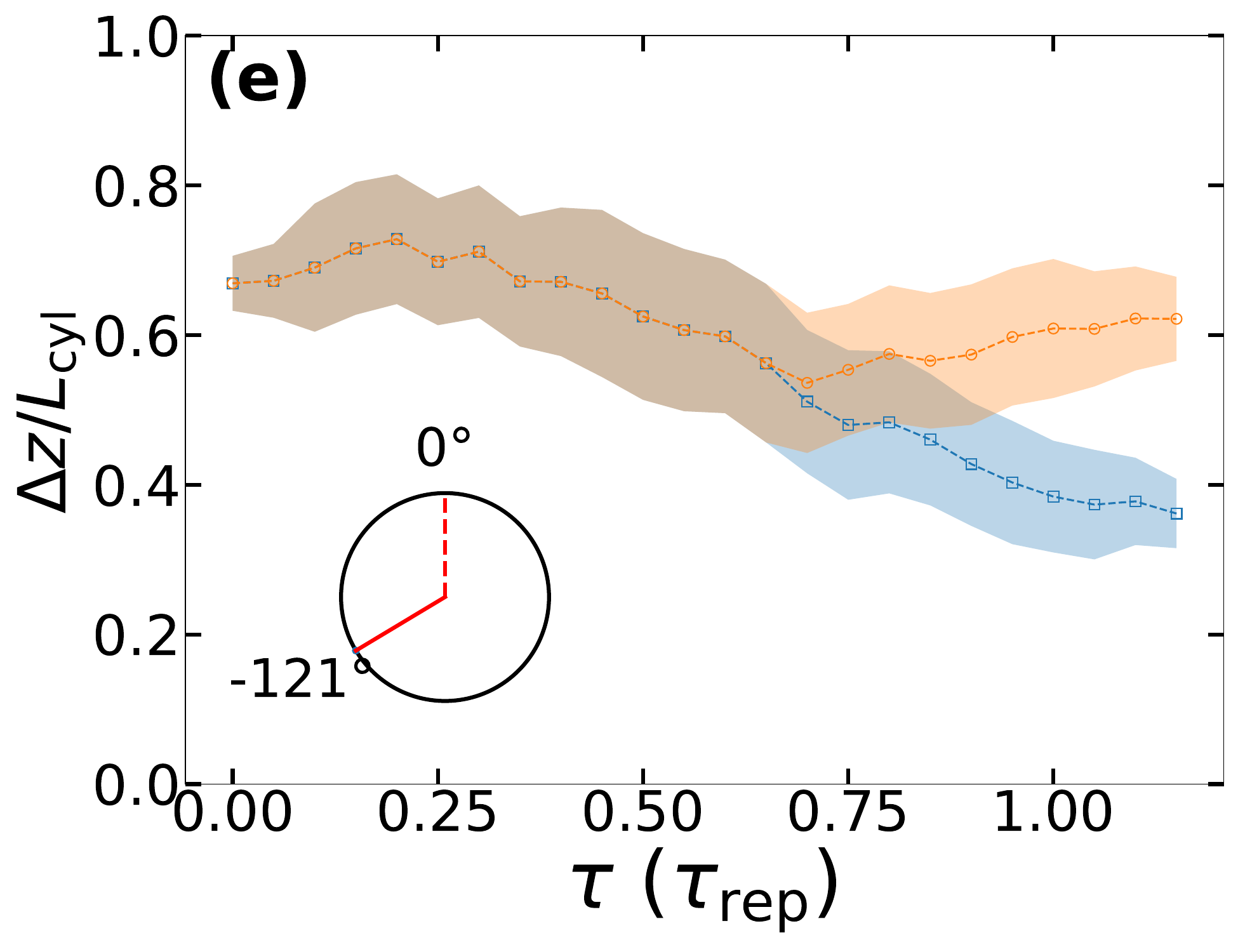}
    \end{subfigure}
    \hfill
    \begin{subfigure}[b]{0.22\linewidth}
        \centering
        \includegraphics[width=\linewidth]{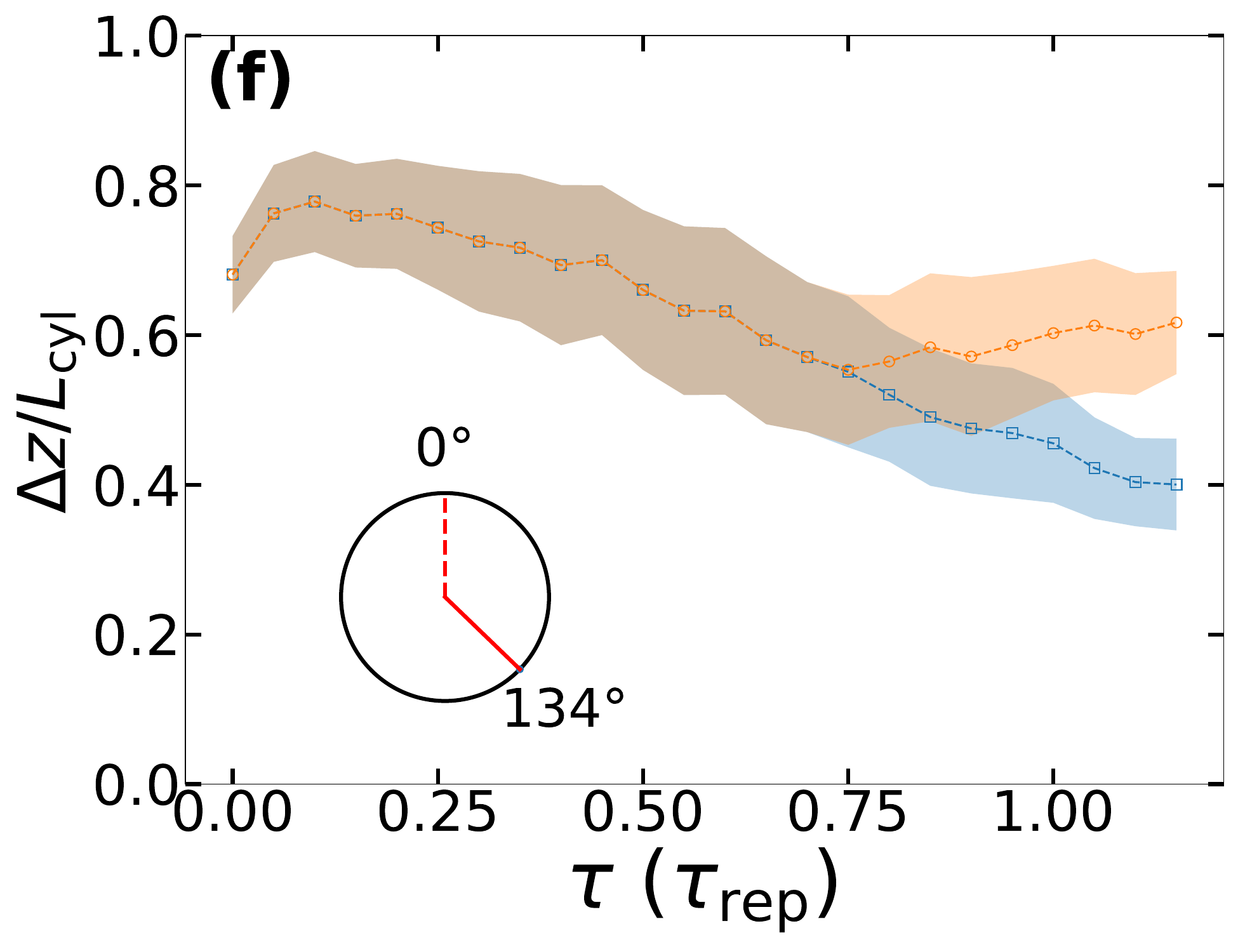}
    \end{subfigure}
        \hfill
    \begin{subfigure}[b]{0.22\linewidth}
        \centering
        \includegraphics[width=\linewidth]{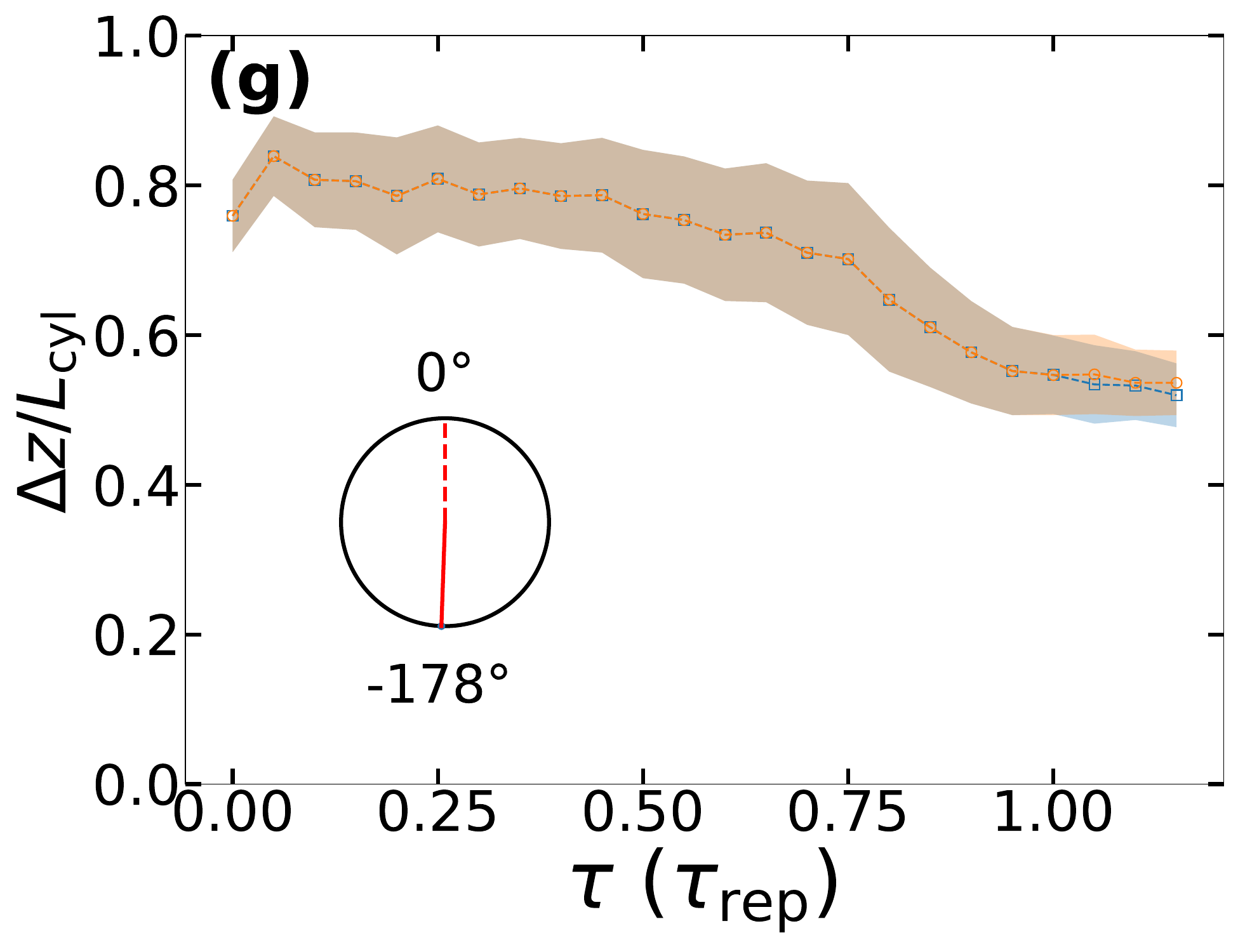}
    \end{subfigure}
        \hfill
    \begin{subfigure}[b]{0.22\linewidth}
        \centering
        \includegraphics[width=\linewidth]{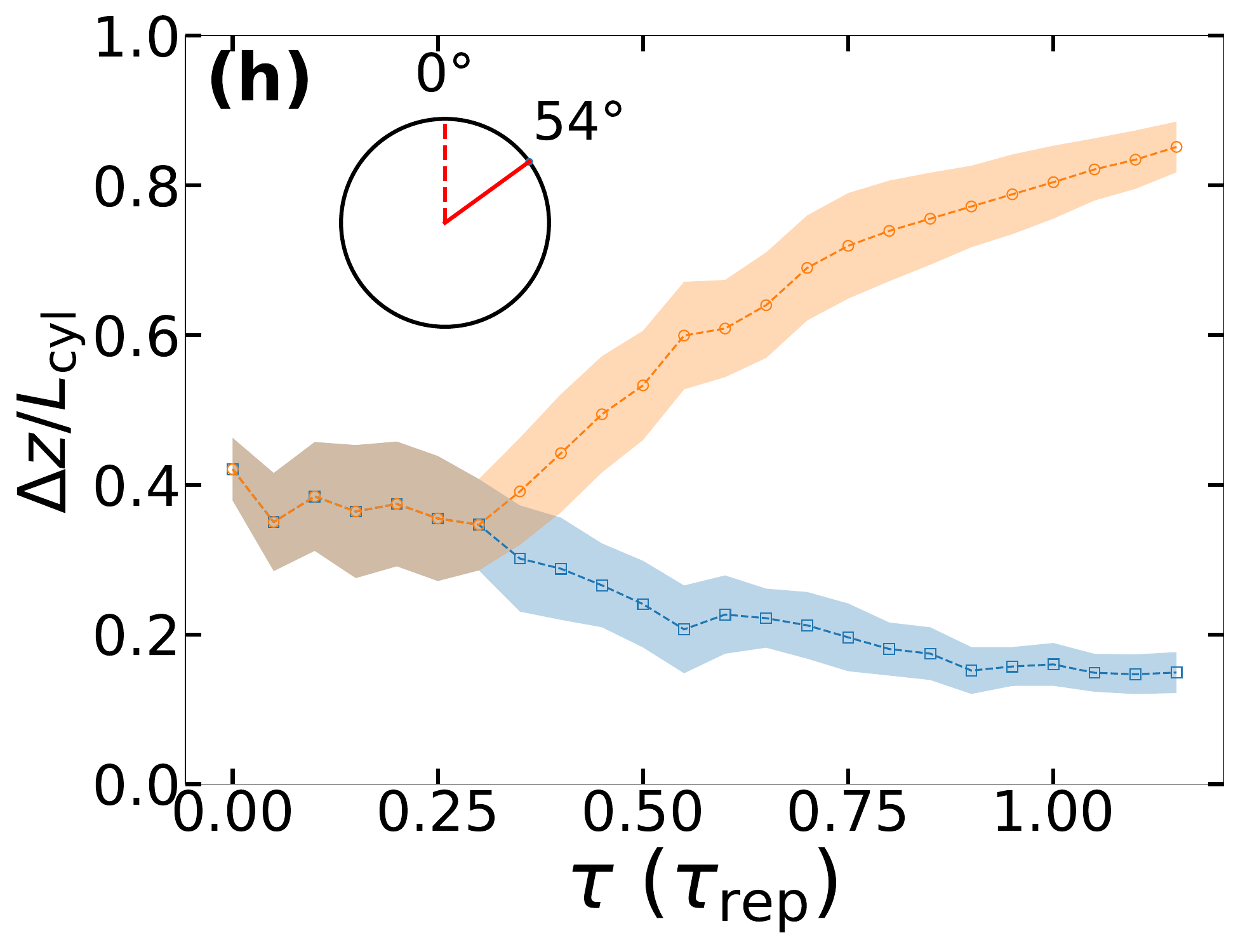}
    \end{subfigure}    
    
    \caption{{\bf Loci trajectories from our simulations:} We plot the trajectories 
    of eight different loci on the DNA polymer of {\em C. crescentus} as a function of time, even as replication is in progress. These loci are the same loci whose experimental trajectories are reported in \cite{Harju2025.04.03.647011}. In these plots, distances of each locus were calculated from the position of the old pole.
    The gray shaded region represents the position of the particular locus when it is 
    still part of the mother DNA. Post-replication, each locus splits into two trajectories 
    post replication. Loci which are proximal to ter undergo replication at a later 
    stage of the replication process. 
    The symbols show the mean position averaged over independent runs,
    whereas the shaded region shows the standard deviation about the mean positions.
    The positional data is averaged over the `successful' runs out of 120 independent runs.}
    \label{fig:simulation_locitraj}
\end{figure*}

%

\begin{figure*}[htbp]
    \centering
    \includegraphics[width=\linewidth]{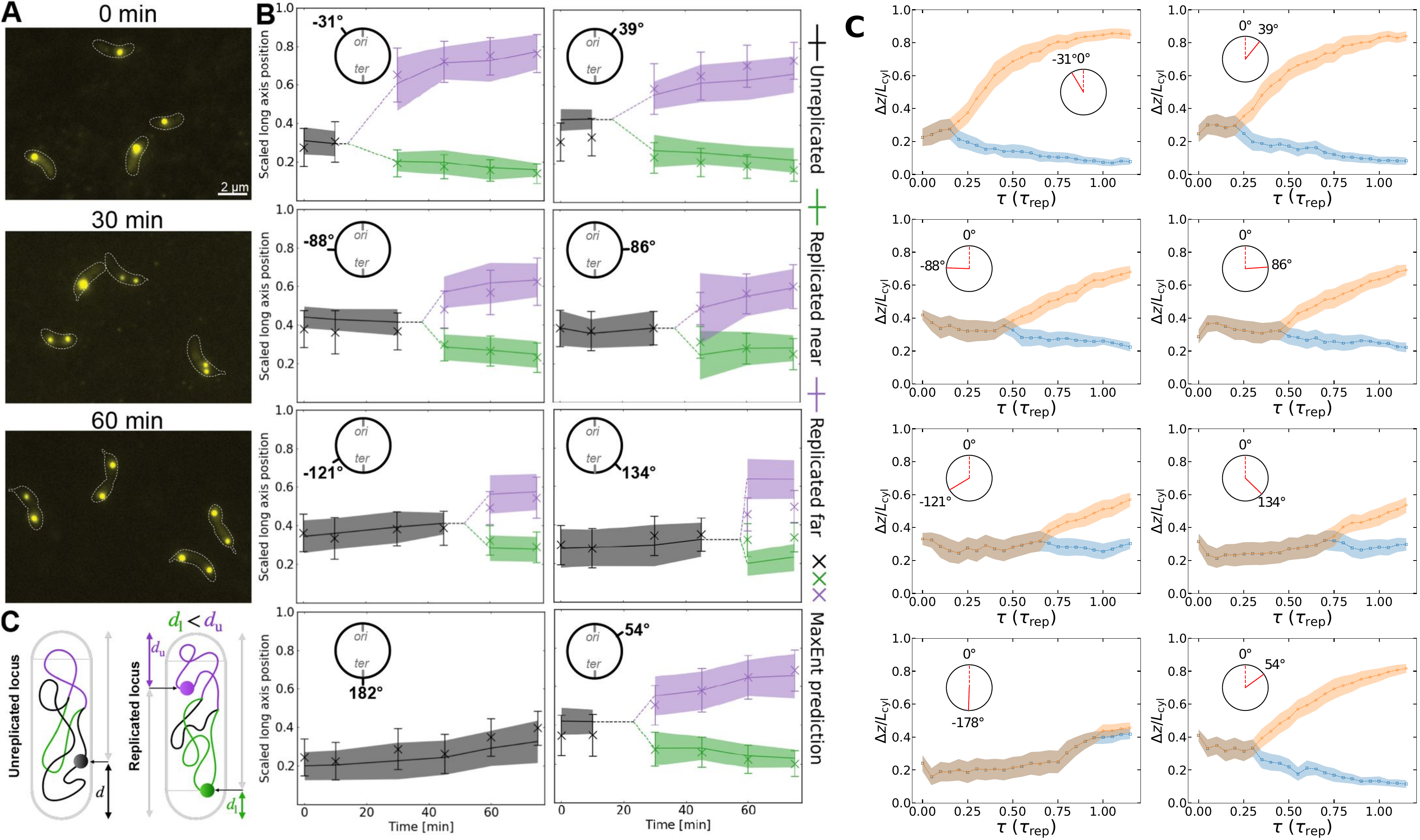}
     \put(-500,68){\color{white}\rule{5mm}{5mm}}
\caption{{\bf Loci Trajectories: Comparing experimental trajectories  [as shown in columns A and B, which are reproduced in this manuscript from \cite{Harju2025.04.03.647011}] and (C) trajectories from our simulations:} (A) Representative microscopy images of cells of strain Tn102 at indicated times after synchronization displaying fluorescently labeled chromosomal loci at -$31^\circ$ in yellow. White dashed lines indicate cell borders. (B) Mean long-axis positions of loci inferred using the 4D-MaxEnt model (crosses; error bars indicate std.), along with experimental measurements (lines, colored border indicates std.). Dashed lines indicate the predicted time of replication based on a linear replication speed. The 4D-MaxEnt inferences have been corrected for the bias due to the resolution of the experiments. (C) Trajectories of monomers corresponding to the loci shown in column B, obtained from our simulations. 
The directed trajectories emerge from the modeled interactions between ParA and ParB monomers.  For all unreplicated loci, we consider their distance to the nearest pole.  For replicated loci, we define $d_l$ as the minimum distance between either locus and either pole, and measure the distance to the other locus from this pole (refer to the text for details). As in column (B), the colored border indicates the standard deviation in the position of loci calculated from the `successful' runs out of the $120$ independent runs with $r_c=1.5a$ and with $100$ ParA-monomers. {\em The figure captions for columns A and B have been reproduced verbatim from \cite{Harju2025.04.03.647011} with minor modifications, especially removal of the references to SI/Appendix of the original paper.} For the experimental data presented in (B), the doubling time of the {\em C. crescentus} cell lines is $ \ge 80$ minutes, 
as the $182^{\circ}$ locus has not been reproduced at the end of $75$~mins. }
\label{fig:ExpAndSim_locitraj}
\end{figure*}
First, it is difficult to unambiguously distinguish the old and new poles in microscopy images containing many cells within the focal plane at a given instant. Second, the time at which a particular locus is replicated and the longitudinal position of a locus undergoing replication cannot be identified precisely from the microscopy images. At the onset of replication, a tagged locus on the mother chromosome appears as a single fluorescent spot. As replication proceeds, the locus is duplicated on the two daughter chromosomes, but the two corresponding spots can be resolved only after they have separated sufficiently for the microscope to distinguish them. Consequently, the apparent onset of locus duplication introduces uncertainty in both the time at which replication occurred, 
and the position of the locus at the time of replication. Third, during replication it is not possible to unambiguously determine which daughter chromosome remains proximal to the old pole and which undergoes directed transport toward the new pole. Thus, the experimental measurements provide the position of a single tagged locus before it separates and the positions of the two daughter loci after separation, but do not directly identify their individual trajectories during the intervening period. Furthermore, the locus position is typically measured relative to the nearest cell pole rather than a consistently identified old or new pole.

These limitations are completely circumvented in our simulations. We have the 
(i) position of all the loci of the mother DNA polymers, (ii) the positions of 
the partially replicated daughter DNA polymers and (iii) the position of the RFs 
at all times of the simulation, even as replication is in progress. 

A one-to-one mapping between the experimentally tagged loci on the chromosomal arms and the corresponding monomer indices in the polymer model is established, as described in SI--VI. This mapping enables a direct comparison between the experimentally observed locus trajectories and those obtained from our simulations. The trajectories 
of the eight loci which were tracked in \cite{Harju2025.04.03.647011}
are shown in Fig.~\ref{fig:simulation_locitraj}. We calculated the distance of each locus
from the old pole, and the distances are shown in units of the instantaneous length 
$L_\mathrm{cyl}$ of the cylinder. Later we also provide the positions of the same loci, but 
plotted using the convention of measurements of the experimental protocol to enable 
direct comparison between experimental trajectories and those obtained from simulations.

{\em Trajectories of loci when distance is measured from the old pole:} 
The trajectories shown in Fig.~\ref{fig:simulation_locitraj} give a simple interpretation of the underlying dynamics of the spatio-temporal organization of the replicating chromosomes. Initially, i.e., 
just at the start of replication,  the two 
arms of the mother chromosome are extended longitudinally along the axis of the cylinder.
The {\em oriC} proximal loci such as $-31^{\circ}$ and  $39^{\circ}$ are positioned 
close to the old pole of the cell (cylinder), where $0$ on the y-axis 
is the position of the old pole. Note that the \emph{ter}-proximal loci  $-121^{\circ}$,
 $134^{\circ}$ and $178^{\circ}$ are positioned near the new pole. Moreover,  the loci 
 at the middle of the two arms, i.e., the $-88^{\circ}$
 and $86^{\circ}$ loci are positioned near the middle of the cylinder.
 
Replication of the $-31^{\circ}$ and $39^{\circ}$ loci occurs close to the old pole. Thereafter,
the corresponding loci of DNA-1 remain near the old pole, whereas the corresponding loci 
of DNA-2 move to the new pole along with the  {\em oriC} of DNA-2. The locus $54^{\circ}$
of the mother DNA is positioned between the old pole and the middle of the cylinder,
gets replicated after the replication of $-31^{\circ}$ and $39^{\circ}$, and follows trajectories
similar to these. The trajectory of the $-88^{\circ}$ and $86^{\circ}$ of the mother DNA is 
instructive: they start out from the middle of the cylinder, are pushed towards the new pole,
but then return to the middle of the cylinder as the monomers of DNA-2 occupy the new pole.
The replication of these two loci occurs near the middle of the cylinder, and then the 
loci of the daughter DNAs segregate and occupy two separate halves of the cylinder. 

The trajectory of the $-178^{\circ}$ loci (which is nearly at the \emph{ter}) loci also 
illuminating.  It remains far from the old pole and maintains its position even as DNA-2 starts occupying the regions close to the new pole due to entropic repulsion with DNA-1. Finally, entropic repulsion between the monomers of DNA-2 and the still-unreplicated mother DNA drives the $-178^{\circ}$  to the middle of the cylinder, where it eventually gets replicated. Loci $-121^{\circ}$ and $134^{\circ}$ precede the \emph{ter} in a similar transition. 

{\em Replisomes:} These trajectories are also consistent with the data of the trajectory of the replisomes provided in the previous subsection: initially, the two replisomes are proximal to each other near the old pole, but as more of the mother gets replicated, the two arms from each of DNA-1 and DNA-2 also entropically move apart from each other even as DNA-1  and DNA-2 segregate to two halves of the cylinder.

{\em Trajectories of loci using convention of experimental measurements:} We detail the convention that was used in \cite{Harju2025.04.03.647011}  to plot the trajectories of the tagged loci. At various stages of the cell cycle, the pole that is closest to the locus (or any one of the loci of the daughter DNAs) is identified, and all distances are measured from that pole. Thus, in the course of measurement, the pole from which distances are being measured can change from (say) old pole to new pole or vice-versa. As an example, refer to the trajectory of locus $86^{\circ}$ shown in Fig.\ref{fig:simulation_locitraj}(d). At the early stages of the simulation, the locus from the mother is closer to the old pole,  but then moves  closer to the new pole. Thereby, initially the distance from the pole will be measured from the old pole and will be less than $0.5 L_\mathrm{cyl}$, but thereafter the distance will be measured from the new pole and will remain less than $0.5 L_\mathrm{cyl}$. 

The trajectories of the loci using this convention, reproduced from \cite{Harju2025.04.03.647011}, are shown in Fig.~\ref{fig:ExpAndSim_locitraj} (A and B) without any image clipping/processing. The distance-measuring convention as well as the color code to denote the mother and the two daughter DNAs are also shown in schematics at the  bottom of subfigure~A. The continuous lines and shaded regions show the experimentally measured mean trajectories and standard deviation, respectively. The symbols (crosses) and error bars indicate the theoretical calculations using the 4D-MaxEnt model (maximizing entropy model) by the co-authors, who use time-resolved Hi-C data at six different stages of the cell cycle as input to their simulations.

\begin{figure*}
    \centering
    \includegraphics[width=0.32\linewidth]{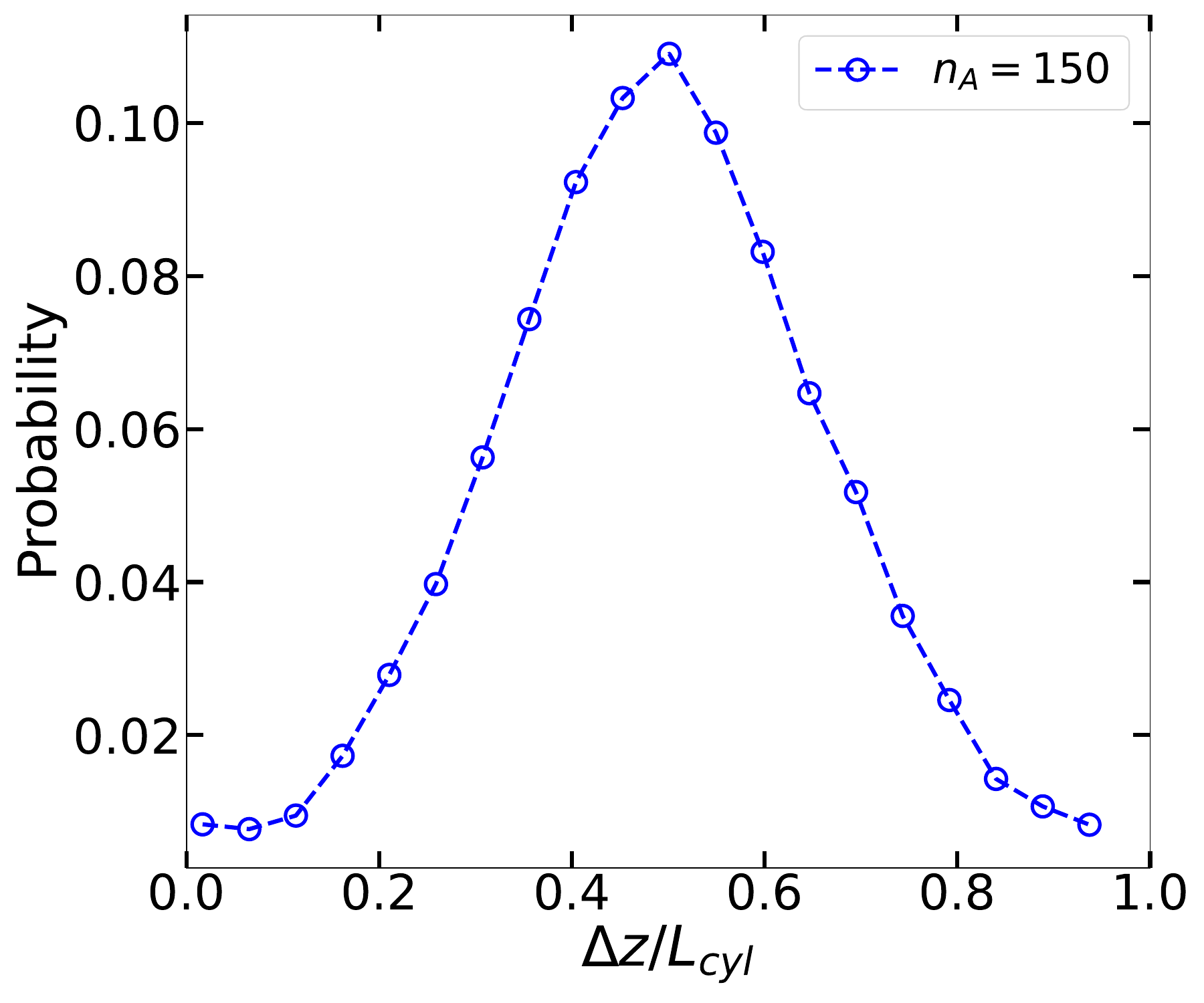}
    \put(-120,80){\large\textbf{(a)}}
    \includegraphics[width=0.32\linewidth]{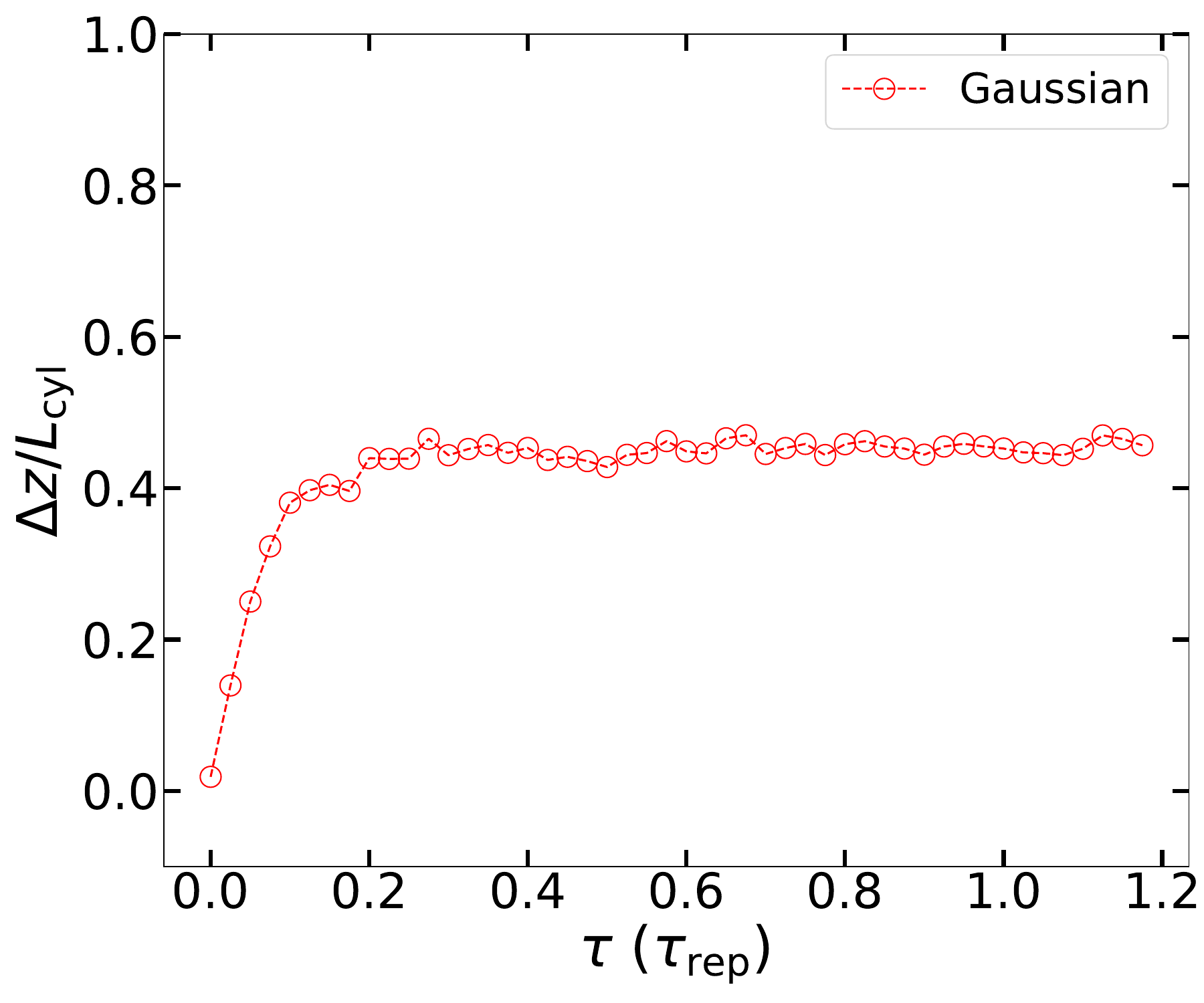}
    \put(-120,80){\large\textbf{(b)}}
    \includegraphics[width=0.32\linewidth]{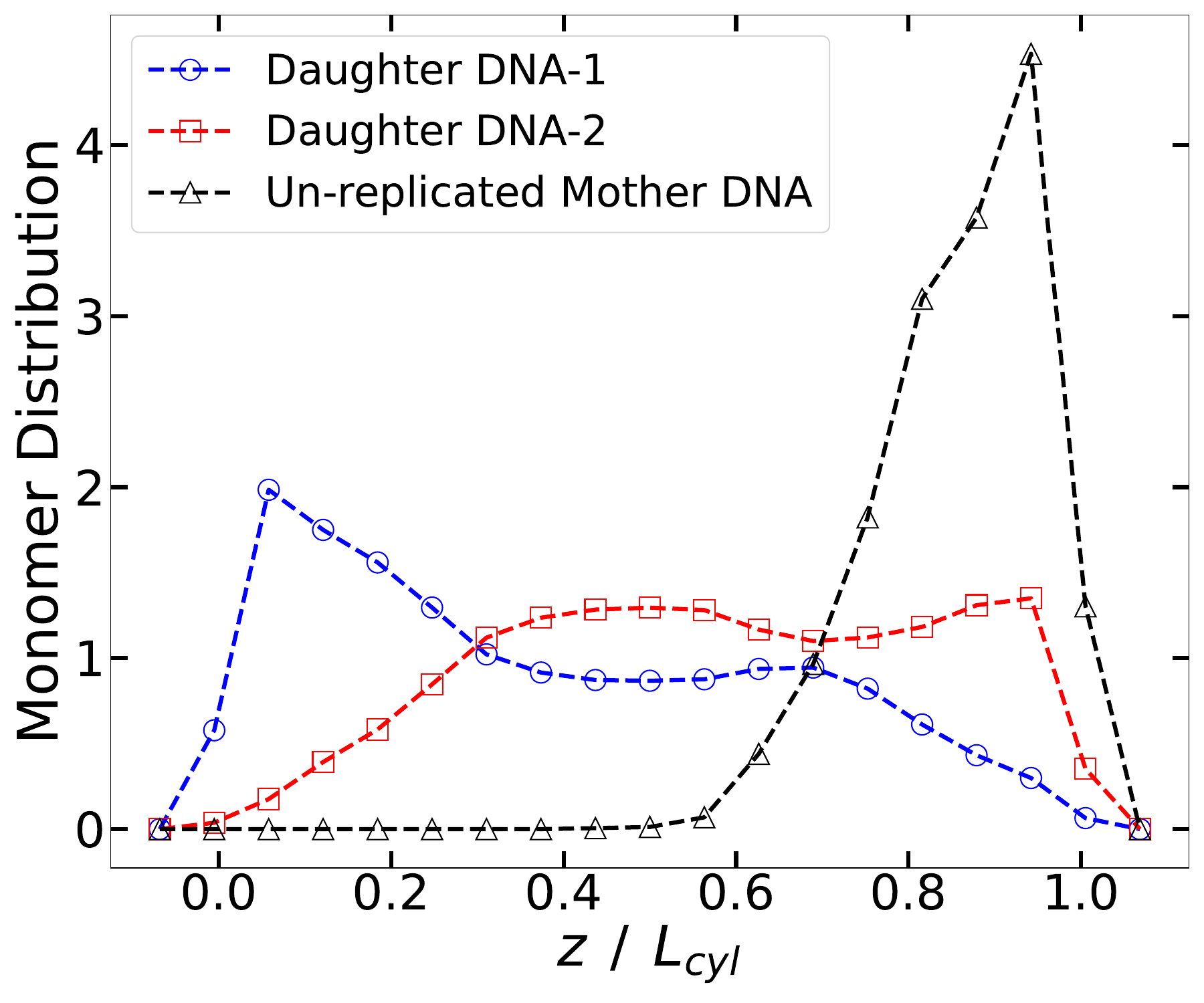}
    \put(-125,80){\large\textbf{(c)}}
    \caption{ \textbf{(a)}  To investigate the role of the ParA gradient, we modify its distribution to peak near the center of the cell. The probability of finding ParA along the long axis of the cell is plotted as a function of the position normalized by the cell length at the initiation of replication. \textbf{(b)} We have plotted simulation data of the distance between two \emph{oriS} averaged over 120 runs, where we have the ParA distribution as shown in figure~\ref{fig:gaussian_parAdist}. In these simulation runs, we have 150 ParA monomers, and the cutoff radius of the interaction potential between ParA--ParB monomers is $r_c^P=0.8a$. \textbf{(c)} distribution of the Monomer density of the mother, daughter DNA-1 and daughter DNA-2 at time $\tau=0.9\tau_{rep}$.  }
    \label{fig:gaussian_parAdist}
\end{figure*}

The Hi-C data used as input to their simulations were obtained at times $0$, $ 10$, $30$, $45$, $60$, and $75$  minutes after the beginning of replication in a synchronized cell population. 
Refer\cite{Harju2025.04.03.647011} for other details of the simulation and experimental protocol.
Note that at $75$ minutes, loci $182^\circ$ which is just $5^\circ$
away from the ter-loci at $-178^\circ$, is still  not replicated. We have shown 
the trajectory for the \emph{ter}. It is germane for the reader to note that our simulation runs 
continue beyond the time of replication of the {\em ter}, which is at $\tau_{rep}=10^5 \tau$.
 
We used the conventions specified in the schematic of Fig.~\ref{fig:ExpAndSim_locitraj}A to replot the trajectories obtained from our simulations. This enables a direct comparison of the trajectory data obtained from our simulations with the experimental data and the theoretical calculations presented in~\cite{Harju2025.04.03.647011}. The data shown in  Fig.~\ref{fig:ExpAndSim_locitraj}B does not show the loci position during the interval when the authors of ~\cite{Harju2025.04.03.647011} expect that the loci are getting replicated. We show loci positions at all times in Fig.~\ref{fig:ExpAndSim_locitraj}C, and the quantitative agreement in the two sets of data
is encouraging. We re-emphasize that the trajectories of {\em oriC},  the various loci, the positional distribution of the replisomes, and the segregation and spatio-temporal organization of the mother and daughter chromosomes obtained from our simulations are an emergent consequence of the transient topological modifications introduced by ParA proteins being bound by ParB protein.  We have provided the time-resolved trajectory of the different loci at $20$ different stages of the cell cycle.


\subsection{The Hi-C map  and linear arrangement of chromosome arms: }
In SI--VII, we reproduce the experimental  Hi-C maps from  \cite{le2013high}, the contact map produced from our simulations using  \emph{Model-2} (with extrusion) and \emph{Model-1} (i.e., without extrusion). Note that the extrusion process creates the ring polymer with modified topology with a linear sequence of internal loops.  It is the linear sequence of loops along the chain contour which organizes the two arms of the mother and daughter polymers along the long axis of the cylinder, and gives rise to the two diagonals of the Hi-C map of~\cite{le2013high}. The linear organization of the chromosome is driven by the entropic repulsion between in the internal loops. The loops entropically avoid each other and occupy different segments of the cylinder along the long axis; refer to~\cite {mitra2022topology} for details.

In SI--VIII, we also provide more detailed positional data of the arrangement of monomers on the  arms of the  mother DNA and the partially replicated daughters at six different stages of replication. In each of the six stages of replication, i.e., when $10\%$, $20\%$, $40\%$, $60\%$, $80\%$ and $90\%$ of replication is complete, we give the mean position of the monomers that belong to the mother and the daughter chromosomes along the long axis. The mean position 
of every alternate monomer along the chain contour is calculated over the `successful' runs of the $120$ runs. Thereby, we also obtain the relative positions of the segments of the mother and daughter chromosomes at different stages of the replication cycle. Initially, the daughter chromosome segments occupy regions proximal to the old pole within the cylinder. But at intermediate stages, large parts of daughter DNA-2 occupy the regions closer to the new pole, where it overlaps with the mother DNA, and the replisome starts moving towards the center of the cylinder. At the last stages of replication, the remaining segments of the mother get localized towards the middle of the cylinder, whereas the two replicated daughter DNAs occupy the two halves of the cylinder, with the two oriCs ($0^\circ$) at the two poles.


\subsection{ If the ParA concentration is peaked at the cell center?}
Previous experiments have also investigated the role of the concentration gradient of ParA in the directed transport of \emph{oriC} and the segregation of daughter chromosomes. In one of these studies, the position of the maximum of the ParA concentration was shifted from the new pole toward the cell center, and the resulting effects on the spatio-temporal organization of the mother and daughter chromosomes were examined \cite{shebelut2010caulobacter}. When the ParA concentration was maximal at the cell center, the \emph{oriC} of daughter-2 remained localized near the cell center, and the two daughter chromosomes failed to segregate.

In our simulation protocol, we also modified the ParA monomer density distribution. In lieu of the concentration gradient observed under normal conditions, we impose the distribution shown in Fig.~\ref{fig:gaussian_parAdist}(a). During replication, every $n_p$ iterations, when the ParA--ParB is switched off, the ParA monomer is redistributed according to this imposed profile.

In this scenario, our simulation shows that the \emph{oriC} of Daughter-2 initially moves away from the old pole but becomes localized near the cell center and does not proceed toward the new pole. Its trajectory is shown in Fig.~\ref{fig:gaussian_parAdist}(b), consistent with what is known from experiments. This localization results from the continued formation of ParA--ParB interactions with ParA-loaded chromosome segments located near the cell center. These segments initially belong to the mother chromosome and subsequently to the daughter chromosomes as replication proceeds. The resulting attractive ParA--ParB interactions locally oppose the entropic tendency of the daughter chromosomes to segregate, thereby preventing displacement of \emph{oriC} toward the new pole.


To further characterize the resulting chromosome organization, we show the monomer density along the longitudinal axis of the cell in Fig.~\ref{fig:gaussian_parAdist}(c). The density profiles demonstrate that the daughter chromosomes remain substantially overlapped even at $t=0.9\tau_{\mathrm{rep}}$, indicating that chromosome segregation is strongly inhibited. In particular, both daughter chromosomes exhibit enhanced monomer density near the cell center, while DNA-1 also extends substantially into the half of the cell proximal to the new pole, as the mother DNA and the RF remain proximal to the old pole. These density profiles are consistent with the localization of \emph{oriC} near the cell center as seen in experiments.

\section{Conclusion and Discussion} \label{sec:conclusions}
In summary,  we show with our coarse grained simulations with \emph{Model-2} (and \emph{Model-1})  that the spatio-temporal organization of different chromosomal loci, the rapid transport of the {\em oriC} towards the new pole and accompanying segregation of the daughter chromosomes, and the relocalization of the replisomes from the old pole to the cell center and spatial separation of the two replisomes are likely to be emergent phenomenon driven by the combination of entropic effects due to consequences of topological modifications by ParABS interactions, and enthalpic effects of ParB monomers in a ParA concentration gradient. Moreover, the modified topology created by extrusion on the daughter arms contributes to maintaining the linear arrangements of arms, as can be inferred from the Hi-C map of \emph{C. crescentus}.

While previous modeling studies~\cite{Sumitabha2026,Harju2025.04.03.647011} predict the (temporal) organization of bacterial chromosomes using the Hi-C maps of {\em E. coli}, {\em B. subtillis} and {\em C. crescentus} as inputs to their simulations, we have adopted a very 
distinct approach in this paper. We seek to identify the minimal model in 
which local protein-mediated interactions modify the topology of replicating chromosomes. Thereby driving the spatio-temporal organization of the mother and the daughter chromosomes. The topological modifications do evolve in time as replication is in progress, and result from the transient interactions between the ParABS system of proteins or 
due to extrusion of the arms of the daughter chromosomes. Thus we focus on the spatially and 
temporally local actions of proteins on the DNA, and obtain directed motion of loci and suitable reorganizations of the DNA segments, thereby making the chromosomes suitably positioned for 
the start of cell division.  

We have investigated the effects of varying the concentration as well as the axial distribution of ParA monomers on chromosome organization. We have also compared chromosome organization in the presence and absence of extrusion. These studies allow us to disentangle the effects of two distinct protein-mediated mechanisms and identify their respective contributions to chromosome organization. We propose that these two systems constitute important drivers of chromosome organization at the $\mu$m length scales considered here.

Other chromosomal interactions, as inferred from Hi-C maps, are likely to contribute to chromosome organization at smaller length scales and  also play important roles in gene regulation. For example, in our previous work, we showed that additional internal loops in the \emph{ter}-proximal region of the segregated ToMo polymers (as plausible models of {\em E. coli} chromosomes) enhance their demixing near the midcell region~\cite{mitra2022topology}. Such topological modifications, plausibly brought about by MatP proteins~\cite {Karoui2008}, could contribute to the efficient separation of daughter chromosomes during cell division.

Thus, rather than providing a complete description of all chromosomal interactions, our model identifies how ParABS-mediated interactions and chromosome extrusion can contribute to chromosome organization and segregation within a common coarse-grained framework. A similar minimal-modeling strategy has been used in our previous studies to investigate the spatio-temporal organization of \emph{E. coli} chromosomes under fast- and slow-growth conditions~\cite{dna1Mitra2022,dna2Pande2024}.

This approach is complemented by our studies of topologically modified (ToMo) polymers, which focused on individual aspects of their emergent behavior without incorporating the multiple active processes relevant to chromosome dynamics~\cite{Kingkini2025,Bhandarkar2026,doshi2026kineticssegregationtopologicallymodifiedring}. Such simplified systems allow the effects of individual topological features to be isolated and their underlying mechanisms to be examined without the additional complexity introduced by simultaneously occurring nonequilibrium biological processes. In particular, these studies demonstrated how the size and position of internal loops in ring polymers can modulate entropic interactions between different polymer segments and thereby control their organization under confinement. Insights from these minimal models provide a basis for interpreting the more complex chromosome simulations considered here, in which several active processes operate simultaneously.

Whether the principles explored in Ref.~\cite{Kingkini2025} can be extended, within similarly minimal models, to investigate the spatio-temporal organization of chromosomes in eukaryotic cells remains an interesting direction for future work.

\section*{Acknowledgements} \label{sec:acknowledgements}
The authors acknowledge useful discussions with Arieh
Zaritsky and Conrad Woldringh. AC, with DST-SERB (IN) identification No. SQUID-1973-AC-4067, acknowledges funding by DST-SERB (IN)  project CRG/2021/007824. AC is grateful to Biosantexc research and mobility grants which enabled him to visit ENS-Lyon and ENS-Paris Saclay for useful discussions. AC is also grateful for a Fulbright-Nehru grant from USIEF which enabled him to host and extensively discuss with Prof. Jie Xiao. AC also acknowledges several discussions in the meetings organized by ICTS, Bangalore, India, and the use of the PARAM-BRAHMA computing facilities. The authors acknowledge useful discussions with Debarshi Mitra, Shreerang Pande and Kingkini Roychoudhury. \\

\section{Data Availability}
The data presented in this work were generated and analyzed using HOOMD-blue and other homegrown codes that are freely available \cite{Sathe_Ccrecentus_SS_AC_2026}.

\section{Author Declarations}
\subsection{Conflict of Interest}
The authors have no conflicts to disclose.

\subsection{Author Contributions}
AC designed the initial project problem and supervised the research project. SS performed the simulations and analysis, and also designed subsequent investigations which 
allowed direct comparison with experimental results. Both the authors contributed to the interpretation of the results and writing of the manuscript. 

\section{Appendix} 
\label{sec:appendix}

In particular, it is not feasible to estimate the effective friction coefficient, $\zeta$, of a coarse-grained bead representing approximately $4$ kbp of chromosomal DNA together with its associated DNA-binding proteins. Moreover, the soft interaction potential employed in the simulations permits bead interpenetration and chain crossing, thereby mimicking the action of topoisomerase that relieves topological constraints in vivo. Although the order of magnitude of the bead sizes and repulsive strength are motivated by estimates from bacterial ~\cite{messelink2021learning,Harju2025.04.03.647011} and eukaryotic systems \cite{kadam2023predicting}, their quantitative relevance to a coarse-grained bead representing $4$ kbp of \emph{C. crescentus} DNA remains unverified. Consequently, we restrict our comparison between simulations and experiments to dimensionless measures and relative time scales.
\bibliography{bibliography}

\end{document}




\begin{center}

{\Large\bfseries ParABS system induces topological modifications to spatio-temporally organize its mother and daughter chromosomes in the replicating \emph{C. crescentus} cell: Supplementary Information  }

\vspace{0.5cm}

Shailesh Sathe and Apratim Chatterji


\emph{Department of Physics, \\ Indian Institute of Science Education and Research  (IISER-Pune), \\  Dr. Homi Bhabha Road, Pune, India-411008.}

\end{center}

\vspace{0.5cm}

\section{Simulation Model}
\subsection{The bead-spring model of the DNA polymer}
The polymer ring is represented as a sequence of spherical beads connected by harmonic springs. Thus we use a coarse-grained representation of a
\emph{C. crescentus} chromosome, where we choose to have $N=1000$ monomeric beads in the ring polymer. Thereby, each monomer represents $4$ kilo-basepairs (kbps) of DNA segments, as the {\em C. crescentus} chromosome has $4$ million base pairs. Using this simplified description and additional features. we can capture the essential features of chromosome organization. This simplified model of the chromosome allows us to probe the role of entropic and topological effects under cylindrical confinement. 

We use a harmonic spring potential $U_{spring}$  to model the connectivity between adjacent beads along the chain contour. The expression for $U_{spring}$ is,
\begin{equation}
    U_{spring}=\frac{1}{2} \kappa(r- a)^2,
\end{equation}
where $a$ is the equilibrium bond length. We set $a =1$, and measure all other lengths in units of $a$.
We use the value of the  spring constant  $\kappa$ to be $100k_BT/a^2$.
We use  a spherically symmetric soft potential $U(r)$ to model the effective excluded volume (EV) interactions between the beads.  The soft potential $U(r)$ has the form :
\begin{equation}
U_{EV}(r)   =
\begin{cases}
A \, \exp\!\left(-\dfrac{r}{R}\right), & r < r_c \\
0, & r \geq r_c
\end{cases}
\end{equation}
where we choose $A=5 k_B T$, $r_c= 0.8a$ and $R = 0.6a$, unless specified otherwise. 
An estimate of the diameter $\sigma$ of bead is taken to be equal to the range of the soft repulsive potential, i.e. $ \sigma = r_c=0.8a$. The soft potential allows partial overlap between monomers. Moreover, it allows chain crossing which allows the release of topological constraints between any polymer segments.  To avoid discontinuities in the force at $r_c$, the pair potential can be modified according to the selected smoothing mode. We have used the method of ``xplor" (refer \cite{ANDERSON2020109363}) to ensure that the potential and force goes to zero at the cutoff. All of the simulations were performed using HOOMD-blue \cite{ANDERSON2020109363} on a CPU.

\begin{figure}
    \centering
    \includegraphics[width=0.4\linewidth]{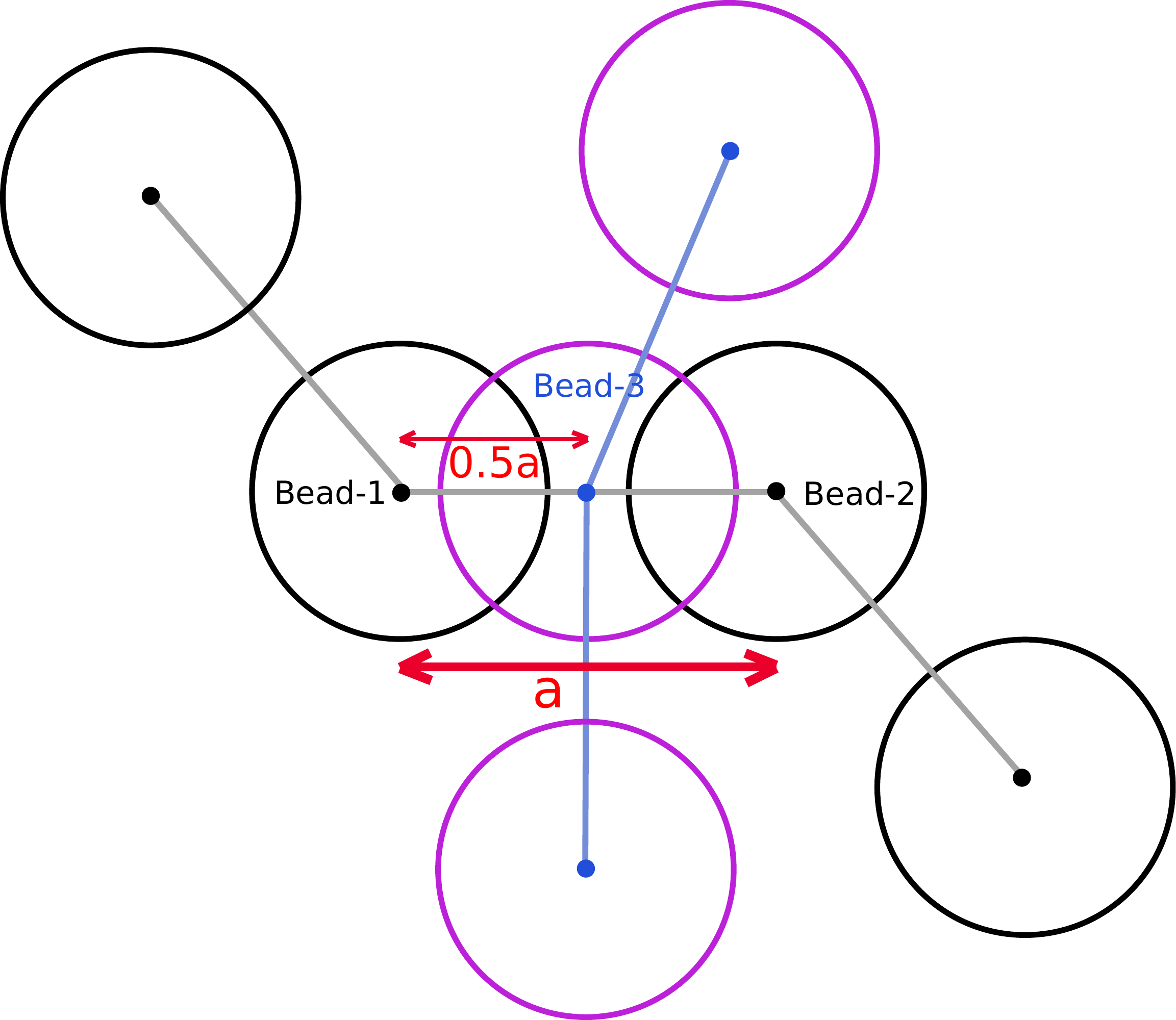}
    \caption{Schematics  showing the overlap of Bead-3 with both  Bead-1 and Bead-2.
    Bead-1 and Bead-2 are adjacent monomers along the contour of a DNA polymer (shown in black),  and thereby bonded by a harmonic spring. However, the center of  Bead-3, belonging to a separate segment of the same or any other DNA polymer (shown in purple), has its center at the exact middle of the line joining the center of Bead-1 and Bead-2. Such a situation may arise when two different segments are crossing each other to release topological constraints. However, in a simulation, the center of Bead-3 need not always be at the center of the line joining Bead-1 and Bead-2.}
    \label{fig:overlap_schem}
\end{figure}

The release of topological constraints is crucial for DNA-segregation as has been demonstrated \cite{dna1Mitra2022,mitra2022topology}. A polymer model that allows for the release of topological constraints incorporates the role of topoisomerases in \emph{C. crescentus} cells. Refer to Fig.~\ref{fig:overlap_schem} for a schematic of two chains passing through each other. For two chain segments to cross each other, a bead (say Bead-3) of a polymer segment has to be in between two adjacent monomers (say Bead-1 and Bead-2) belonging to another polymer segment.  Next, we provide an estimate of the energy barriers that must be overcome for two chains to cross in our model.

Suppose that distance between Bead-1  and Bead-2 is $a$, so that the spring is not stretched. Suppose that the center of Bead-3 lies at the midpoint of the line joining the centers of Bead-1 and Bead-2, as shown schematically in Fig.~\ref{fig:overlap_schem}, such that the distance between the center of  Bead-3 from Bead-1 and Bead-2 is $0.5a$. In this configuration, the total overlap energy of the triplet of monomers is $2.88 k_BT$.

However, if the distance between Bead-1  and Bead-2 is $1.2 a$, so that the spring between the two monomers is stretched, the harmonic energy cost is $\frac{1}{2} \kappa (0.2)^2 = 2k_BT$. If the center of Bead-3 sits at the center of the line joining the centers of Bead-1 and Bead-2, such that its distance from each is $0.6a$, then the overlap energy of the triplet of monomers is $1.5 k_BT$. In this scenario, the total overlap energy of the triplet of monomers is $3.5 k_BT$. Refer to the schematic shown in~Fig.\ref{fig:overlap_schem}. 

We use the Langevin equation of motion to describe the dynamics of the polymer.
    
   \begin{equation}
    m \frac{d\vec{v}_i}{dt} \;=\; \vec{F}_c \;-\; \Gamma \vec{v}_i \;+\; \vec{F}_r(t),
\end{equation}

\begin{equation}
    \langle \vec{F}_r(t) \rangle \;=\; 0,
\end{equation}
\begin{equation}
    \langle F_{r,\alpha}(t)\, F_{r,\beta}(t') \rangle 
    \;=\; 2 \Gamma k_B T \, \delta_{\alpha\beta}\,\delta(t-t'),
\end{equation}
where $ \vec{v}_i$ is the velocity of the $i$-th monomer, $\vec{F_c}$ is the force due to all the conservative potentials, $\Gamma$ is the viscous damping constant for the drag force $\Gamma \vec{v}_i$ acting on the $i$-th bead, and  $F_r$  is a random force which is delta-correlated in time, from a uniform random number generator between $-1$  and $+1$.
All of the simulations were performed using HOOMD-blue \cite{ANDERSON2020109363} on a CPU.  As established in Ref.~\cite{Dunweg1991-zi}, random forces generated from a uniform distribution, with an appropriate scaling chosen to satisfy the fluctuation-dissipation theorem, reproduce the correct statistical behavior in the long-time limit.  The equations of motion are numerically integrated using the Velocity-Verlet algorithm. The value of viscous damping force constant is $\Gamma=m/\tau$, where $\tau= a \sqrt{m/k_BT}$, where  $a=1,m=1, \tau =1$ and and we use timestep $\Delta t =0.01\tau$.   

We start our simulations with the equilibration of an $N=1000$ monomer ring polymer 
 when we simulate using {\em Model-1},  and  with the modified architecture shown in Fig.~$1$(a) of main text when we run simulations using {\em Model-2}. 
 When we use {\em Model-2},  we additionally introduce springs,$U_{spring}$, between monomers ($166$) and ($832$) to cross-link them, and between monomers ($333$) and ($666$). This ensures a linear arrangement of loops (see Ref.~\cite{mitra2022topology}), which helps keep the left and right arms of the chromosome overlapped along the length of the cylinder. We use insights from our previous study \cite{mitra2022topology}. However, we release the additional 
CLs of {\em Model-2} at the start of replication and reintroduce them through the extrusion process in daughter DNA-1 and daughter DNA-2. Details are described in a subsequent subsection.  The equilibration times and other details related to initialization are given in the next subsection.





\subsection{Replication Scheme} \label{Replication}
 Replication initiates at the \emph{oriC} locus (at monomer $1$) and proceeds bidirectionally toward the \emph{ter} locus (monomer $500$). The mother chromosome contains $1000$ monomers, indexed from $1$ to $1000$, with \emph{oriC} at monomer $1$ and the two arms extending to monomer $500$. We equilibrate the mother DNA polymer for $10^6 \tau$ before the start of the simulation. Once the simulation starts, we store the position of all the monomers (belonging to the mother DNA polymer, daughter DNA-1, and DNA-2) after $400 \tau \equiv 4 \times 10^4$ iterations. We run the simulation for another $2 \times 10^4 =200 \tau$ iterations before the initiation of replication.

 The two replication forks (RFs) are initially identified with monomers with indices  $1$ and $1000$ 
 of the mother DNA polymer. At the initiation of replication, two new monomers are introduced in lieu of each of the two monomers numbered $1$ and $1000$ of the mother DNA. The new monomers, referred to as monomers $1$ and $1'$ (and $1000$ and $1000'$), belong to daughter DNA-1 and daughter DNA-2, respectively, and are connected by springs to monomers $2$ and $999$ of the mother DNA polymer. Monomers with indices $2$ and $999$ are the new positions of the two RFs, respectively. After $n_R=20,000$ Langevin iterations, corresponding to $200\tau$, during which the polymer is allowed to relax  locally, the  monomers $2$ and $999$ are replicated. Thus, each of these two monomers of the mother DNA is replaced by two new monomers. The RFs then advance to monomers $3$ and $998$, respectively. This procedure is repeated every $n_R$ iterations (unless stated otherwise). Thus, each RF traverses $499$ monomers before reaching the \emph{ter} locus over a replication time of  $\tau_{rep} = N/2 \times 200 \tau$.

More generally, when the RFs are located at monomers $i$ and $1000-i+1$, replication of these loci introduces two daughter monomers, one on each daughter DNA polymer. For example, in
Fig.~$1$(c), we show RFs located at monomers $9$ (and $992$). Replication of these loci generates the corresponding daughter monomers ($9$ and $9'$),  while maintaining bond lengths of approximately $a$ with the neighboring monomers.  The neighbouring monomers are $8$ of daughter DNA-1, $8'$ for daughter DNA-2 and monomer $10$ of mother DNA.  The RFs then advance to monomers $10$ and $991$, respectively, and the process is repeated. The replication procedure is analogous to the train-and-track model. Since replication conditions in bacterial cells depend on the growth environment, the replication time in the simulations can be rescaled to the experimentally relevant timescale. In our simulations, two monomers are replicated every $200\tau$, giving a total replication
time of $\tau_{rep} = N/2 \times n_R = 10^5 \tau = 10^7$ iterations.

We do not model the replication fork explicitly. Instead, we identify the RF position with the loci immediately ahead of the replicated region, i.e., the $(i+1)$th monomer on each arm of the mother chromosome. Thus, the RFs advance by one monomer along each arm after every replication interval.

\subsection{Tethering of oriC by Pop-Z to the poles} \label{Popz_modelling}
 
 The mother DNA polymer is confined within a cylinder. At the start of the simulation, i.e., at
 time $t=0$, the length of the cylinder is $L_\mathrm{cyl}=28a$, and the diameter of cylinder is $D_{\mathrm{cyl}}=8a$.   The monomer with index $1$ is labeled as the {\em oriC} and the $500-$th monomer is the {\em ter}. We tether monomer-$1$ to one end of the cylinder, which we 
 refer to as the old pole, using a harmonic potential $U_{spring}^{Pole}$ with $\kappa=100k_BT$   between monomer-$1$ and the old pole, where the equilibrium length of the spring is $a$. 

 During replication, the mother DNA is replicated to form two daughter DNA molecules. In our model, one of the replicated \emph{oriC} loci remains tethered to the old pole, while the second \emph{oriC} undergoes directed transport toward the new pole. To model the tethering  of the second \emph{oriC} to the new pole by the {\em popZ}, we introduce a virtual particle at the new pole that attracts the second \emph{oriC} and nearby ParB monomers once they approach the virtual particle. The virtual particle has no repulsive interactions with monomers belonging to either the mother or daughter DNAs. We implement this attraction using the potential $U(r)$:
 
\begin{equation}
U_{Pole}(r) =
\begin{cases}
A' \, \exp\!\left(-\dfrac{r}{\rho'}\right), & r < r_c' \\
0, & r \geq r_c'
\end{cases}
\label{soft_attraction_pop}
\end{equation}
where $r$ is the distance between the virtual particle at the new pole and the ParB monomers of daughter DNA-2, and the cutoff for this attractive interaction is $r_c'= 1.5a$. We choose $A'=-30 k_BT$, $\rho' = 0.8a$. This potential tethers the second {\em oriC} to the new pole.

\subsection{Modeling Walls and Increase in Cylinder Length} \label{Confinement}

We confine the ring polymer with modified architecture within cylindrical confinement with 
an aspect ratio of $3.5$,  as seen in \emph{C. crescentus} cell. The diameter of the cylinder is $8a$ simulation units, which is smaller than the  radius of gyration $R_g$ of the unconfined ring polymer in free space. We implement the repulsive wall potential as follows:

  \begin{equation}
    U_{wall}= \sum_{i=1}^{N_{Particles}} \sum_{\omega \in walls} U_\omega (d_i),
\end{equation}
where the summation is over all particles present within the cylinder. 
The potential $U_\omega(d_i)$ represents a repulsive wall potential acting on particle $i$ when its distance $d_i$ from the wall surface is smaller than a prescribed cutoff distance $d_{\mathrm{cut}}=2^{1/6}.\frac{\sigma}{2}$, where $\sigma=0.8a$.  We use a shifted Lennard-Jones potential for $U_{\omega}(d_i)$, with the shift chosen such that both the potential and the force continuously vanish at $d_{\mathrm{cut}}$.  Specifically, the form of the potential is
\begin{equation}
    U_{\omega}(d_i)=4\epsilon\left[\left(\frac{\sigma}{d_i}\right)^{12}-\left(\frac{\sigma}{d_i}\right)^6\right]+\Delta V(d_i) -  V_{\omega}^c,
\end{equation}
where the value of  $V_{\omega}^c$ is the value of the sum of the first two terms at $d_{cut}$ and 
\begin{equation}
    \Delta V(d_i)=-(d_i-d_{cut}) \frac{\partial V_{LJ}}{\partial d}(d_{cut}).
\end{equation}
The resulting force acts along $\vec{d}_i$, defined as the vector from the closest point on the wall surface to the center of particle $i$. 

 In our simulation, the length of the cylinder increases every $n_R$ iterations
 as replication progresses. We increase the cylinder length according to an exponential growth protocol during replication. By the end of the replication, the cylinder length has increased to twice its initial value, while its diameter remains fixed. The length 
 of the cylinder does not increase once replication is complete,  i.e., time $\tau_{rep}$
 is reached. After completion of the replication process, we further evolve the system for an additional \(2\times10^4\tau\). Thus, we implement:

\begin{equation}
L(t) =
\begin{cases}
L_\mathrm{cyl} \, \exp(k.t), & t \leq \tau_{rep} \\
2L_\mathrm{cyl}, & t > \tau_{rep}
\end{cases}
\end{equation}
where 
\begin{equation}
    k= \frac{log(2)}{\tau_{rep}}
\end{equation}

\subsection{Modelling ParABS system} \label{ParABS}


We consider four categories of monomers in our simulations: ParA, ParB, normal DNA monomers belonging to the mother or daughter chromosomes, and PopZ. All monomers except PopZ interact with each other through a soft repulsive potential to mimic excluded-volume (EV) interactions. Since we do not explicitly model the molecular interactions between ParA and DNA (and ParB on {\em parS} sites and DNA), specific monomers are chosen and designated as  ParA-loaded sites. These are then referred to as ParA (and ParB) monomers. The ParA monomers are assigned additional properties that allow them to transiently bind through attractive interactions with ParB monomers as described below. ParA monomers can lose their special identity upon completion of transient binding, and become normal monomers, while other normal monomers concurrently become ParA monomers. The two monomers on either side of \emph{oriC}, on the other hand, are designated as ParB monomers and remain so throughout the simulations.

We vary the number $n_A$ of ParA monomers in our simulations, using $n_A=100$ as the default value and considering $n_A=70$, $130$, and $160$ as additional cases. The ParA monomers are distributed to generate a concentration gradient along the cylinder, extending from the new pole toward the other pole. At $t=0$, the ParA monomers are distributed over of the cylinder closest to the new pole, with $L_{\mathrm{cyl}}=28a$. This region is divided into 11 bins of equal width, and the number of ParA monomers in each bin at a distance $z_i$ from the new pole is determined according to the prescribed initial distribution. The initial distribution of ParA monomers on the mother DNA before the start of the simulations is given by:
\begin{equation}
N(z_i)=N_0 e^{-z_i/6.6a},
\qquad
\sum_{i=1}^{11}N(z_i)=n_A .
\end{equation}
The mother-DNA monomers present in each bin are identified, and $N(z_i)$ monomers are selected at random from each bin and designated as ParA monomers.

\begin{figure}[htbp]
    \centering
    \includegraphics[width=0.5\linewidth]{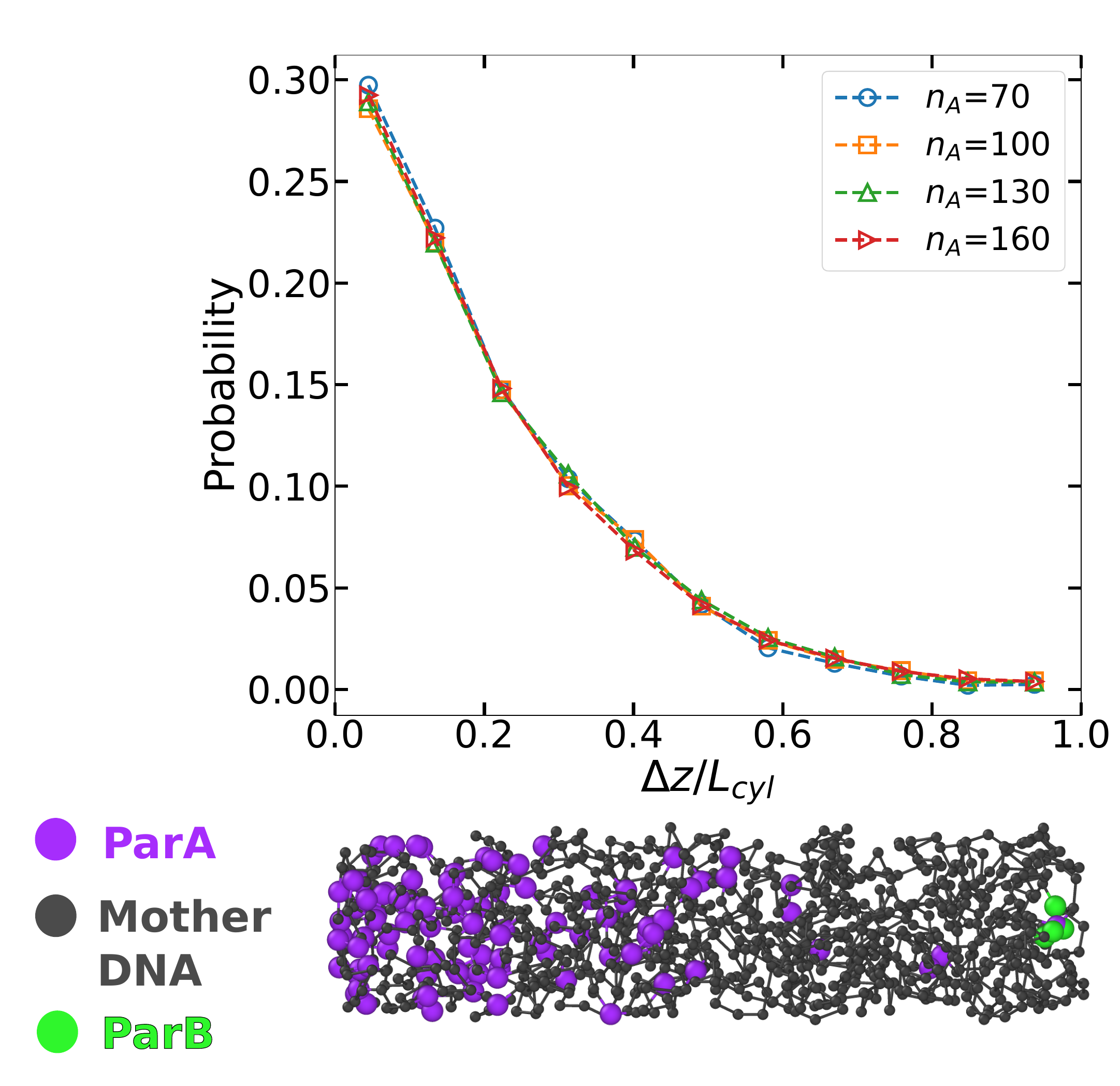}
    \put(-230,200){\large\textbf{(a)}}
    \put(-200,50){\large\textbf{(b)}}
    \caption{ \textbf{(a)} The calculated distribution of ParA monomers along the longitudinal axis of the cylindrical confinement at the beginning of the simulation, where $z$ is the distance from 
    the new pole of the cell. The ParA concentration is highest at the new pole. We show the initial distribution of ParA monomers for different values of $n_A$ (refer to the legend) used in the simulation. For most of the results presented, we use $n_A=100$ unless otherwise mentioned. \textbf{(b)} Representative snapshot of the initial configuration of the monomers of the mother DNA (black circles for normal monomers), showing $100$ ParA monomers (purple). Four ParB monomers are shown in green at the old pole, illustrating their spatial distribution prior to the start of the simulation.}
    \label{fig:ParA-Bgradient}
\end{figure}

The calculated distribution of the $n_A$ ParA monomers along the longitudinal axis of the cylinder is shown in Fig.~\ref{fig:ParA-Bgradient}(a) for different values of $n_A$. A representative initial configuration of the polymer inside the cylindrical confinement is also shown in Fig.~\ref{fig:ParA-Bgradient}(b), for $n_A=100$. To account for the effective affinity between ParA and ParB, we introduce a specific pairwise interaction between ParA and ParB monomers. This interaction is modeled as the sum of an attractive potential, representing the effective ParA--ParB affinity, and a short-range repulsive potential that prevents overlap between the interacting particles. The total ParA--ParB interaction potential is therefore given by:
\begin{equation}
U(r) = U_{\mathrm{attractive}}(r)+U_{\mathrm{repulsive}}(r),
\label{ParAParBinteraction}
\end{equation}
where
\begin{equation}
U_{\mathrm{attractive}}(r) =
\begin{cases}
A_{\mathrm{att}} \exp\left(-\dfrac{r}{\rho_{\mathrm{att}}}\right), & r < r_c^P,\\
0, & r \geq r_c^P,
\end{cases}
\label{soft_attraction}
\end{equation}
and
\begin{equation}
U_{\mathrm{repulsive}}(r) =
\begin{cases}
A_{\mathrm{rep}} \exp\left(-\dfrac{r}{\rho_{\mathrm{rep}}}\right), & r < r_c^P,\\
0, & r \geq r_c^P.
\end{cases}
\label{soft_repulsive}
\end{equation}

We choose $A_{\mathrm{att}}=-20 k_BT$, $\rho_{\mathrm{att}}=0.6a$, and $A_{\mathrm{rep}}=15 k_BT$,  and $\rho_{\mathrm{rep}}=0.2a$, $r_c^P=1.5a$, giving a short-range repulsive contribution. The ParA--ParB interaction is activated when the separation between a ParA and a ParB monomer becomes smaller than $r_c^{\mathrm{att}}$. The potentials are smoothly truncated at their respective cutoffs to avoid discontinuities. We have varied the range of attractive interaction to assess the effect of a reduced range of attraction. The ParA and ParB proteins will be much smaller in size 
than the size of the bead considered in the simulation.

Every ${n_P}$ iterations, we identify the ParA--ParB pairs formed during the simulation and turned off the corresponding ParA--ParB interactions. Once a ParA--ParB interaction is turned off, the ParA monomer is reclassified as a normal DNA monomer. A different normal DNA monomer is then selected at random and reclassified as a ParA monomer, thereby keeping the total number of ParA monomers in the system fixed. The newly selected ParA monomer may belong to the mother DNA polymer or to either of the two daughter DNA polymers.

To determine the location of the redistributed ParA monomer, we first calculate the distribution $N'(z_i)$ of the remaining ParA monomers along the longitudinal axis of the cylinder after the ParA--ParB interaction is turned off. We then calculate the difference between the initial ParA distribution, $N(z_i)$, and the updated distribution, $N'(z_i)$:
\begin{equation}
N^{\mathrm{diff}}(z_i)=N(z_i)-N'(z_i).
\end{equation}
The resulting distribution represents the spatial deficit of ParA monomers along the longitudinal axis and is used to determine the position at which the redistributed ParA monomer is introduced.

Although the length of the cylinder increases during the simulation, the redistributed ParA monomer is introduced only within a distance $L_\mathrm{cyl}$ from the new pole. We employ a Metropolis algorithm using $N^{\mathrm{diff}}(z_i)$ to select a new position along the longitudinal axis. Once the position is selected, we identify all mother and daughter DNA monomers located within the corresponding spatial bin. One monomer is then selected at random from this set and reclassified as a ParA monomer. Thus, the redistributed ParA monomer is assigned to either the mother DNA polymer or one of the daughter DNA polymers while preserving the underlying identity of the selected DNA polymer. This procedure is repeated sequentially for each ParA monomer that was bound to a ParB monomer.

\begin{figure}
    \centering
    \includegraphics[width=0.32\linewidth]{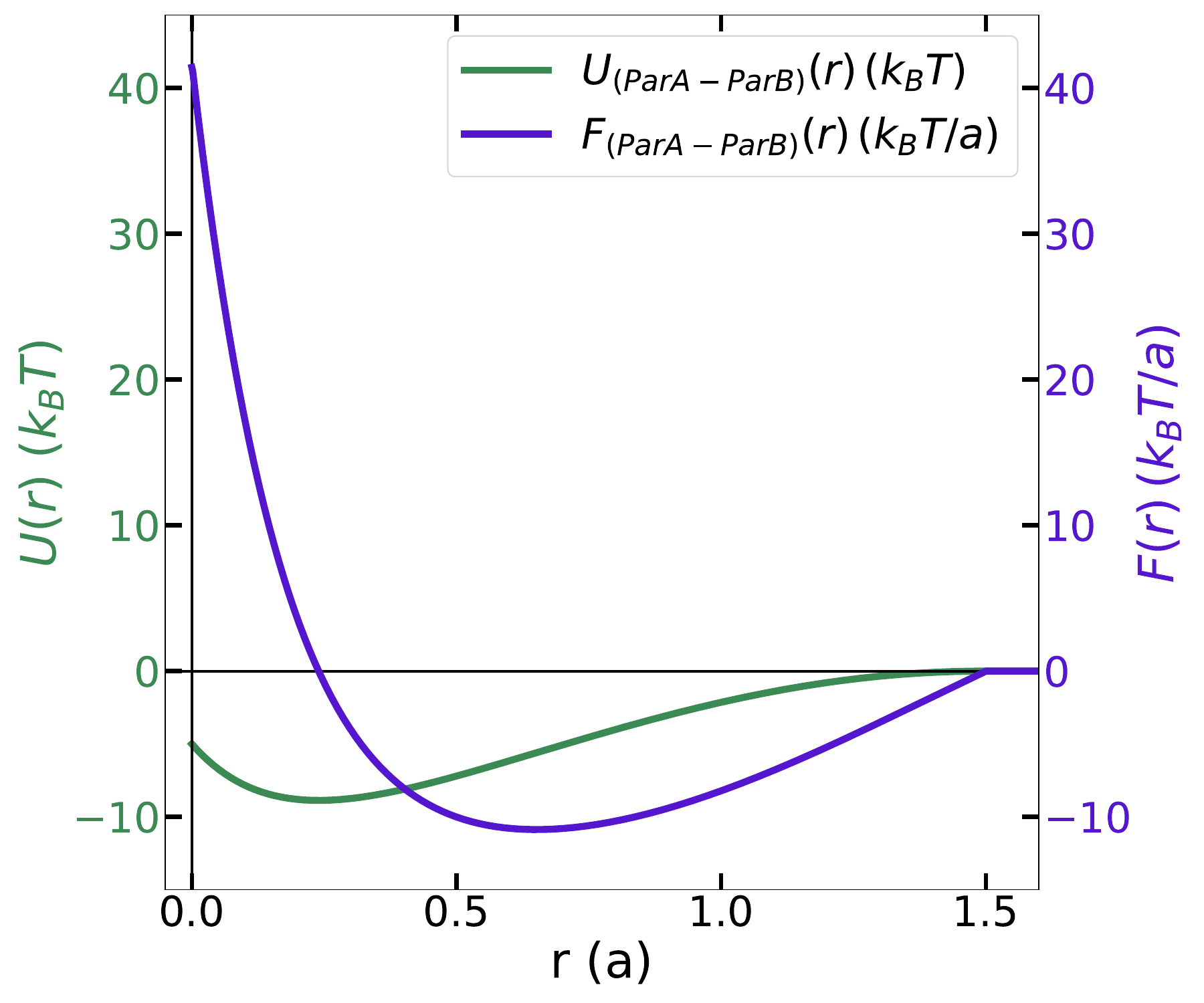}
    \put(-130,100){\large\textbf{(a)}}
     \includegraphics[width=0.32\linewidth]{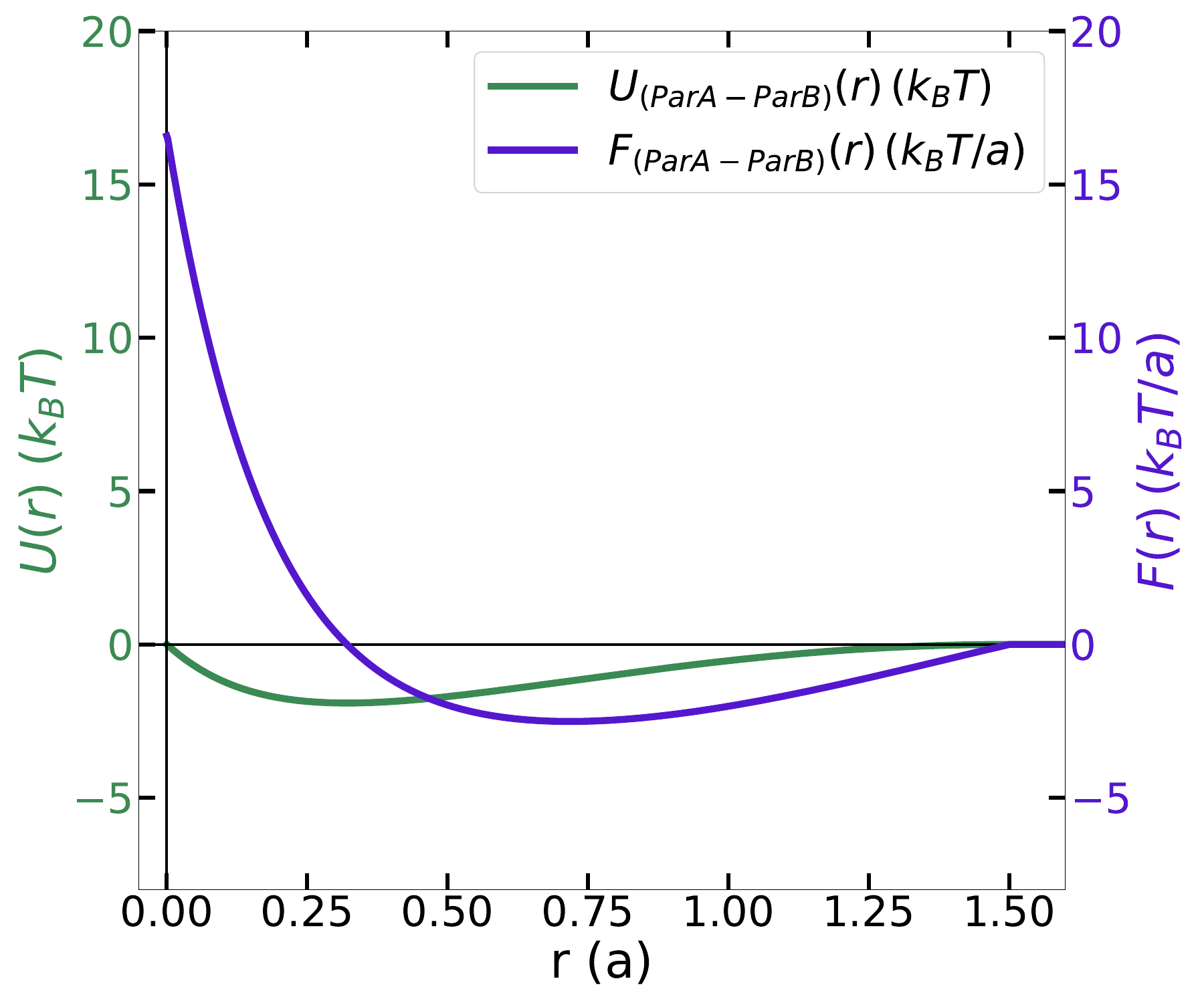}
     \put(-130,100){\large\textbf{(b)}}
     \includegraphics[width=0.32\linewidth]{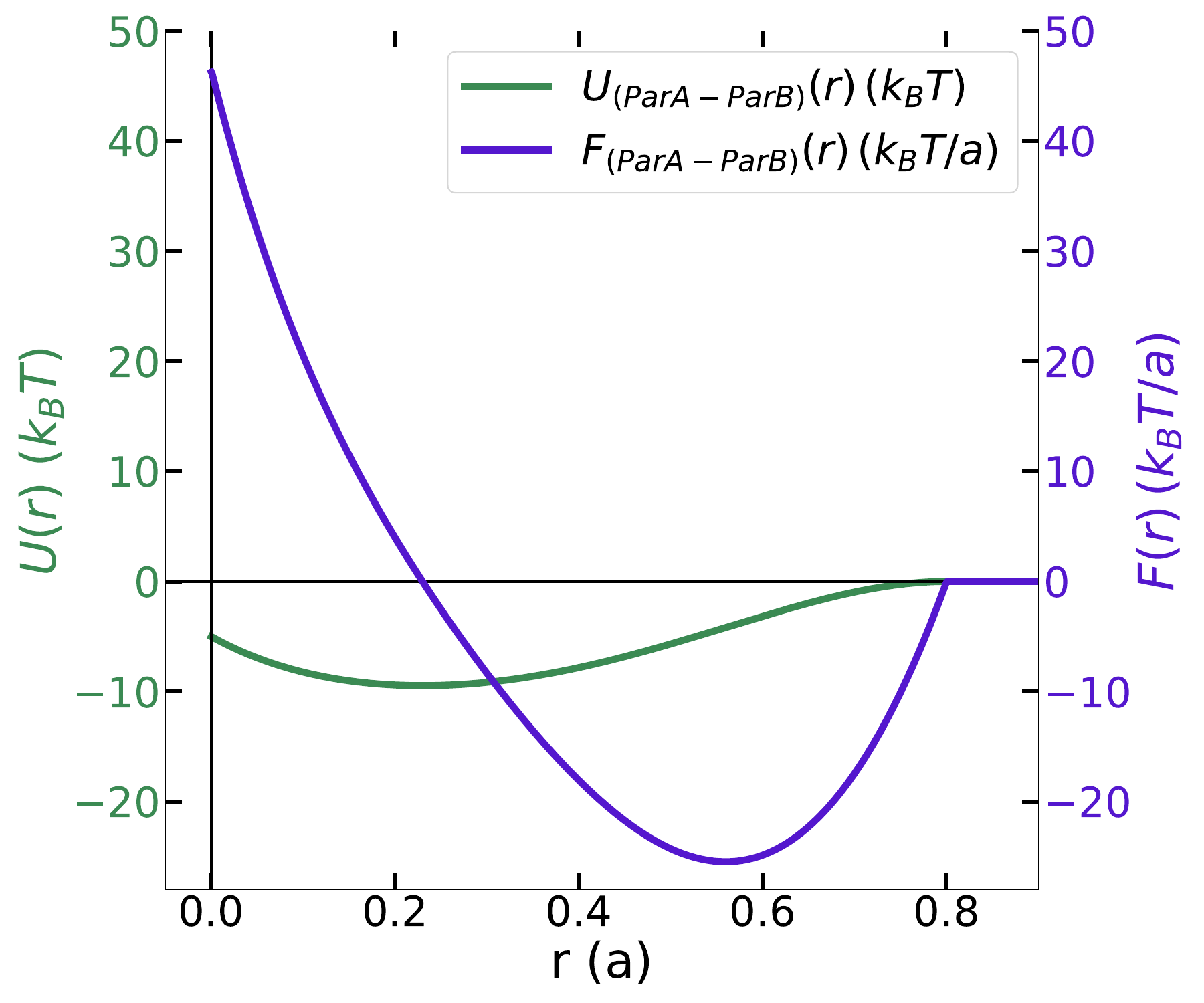}
     \put(-120,100){\large\textbf{(c)}}
    \caption{Interaction potential between ParA-ParB monomers $U_{\{\text{ParA}-\text{ParB}\}}(r)$ (green curves, in units of $k_B T$) and the corresponding interaction force $-\frac{d}{dr}U_{\{\text{ParA}-\text{ParB}\}}(r)$ (purple curves, in units of $k_B T / a$) plotted as a function of separation distance $r$ (in units of $a$).  We use different parameters for different sets of studies, as given below: 
    \textbf{(a)} $A_{att}=-20k_BT$  and $ A_{rep}=15k_BT$ ; $r_c^{att}=r_c^{rep}=1.5a$ ; $\rho_{att}=0.6a$ and $ \rho_{rep}=0.2a$.
    \textbf{(b)} $A_{att}=-5k_BT$ and $ A_{rep}=5k_BT$ ; $r_c^{att}=r_c^{rep}=1.5a$ ; $\rho_{att}=0.6a$ and $ \rho_{rep}=0.2a$.
    \textbf{(c)} $A_{att}=-20k_BT$ and  $A_{rep}=15k_BT$ ; $r_c^{att}=r_c^{rep}=0.8a$ ; $\rho_{att}=0.7a$ and  $\rho_{rep}=0.2a$.
    } 
    \label{fig:ParAParBinteractionpotential}
\end{figure}

\subsection{Modelling an extrusion-like process to create modified polymer architecture}

The Hi-C map of the \emph{C. crescentus} chromosome exhibits prominent characteristic diagonal features, as reported in Ref.~\cite{le2013high}. For reference, we reproduce the experimental Hi-C map in Fig.~S$14$(a) of SI-VII. In our previous work, we proposed that  
the {\em C. crescentus} chromosome is a topologically modified (ToMo) ring polymer due 
to the action of linker proteins. We then investigated the emergent organization of the  
non-replicating ToMo polymer to elucidate the mechanisms underlying the experimentally observed localization of chromosomal loci~\cite{viollier2004rapid}, while maintaining consistency with the experimental Hi-C map~\cite{mitra2022topology}.

Here, we use a similar modification of the topology of the ring polymer for the mother chromosome at the beginning of the simulations. See Fig.~$1$a for a schematic of the topology of the modified 
ring polymer used in this work. The daughter chromosomes progressively acquire the same topology as replication proceeds. We construct this architecture by introducing additional cross-links between selected monomers, as shown in Fig.~$1$a of main text. The simulation is initiated with a single mother chromosome having this particular topology,
whose \emph{oriC} is tethered to the old pole.

\begin{figure}[htbp]
        
    \begin{subfigure}[htbp]{\linewidth}
        \centering
        \includegraphics[width=0.9\linewidth]{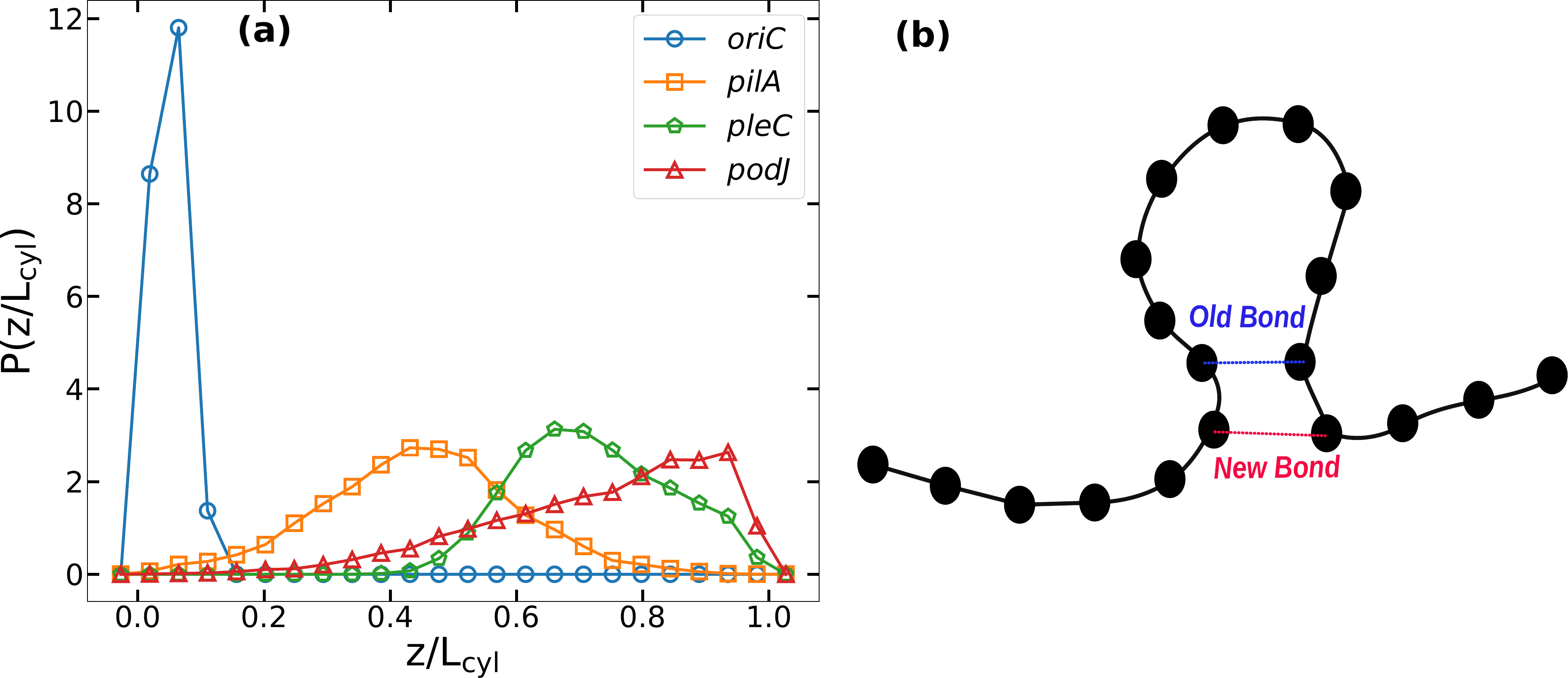}
        
    \end{subfigure}
    \caption{
    \textbf{(a)} Spatial organization of \emph{C.crescentus} chromosome: the figure shows the spatial probability distributions of loci obtained from simulations as shown in ref.~\cite{viollier2004rapid}, respectively. In the simulations, we use the polymer architecture shown in Fig.~$1$(a) with the monomer corresponding to the \emph{oriC} tethered to the cell pole, as is the case in vivo. The x-axis has been scaled by the length of the confining cylinder. The \ emph {oriC} locus
remains tethered to the cell pole and therefore remains localized around the pole. Here, we confine a single polymer architecture within cylindrical confinement, with no replication process.
      \textbf{(b)} Schematic of the extrusion process, where the link 
     between two monomers on the two arms of the chromosome are modeled by harmonic bonds. This
     bond starts from the {\em oriC} proximal regions and moves progressively away from it.}
     \label{fig:extrusionprocess}
\end{figure}

A major challenge is to generate the same topological architecture in the daughter chromosomes during replication. Since extrusion factors are known to be enriched in regions proximal to \emph{oriC}, we model an extrusion-like process to dynamically establish the required cross-links in the daughter chromosomes.

 To implement this process, we remove the first cross-link between monomer-$166$ and monomer-$832$ from the mother chromosome $4400\tau$ after the onset of replication, corresponding to the stage when $4.4\%$ of the mother chromosome has been replicated. We then initiate extrusion in each daughter chromosomes. We model the extruding protein complex as two (imaginary) cuffs linked to each other, initially positioned at monomers $1$ and $1000$ on the two arms of each daughter chromosome. After $400\tau$, the cuffs advance to monomers $2$ and $999$, respectively, and a harmonic bond is introduced between these monomers to represent the link between the cuffs. This creates an extruded segment consisting of monomers $1$ and $1000$. 
The cuffs then advance by one monomer every $400\tau$, generating successive harmonic bonds between monomers $3$ and $998$, and so forth; see Fig.~\ref{fig:extrusionprocess}. Since new monomers are incorporated at the RFs every $200\tau$, the extrusion rate in our simulations is one-half of the replication rate.

 We introduce a second extrusion process by introducing a second harmonic bond between 
monomers-$1$  and monomer-$1000$, when the first extruding bond has moved to a bond between $3$ 
and $998$, i.e. $800 \tau$ after the introduction of the first bond. At the same time we remove the second crosslink between monomer-$333$ and monomer-$666$ from the mother chromosome. The second follows the first extrusion factor, and moves at the same rate as the first one, connecting corresponding monomers from the two arms. The first extruding bond stops once they connect monomer-$166$ and monomer-$832$.  The second 
extrusion  bond pair keeps moving towards the $333-666$  pair of monomers.
But the RFs reach the {\em ter}-loci, and the simulation run ends much before the bond reaches 
the $333-666$ monomer pair.

\section{Results from \emph{Model-1}(Without extrusion and without modified polymer topology)}
\subsection{\emph{oriC} Dynamics}

In this section, we present the results for \emph{Model-1}. We do not include an extrusion-like process in \emph{Model-1}. Moreover, the mother DNA polymer ring is replicated into two daughter DNA ring polymers without any additional cross-links. Effective cross-links arising from transient attraction between ParA and ParB monomers are, however, present.  
 
\begin{figure}[htbp]
\centering
\begin{subfigure}[htbp]{0.49\linewidth}
    \centering
    \includegraphics[width=\linewidth]{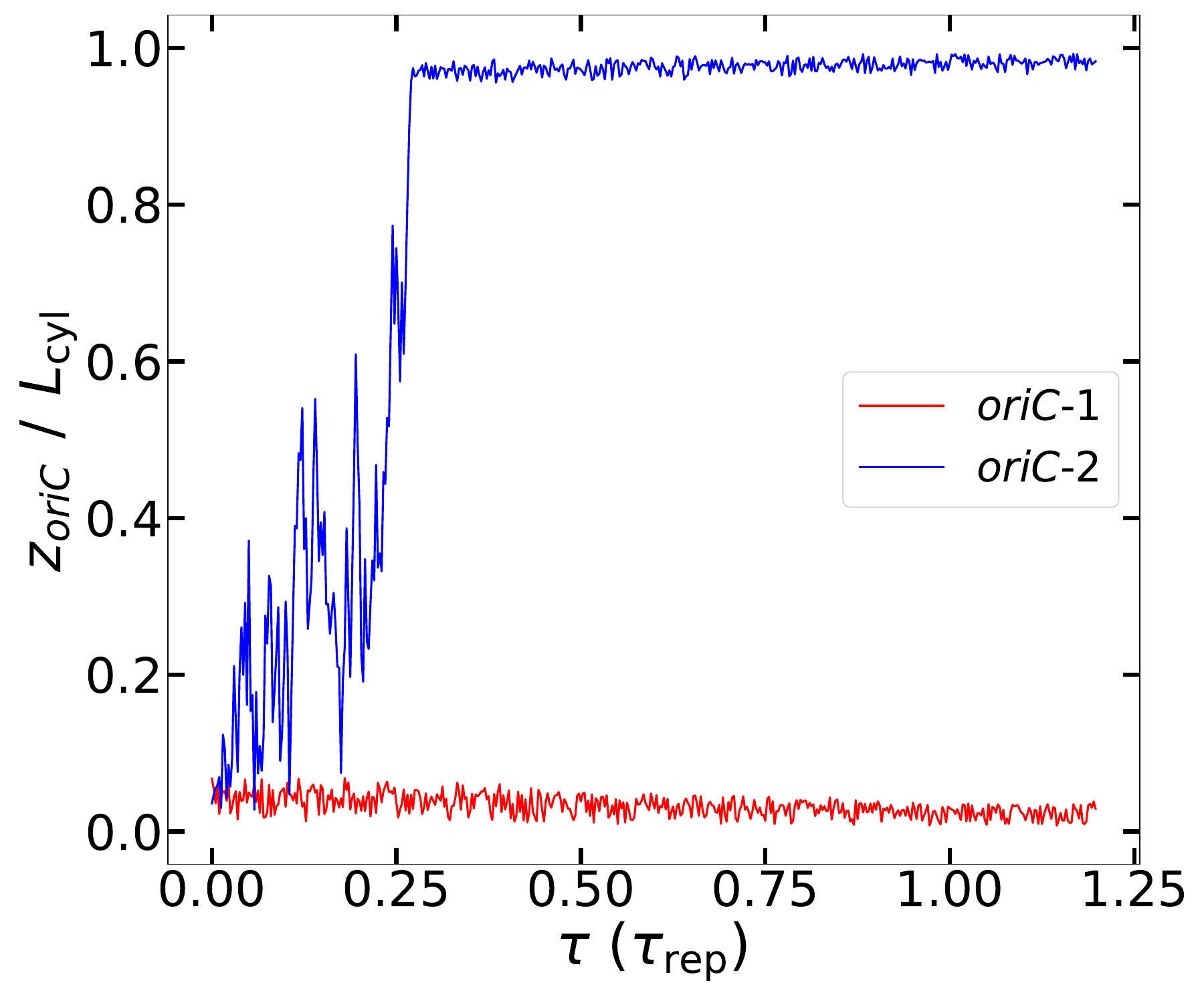}
    \put(-210,180){\large\textbf{(a)}}
\end{subfigure}
\begin{subfigure}[htbp]{0.49\linewidth}
    \centering
    \includegraphics[width=\linewidth]{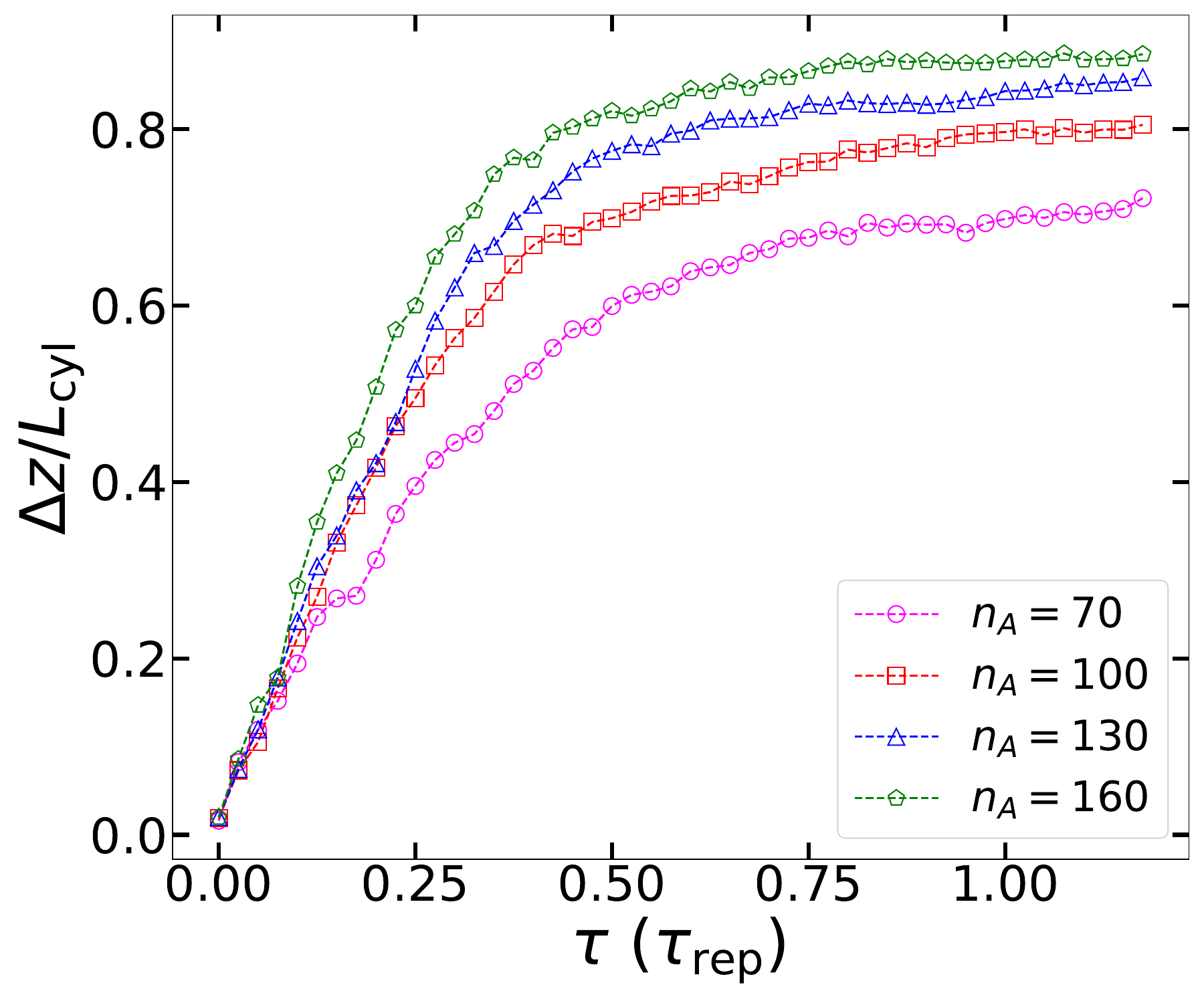}
    \put(-210,180){\large\textbf{(b)}}
\end{subfigure}
\begin{subfigure}[htbp]{0.7\linewidth}
\centering
    \includegraphics[width=\linewidth]{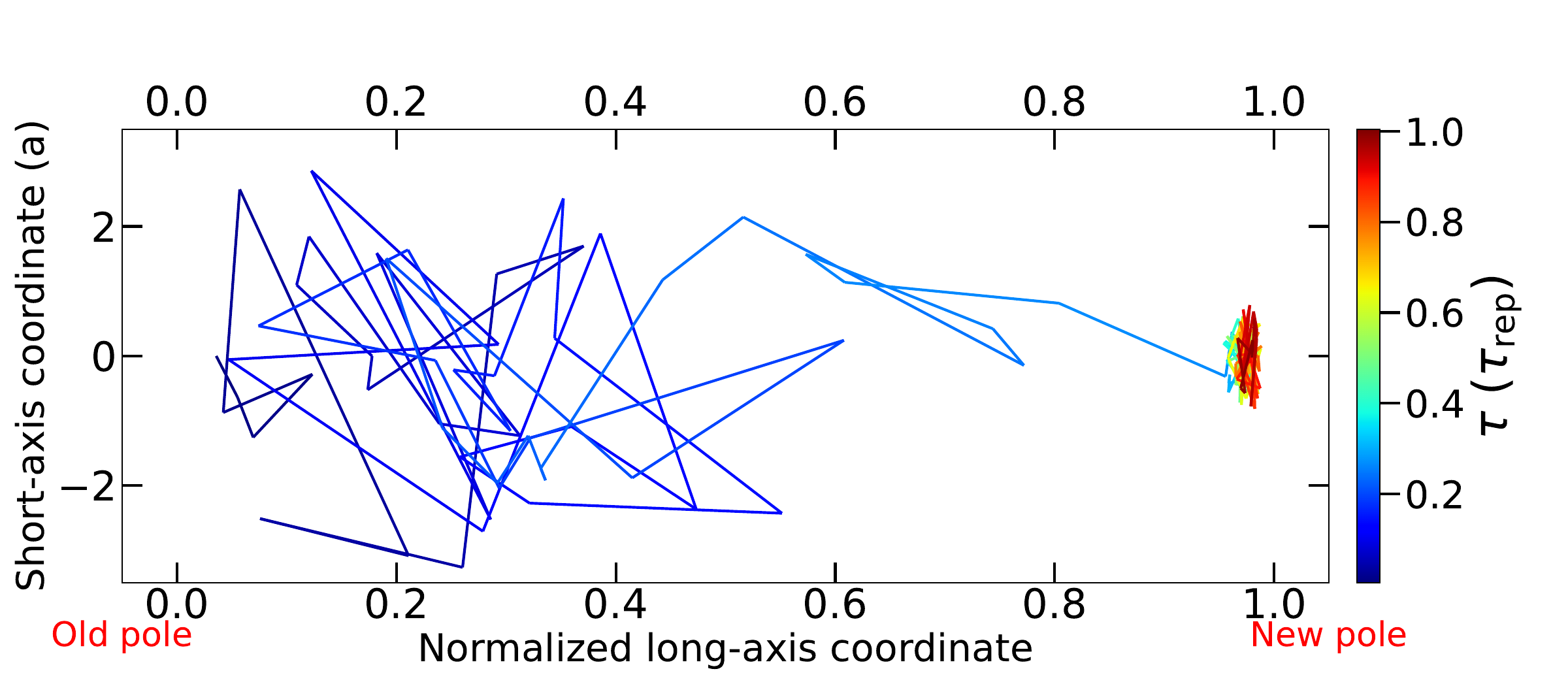}
    \put(-319,112){\large\textbf{(c)}}
\end{subfigure}
\caption{\textbf{\emph{OriC} trajectory:} \textbf{(a)}We plot long axis position, normalized by the instantaneous cell length $\mathrm{z}_{\emph{oriC}}/L_{\mathrm{cyl}}$ of each replicated \emph{oriC} monomer as replication proceeds. The red curve represents the position of the long axis of the tethered \emph{oriC}, normalized by the instantaneous cell length, while the blue curve represents the corresponding position of the second replicated \emph{oriC}. We use a ring polymer without any extra cross-links. \textbf{(b)} {\bf Segregation dynamics for different numbers of  ParA monomers.} Average distance between two replicated \emph{oriCs} ($\Delta \mathrm{z}_{\emph{oriC}}$ )normalized with instantaneous length ($L_\mathrm{cyl}$) with different number of ParA with time. These distances are averaged over $120$ runs.  We add monomers every  $2\times10^4$ iterations, whereas $r_c^P= 1.5a$ and $n_P=10^4$ iterations for the data presented in this plot. \textbf{(c)} We have plotted $2-d$ trajectory of \emph{oriC} that we have observed in our simulations with \emph{Model-1}. We have only shown the trajectory of the \emph{oriC} of Daughter DNA-2, i.e., the one which is not anchored at the old pole and hence is able to get transported to the new pole. In our simulations, we have the positions of the two \emph{oriCs} as soon as the replication begins.}
\label{fig:Oritraj_model1}
\end{figure}

 As shown in the main text (refer Fig.$4$(a) and $4$(b)), the data shown in Fig.~\ref{fig:Oritraj_model1}, obtained using simulations with \emph{Model-1}, reveal the same distinct stages of \emph{oriC} segregation: (1) release from the old pole, (2) polar retraction, (3) early translocation (slow), and (4) late translocation (fast). 

In the main text (refer Fig.~$6$(b)), we showed that  increasing the number of ParA monomers, $n_A$, for simulations using \emph{Model-2} enhances the value of $\Delta \mathrm{z}/L_{\mathrm{cyl}}$ 
as a consequence of the increased success rate of \emph{oriC} translocation. In Fig.~\ref{fig:Oritraj_model1}(b),  as a corresponding analysis using \emph{Model-1}, we plot the normalized longitudinal distance ($\Delta \mathrm{z}/L_{\mathrm{cyl}}$) between the two replicated \emph{oriC} loci as a function of replication time, $\tau_{\mathrm{rep}}$, for different values of $n_A$. The distance is averaged over 120 independent simulation runs. Since the averaging includes both `successful' and `unsuccessful' segregation events, a higher segregation success rate results in a larger average separation between the two \emph{oriC} loci. The trend is consistent with the trend observed in simulations using \emph{Model-2}. 

As in Fig.$5$(b) in the main text, we plot a representative two-dimensional trajectory of one of the \emph{oriC}s for a simulation run in Fig.~\ref{fig:Oritraj_model1}(c) using \emph{Model-1}. We can clearly distinguish different stages of \emph{oriC} dynamics. Soon after the start of replication, till about $0.2\tau_{\mathrm{rep}}$, \emph{oriC} exhibits gradual diffusive motion away from the old pole. It subsequently remains proximal to the old pole and undergoes fluctuations in position around $0.2L_{\mathrm{cyl}}$-$0.4L_{\mathrm{cyl}}$ in the time range $(0.2-0.4) \tau_{rep}$. Starting at around $0.4\tau_{\mathrm{rep}}$, \emph{oriC} undergoes a rapid translocation toward the new pole. Once \emph{oriC} reaches the new pole, it becomes anchored there by the PopZ protein system. 

\subsection{Replisome localization}
\begin{figure}[htbp]
\begin{subfigure}[htbp]{0.329\linewidth}
    \centering
    \includegraphics[width=\linewidth]{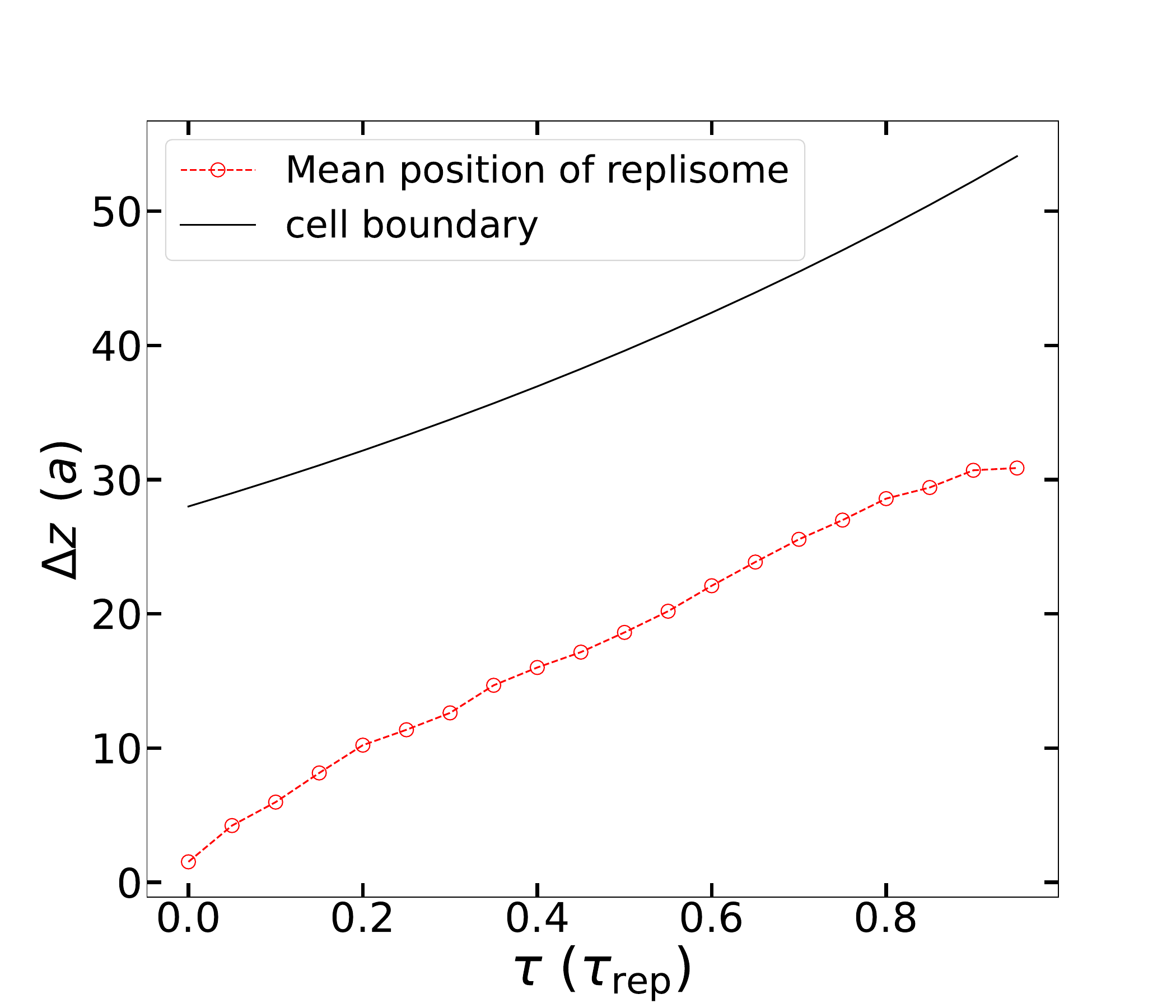}
    \put(-40,95){\textbf{(a)}}
\end{subfigure}
\begin{subfigure}[htbp]{0.329\linewidth}
    \centering
    \includegraphics[width=\linewidth]{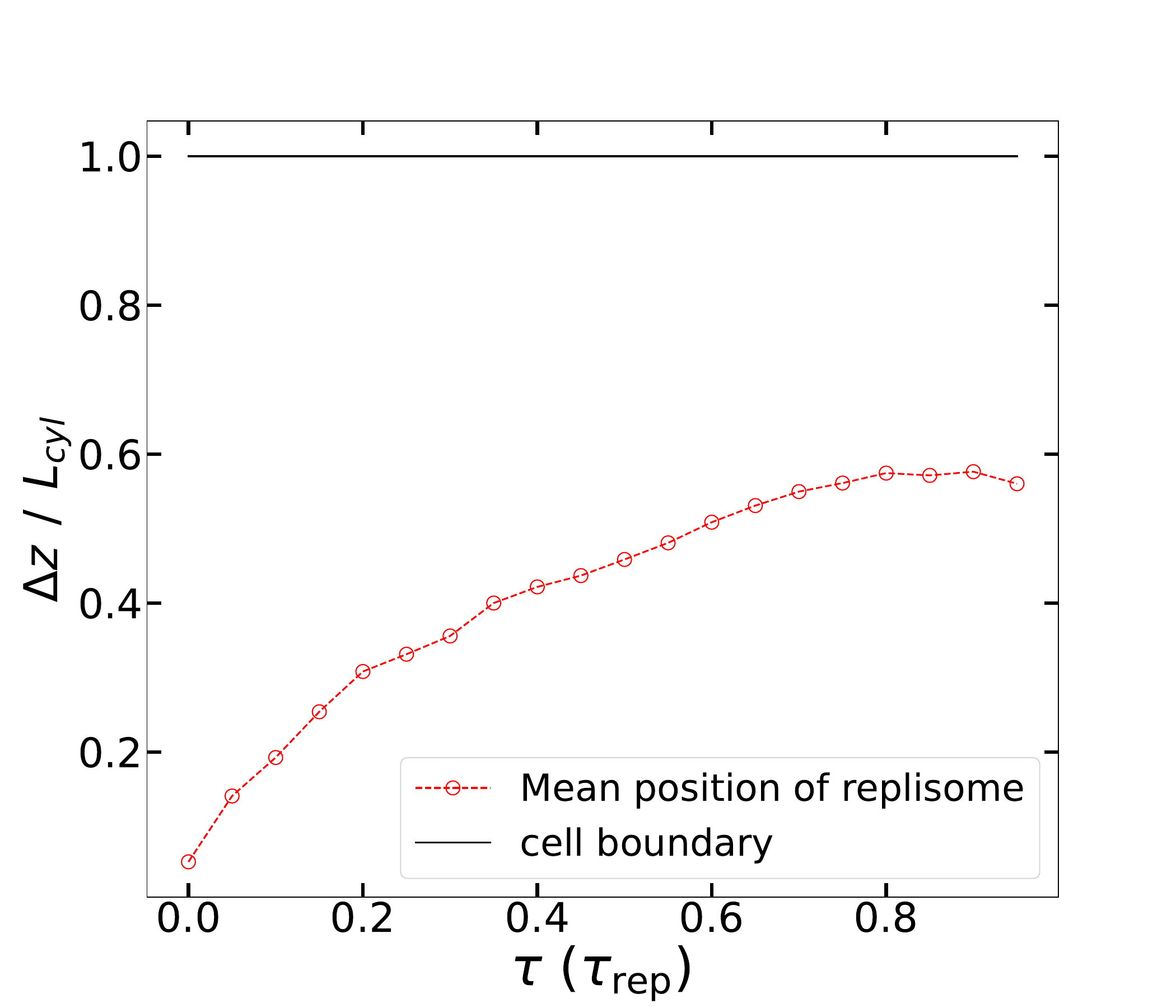}
    \put(-40,95){\textbf{(b)}}
\end{subfigure}
\hfill
\begin{subfigure}[htbp]{0.329\linewidth}
    \centering
    \includegraphics[width=\linewidth]{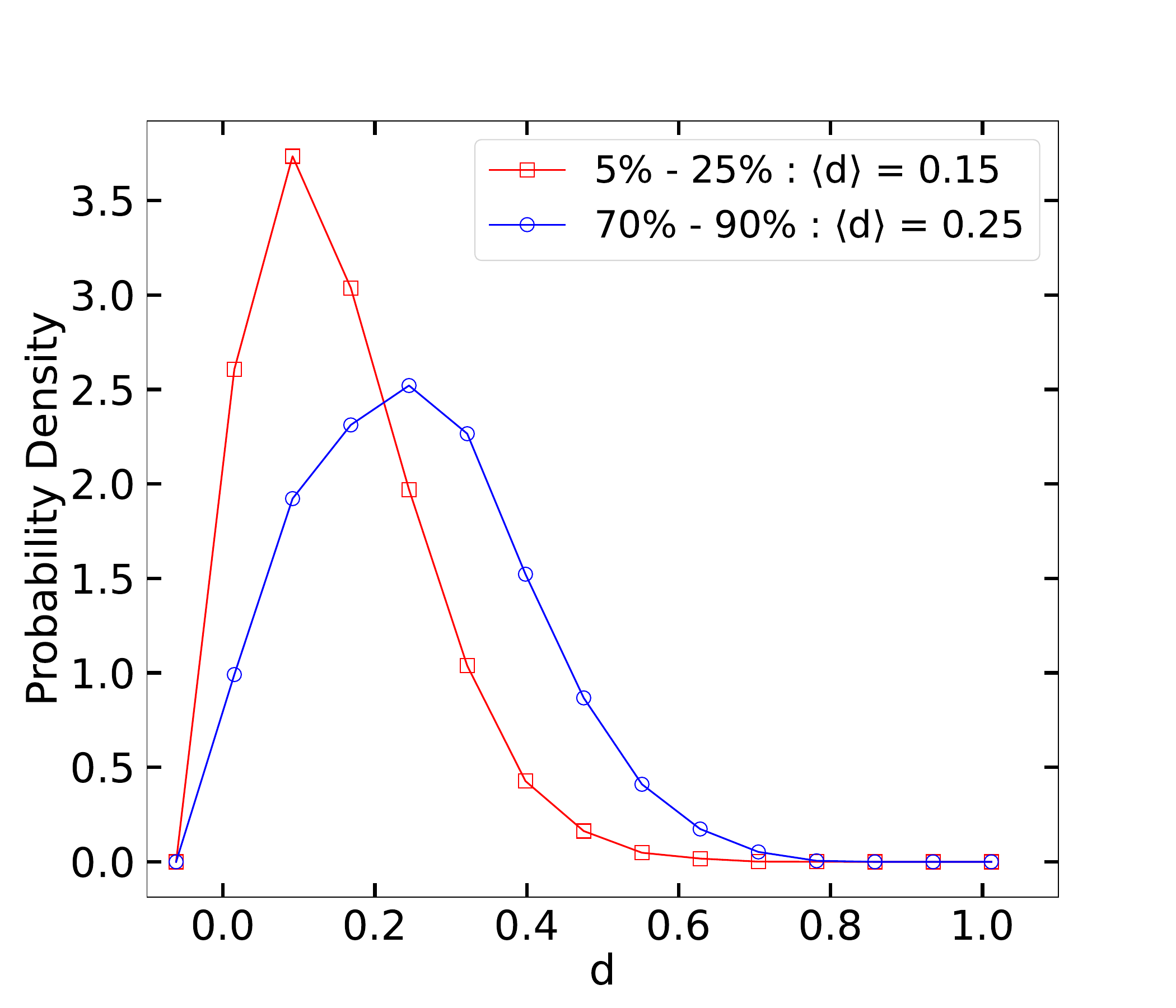}
    \put(-40,90){\textbf{(c)}}
\end{subfigure}
\caption{{\bf Reorganization of replication forks (RFs) in simulations; RFs represent the replisome.} The simulations are performed with $100$ ParA monomers, $r_c^P= 1.5a$ 
  and $n_P =10^4$ iterations. \textbf{(a)} The plot shows the longitudinal mean position $\Delta \mathrm{z}$ of the replisomes (RFs), averaged over their instantaneous positions, as a function of time normalized by the replication time. The distance $\Delta \mathrm{z}$ is measured from the old pole of the cell. Data is averaged over the runs that showed successful {\em oriC} transport to the new pole out of $120$ independent runs. 
\textbf{(b)}  The plot shows the average long axis position $\mathrm{z}$ (normalized with instantaneous length of cylinder $L_{\mathrm{cyl}}$) of the replication fork vs. time in units of replication time.
\textbf{(c)}  Probability density distribution of long axis distance ($d$) between two RFs at ($5\%$ to $25\%$) 
of replication time (Red) and (70\%-90\%) of replication time (Blue). The long-axis distance, $d$, is normalized by the instantaneous cylinder length, $L_{\mathrm{cyl}}$.}
\label{fig:Model1_replisome_traj}
\end{figure}

In our simulations, we do not explicitly model replication forks. However, the position of the replication fork is known throughout the simulations. We plot replication fork trajectories and the distribution of distances between the two replication forks for early and late stages of replication in Fig.~\ref{fig:Model1_replisome_traj}. In Fig.~\ref{fig:Model1_replisome_traj}(a), we plot the mean of the positions of the two replication forks with replication time. We have averaged these positions over only `successful' runs out of $120$ independent runs. In the case of \emph{Model-1},  the replication fork initially is near the old-pole region and gradually moves toward the cell center as replication proceeds, as shown in Fig.~\ref{fig:Model1_replisome_traj}(a). 
For the data shown in Fig.~\ref{fig:Model1_replisome_traj}(b), we use the positions of the replication fork, measured from the old pole and normalized by the instantaneous cell length. 
Consistent with the previous figure, the plot clearly shows the progressive transition of the replication fork from the old-pole region toward the cell center during replication.

As shown in Fig.~8(c) of the main text, the replication fork exhibits a splitting behavior during the late stages of replication. A similar behavior is observed for \emph{Model-1}, as shown in Fig.~\ref{fig:Model1_replisome_traj}(c). To quantify the separation between the two replication forks, we calculate the distribution of their long-axis distance, normalized by the instantaneous cell length, during the $5\%-25\%$ and $70\%-90\%$ intervals of the replication time. The average normalized distance between the two replication forks is approximately $0.15$ during the $5\%-25\%$ interval and increases to approximately $0.25$ during the $70\%-90\%$ interval of $\tau_{\mathrm{rep}}$. This increase indicates that the two replication forks become progressively separated along the long axis of the cell as replication proceeds.

All the results presented for $Model-1$ in Fig.~\ref{fig:Model1_replisome_traj} are analogous to the corresponding results for $Model-2$ presented in Fig.~8 of the main text.

\subsection{Loci-localization} 
We plot trajectories of eight different loci shown in Fig.~\ref{fig:loci_schematics}(b) for the \emph{Model-1} case. We used the conventions specified in Ref.~\cite{Harju2025.04.03.647011} to plot simulation trajectories of eight loci. 

The gray shaded regions indicate portions of the chromosome that have not yet undergone replication at the corresponding time. Thus, the evolution of the loci trajectories also provides a representation of the progression of chromosome replication and the subsequent spatial redistribution of the replicated loci. The trajectories shown here correspond to those presented for \emph{Model-2} in Fig. 10(c) of the main text, and provide additional details on the localization and temporal evolution of individual chromosomal loci.

\begin{figure}[htbp]
\centering
        \includegraphics[width=0.4\linewidth]{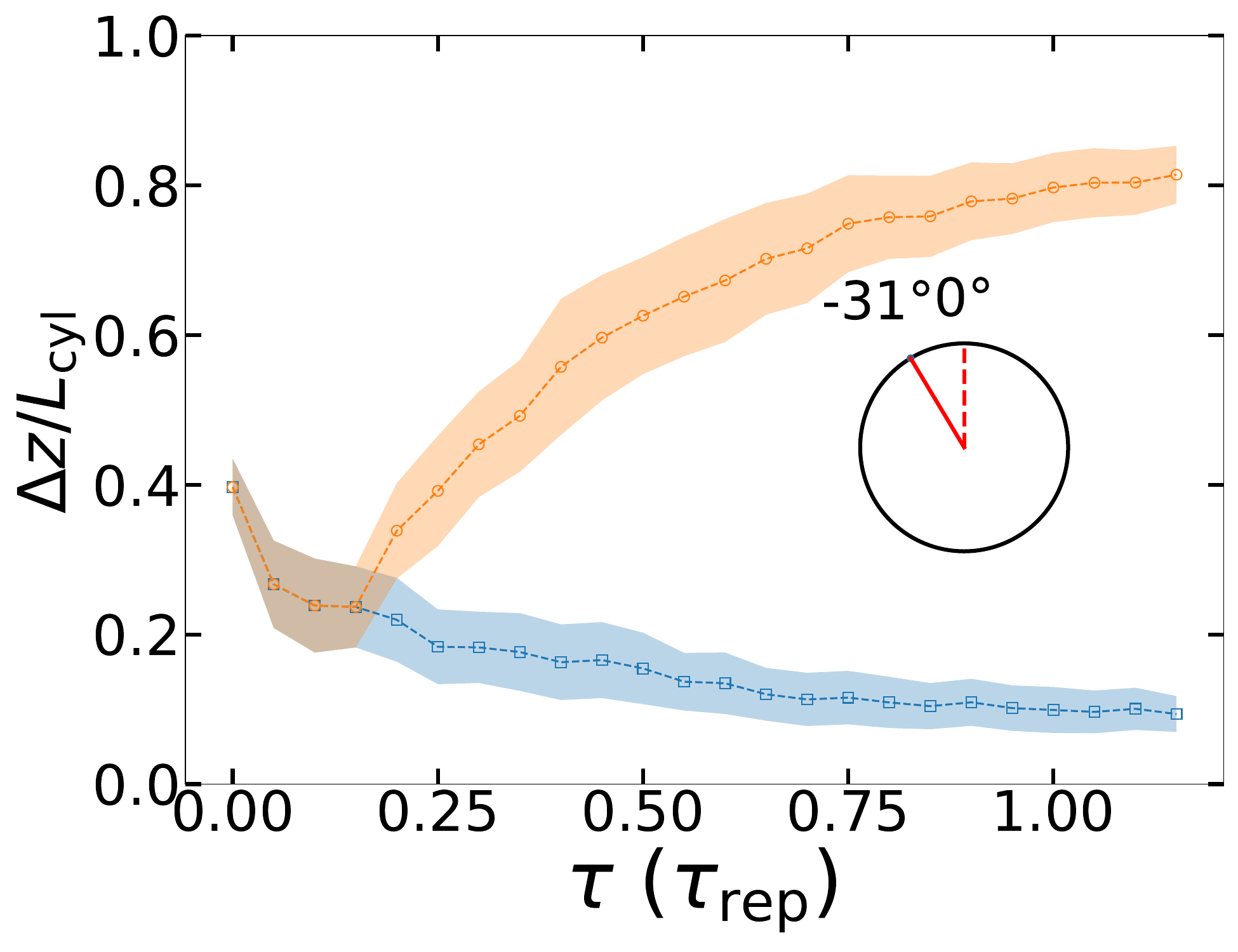}
        \includegraphics[width=0.4\linewidth]{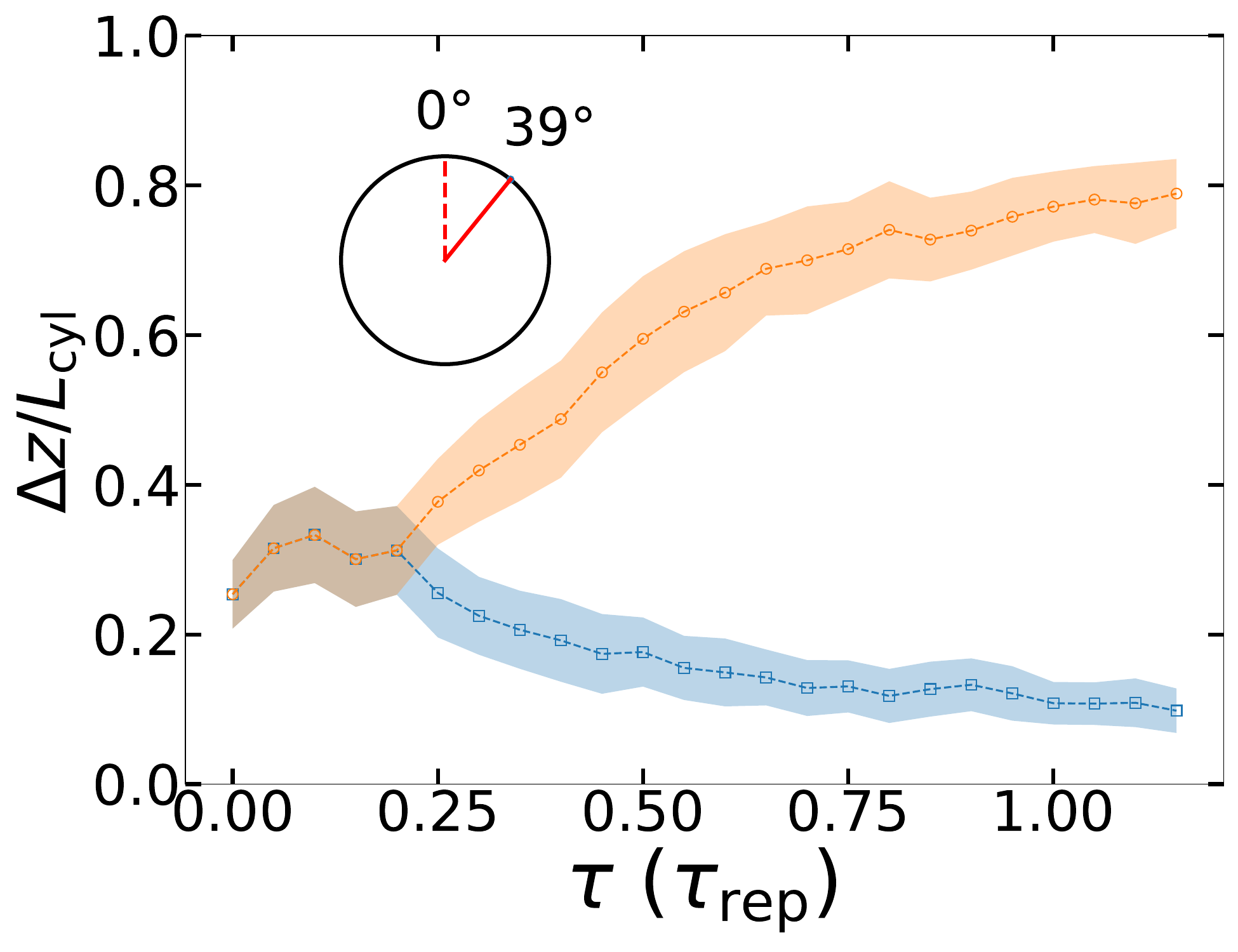}
        \includegraphics[width=0.4\linewidth]{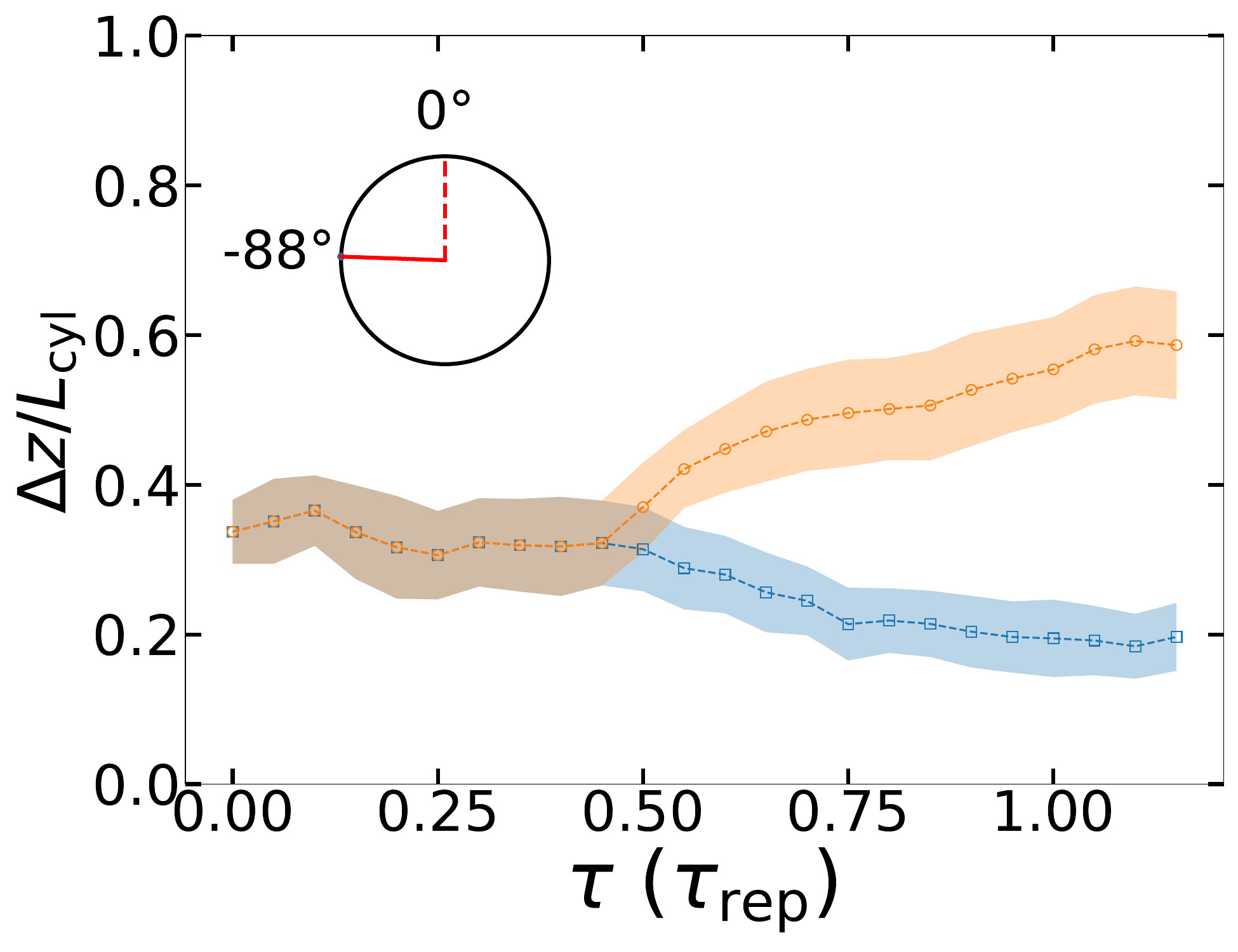}
        \includegraphics[width=0.4\linewidth]{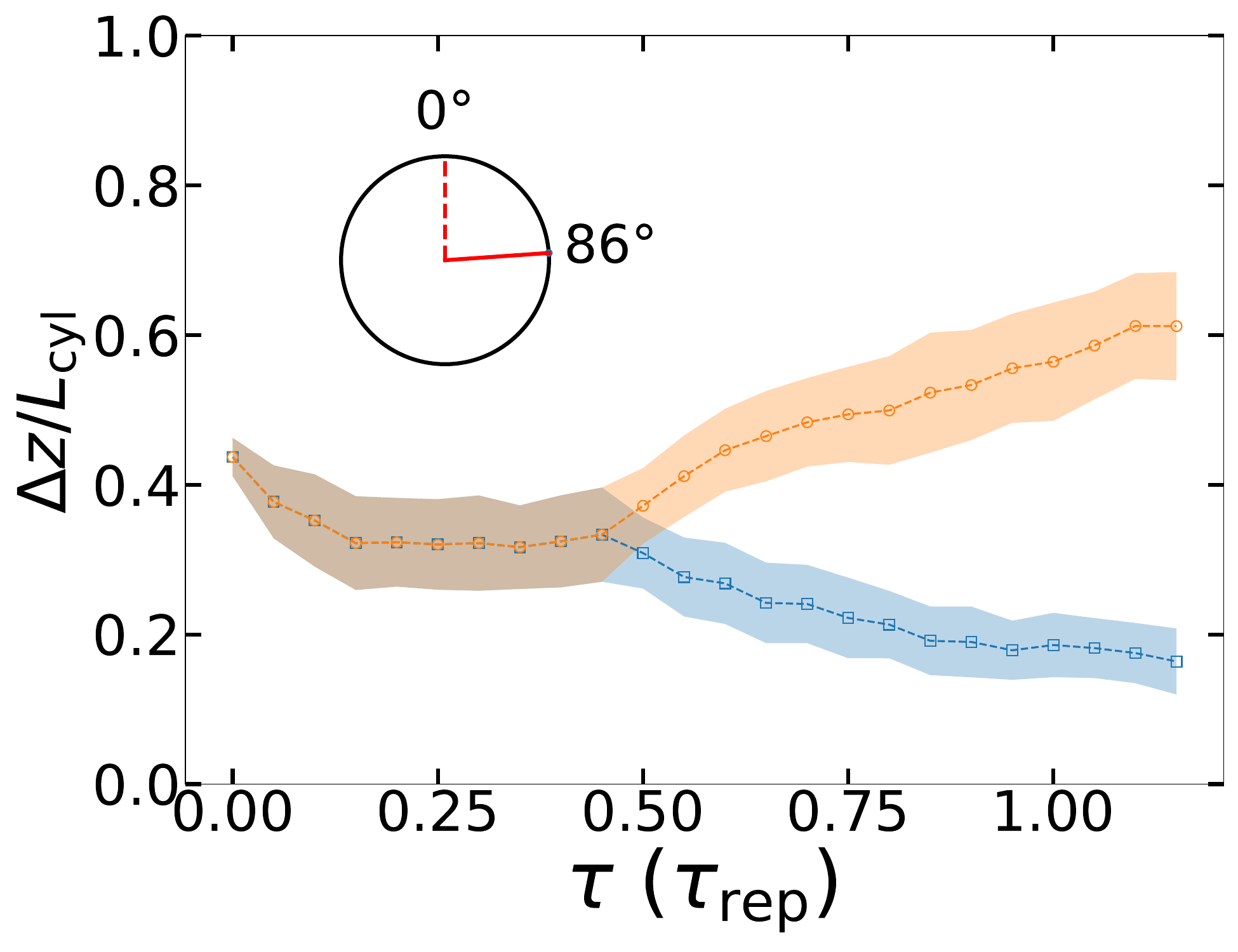}
        \includegraphics[width=0.4\linewidth]{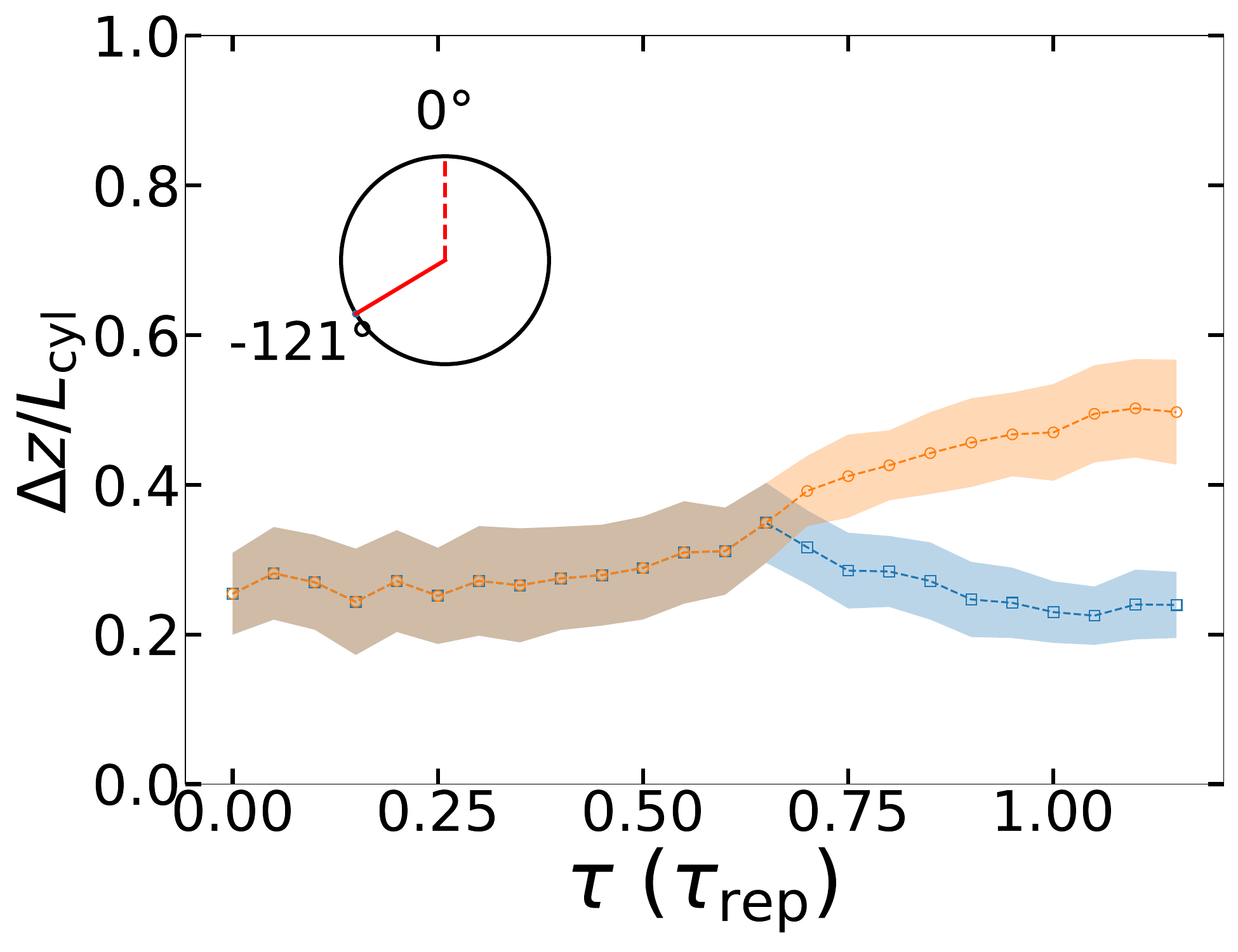}
        \includegraphics[width=0.4\linewidth]{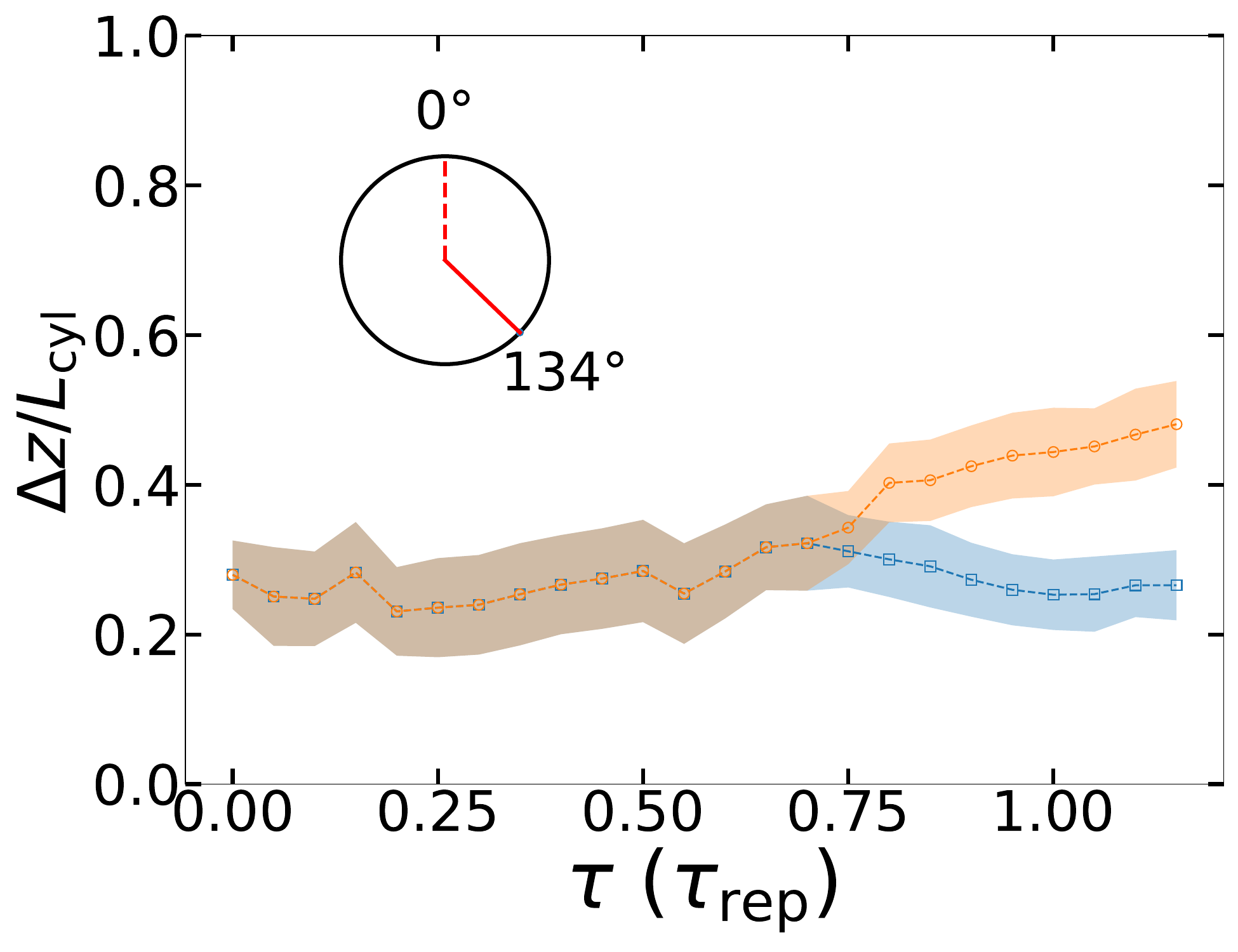}
        \includegraphics[width=0.4\linewidth]{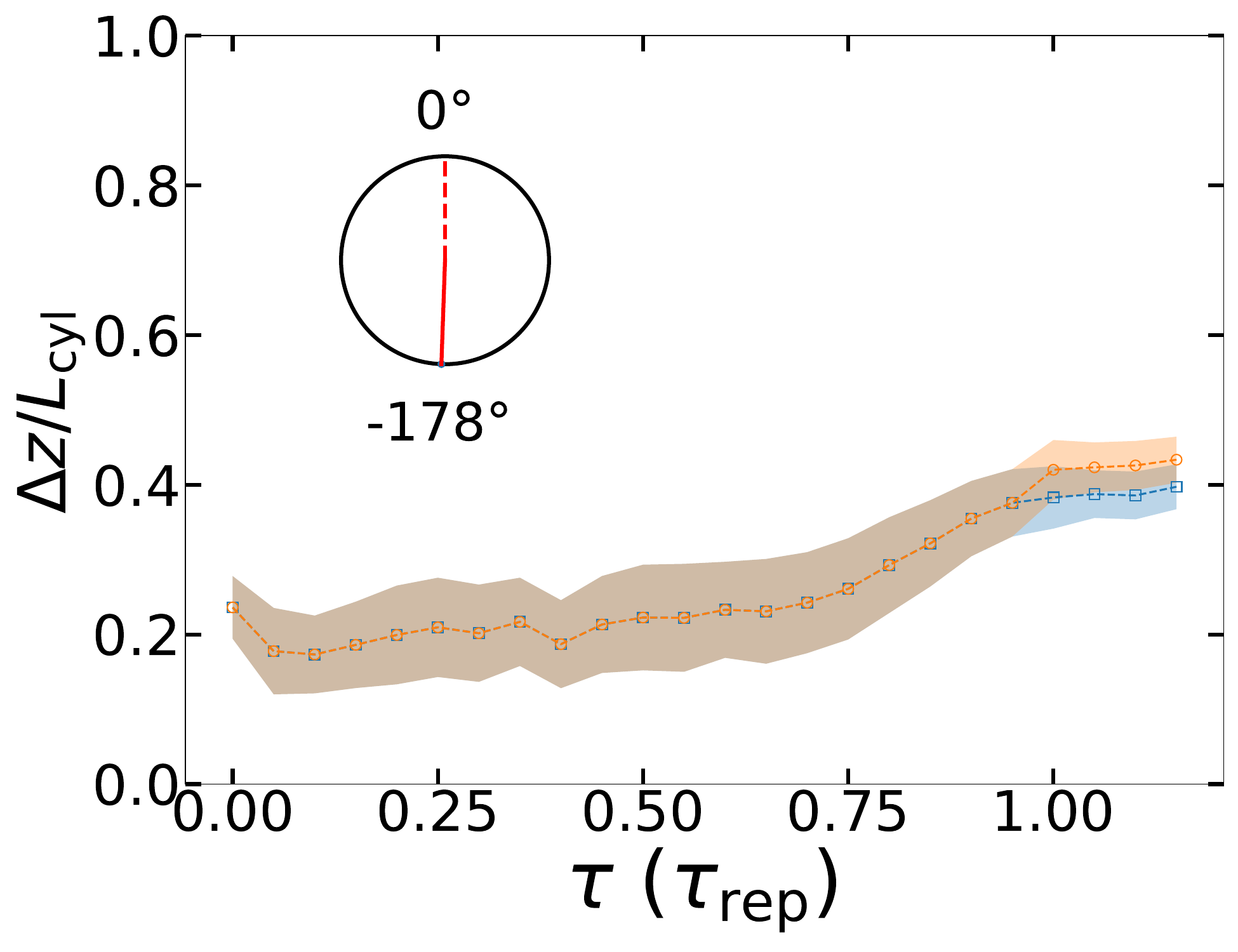}
        \includegraphics[width=0.4\linewidth]{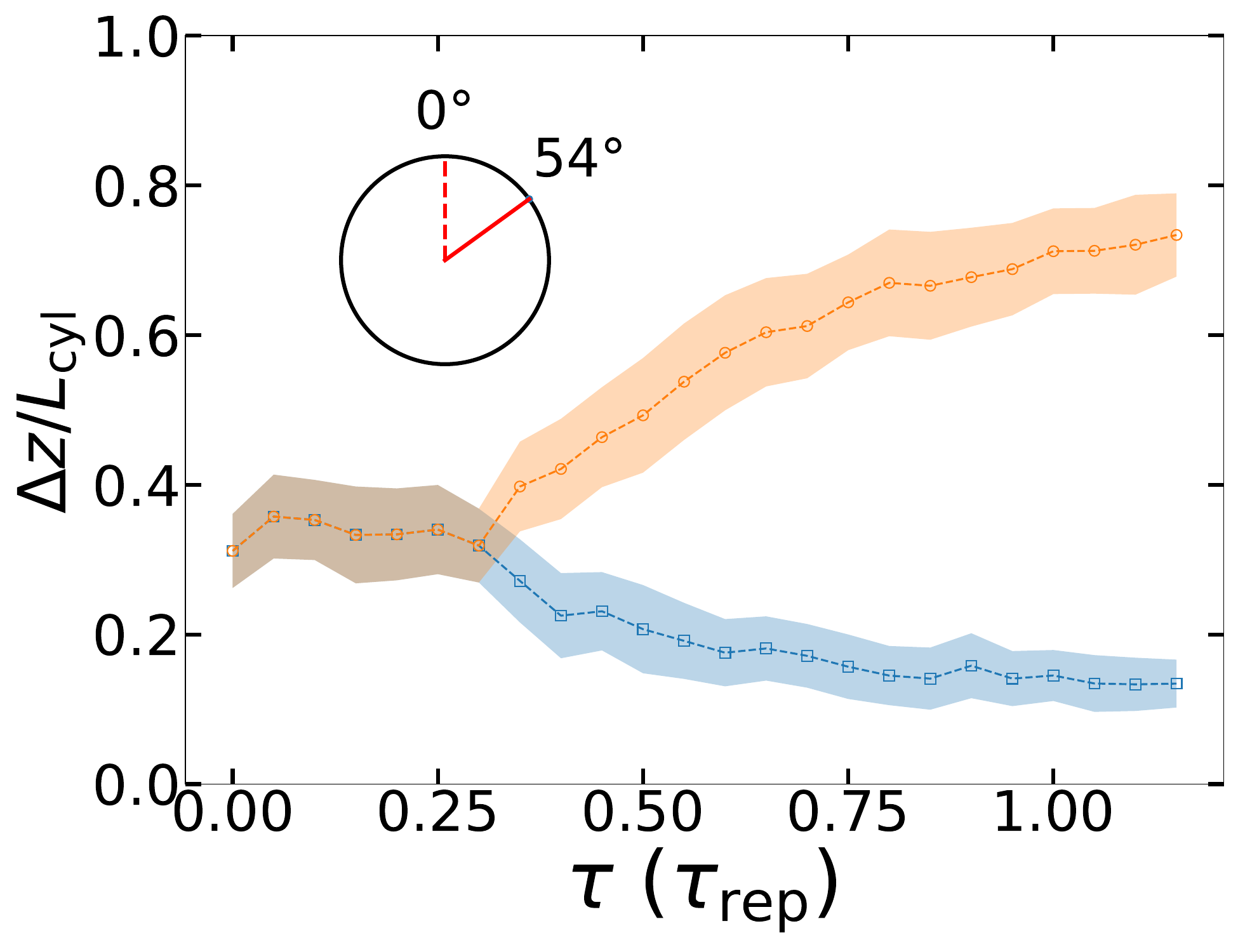}   
    \caption{{\bf Loci trajectories from our simulations:} We plot the trajectories 
    of eight different loci on the DNA polymer of {\em C. crescentus} as a function of time, even as replication is in progress. These loci are the same loci whose experimental trajectories are reported in \cite{Harju2025.04.03.647011}. For these plots, the position of each locus was calculated as follows. At each time point, we first determine the minimum distance, $d_l$, between either pole and any of the loci, where the locus may correspond to either the mother locus or one of the daughter loci. The pole associated with this minimum distance is then taken as the reference pole, and the positions of all loci are calculated as their distances from this pole. The gray shaded region represents the position of the particular loci when it is still part of the mother DNA. Post-replication, each locus splits into two trajectories. Loci which are proximal to ter undergo replication at a later stage of the replication process. The symbols show the mean position averaged over independent runs, whereas the shaded regions show the standard deviation about the mean positions. The positional data is averaged over the `successful' runs out of 120 independent runs.}
    \label{fig:simulation_locitraj}
\end{figure}

\FloatBarrier

\section{\emph{oriC} Trajectory for $ \boldsymbol{n_A=0}$}
\begin{figure}[htbp]
    \centering
    \includegraphics[width=0.5\linewidth]{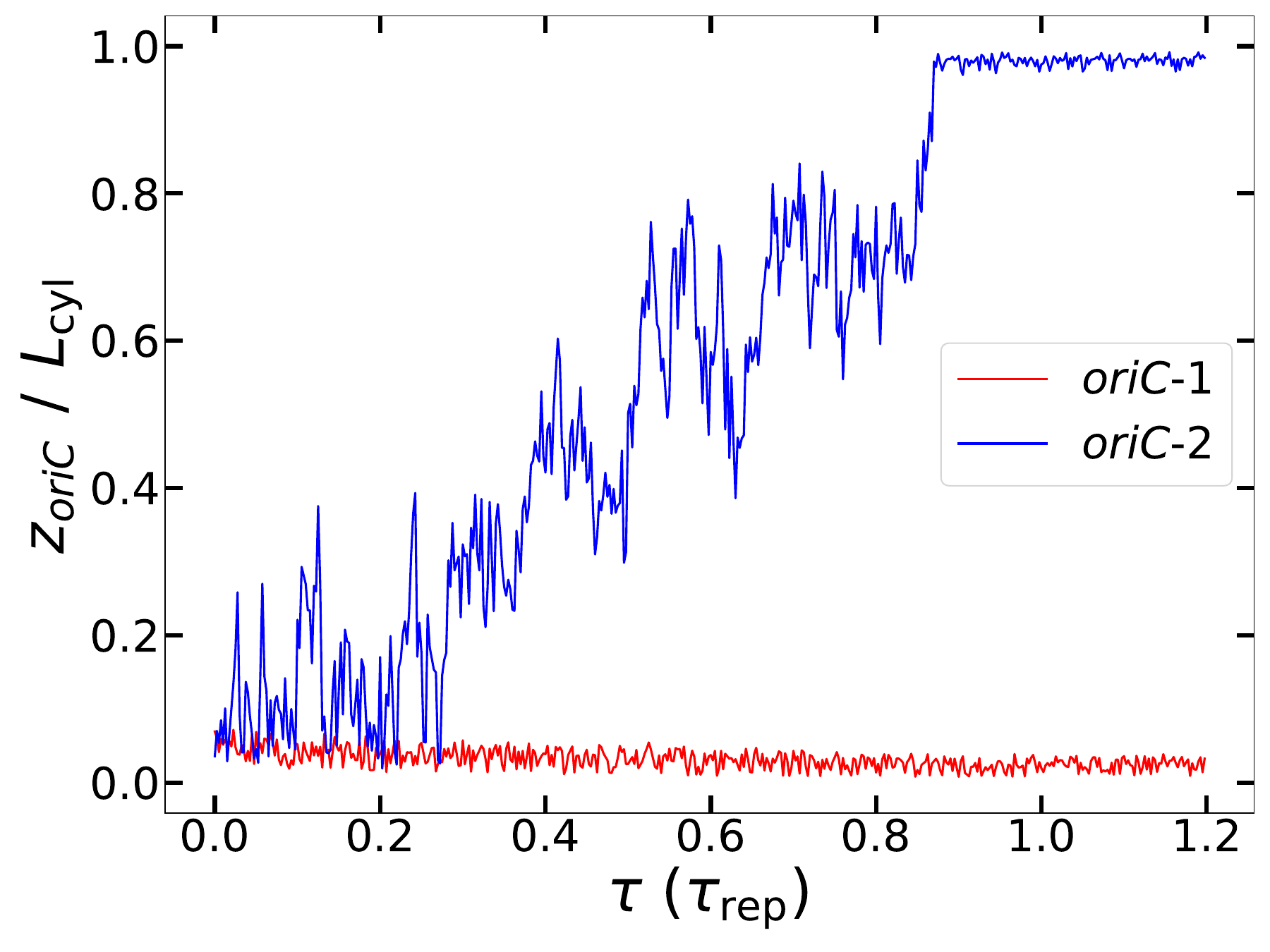}
    \caption{The plot shows long axis position, normalized by the instantaneous cell length, $\mathrm{z}_{\emph{oriC}}/L_{\mathrm{cyl}}$, of each replicated \emph{oriC} monomer as replication proceeds. The red curve represents the normalized long axis position of the tethered \emph{oriC}, while the blue curve represents the corresponding position of the second replicated \emph{oriC}. In this simulation, ParA--ParB interactions are absent since $n_A=0$, i.e., no ParA monomers are present, while the DNA polymer undergoes both replication and extrusion.
 }
    \label{fig:noParA_oritraj}
\end{figure}
\FloatBarrier

\section{Entropy vs enthalpy controlled run plot}
\begin{figure}[htbp]
    \centering
    \begin{subfigure}[t]{\linewidth}
        \includegraphics[width=0.5\linewidth]{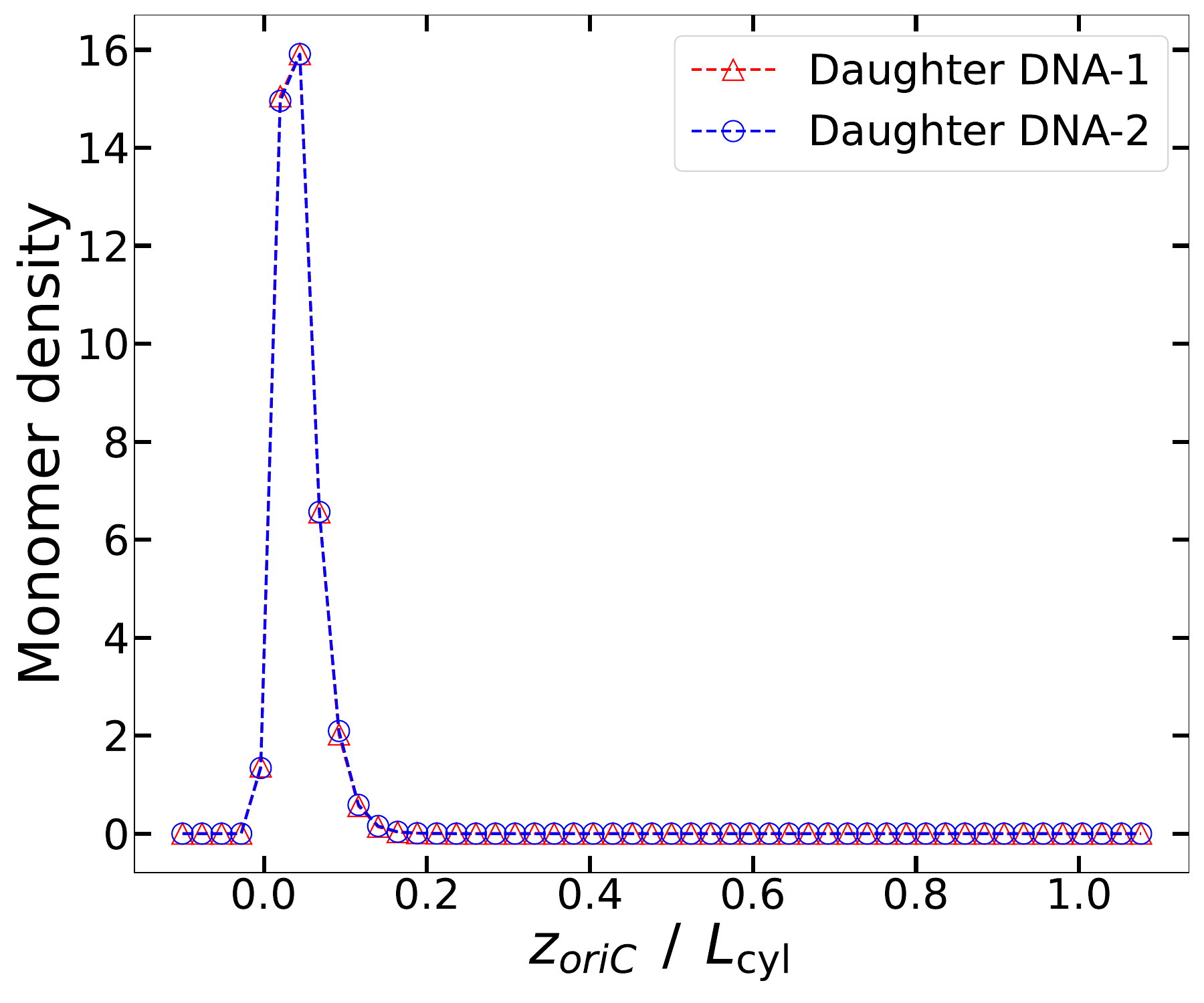}
        \label{fig:noexclusion_mon_distribution}
    \end{subfigure}

\caption{Excluded-volume interactions between all the monomers present in the simulation. Interaction range between ParA--ParB monomers kept fixed at $r_c^P=1.5a$, while the strength of the attractive interaction between ParA and ParB is set to $-5k_BT$ and the strength of the repulsive interaction to $5k_BT$. The plot shows the distribution of each daughter monomer along the long axis, normalized by the cell length, after completion of replication of the mother DNA.}
\end{figure}

\begin{figure}[htbp]
    \centering
    \includegraphics[width=0.49\linewidth]{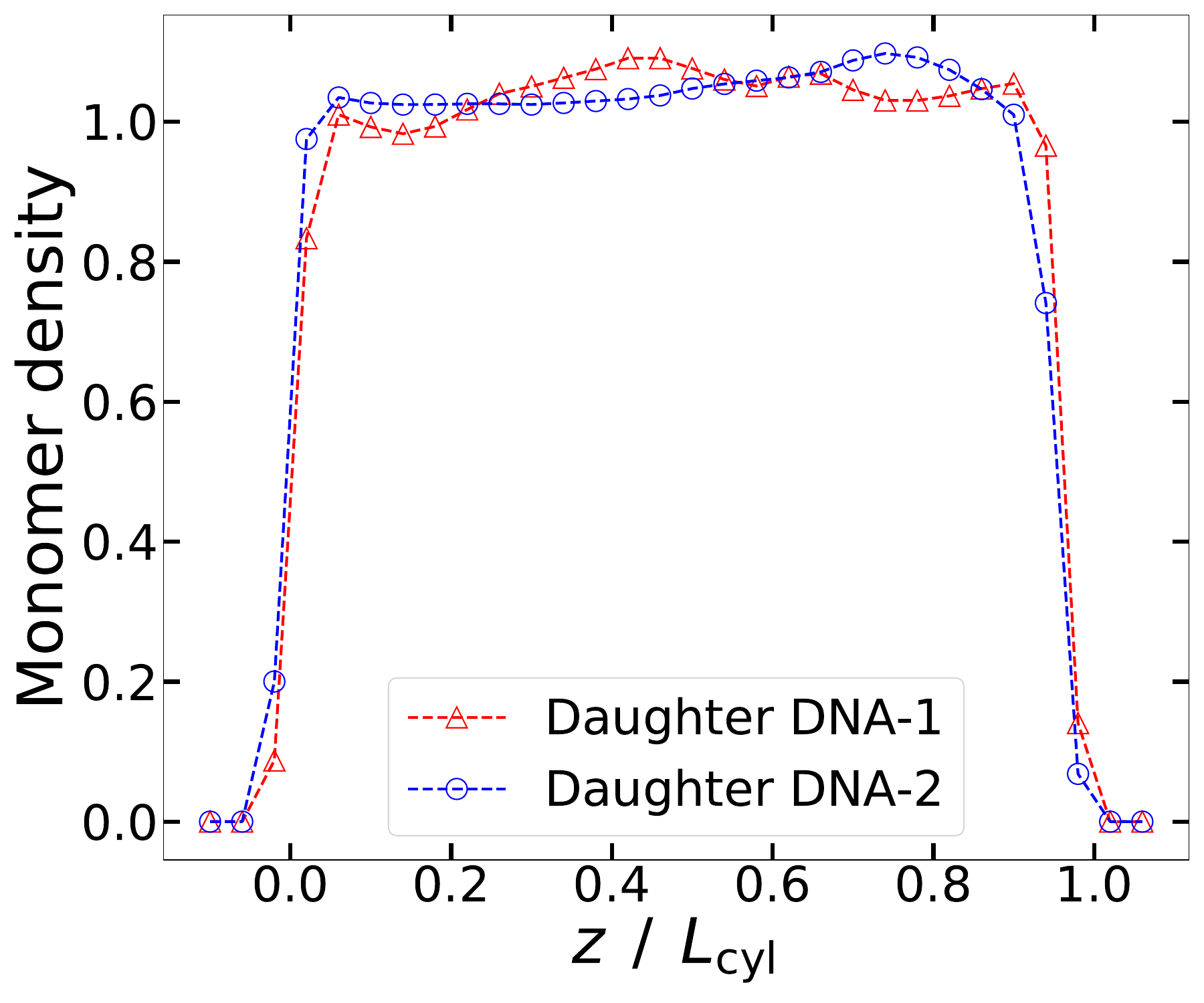}
    \put(-115,78){\large\textbf{(a)}}
    \includegraphics[width=0.49\linewidth]{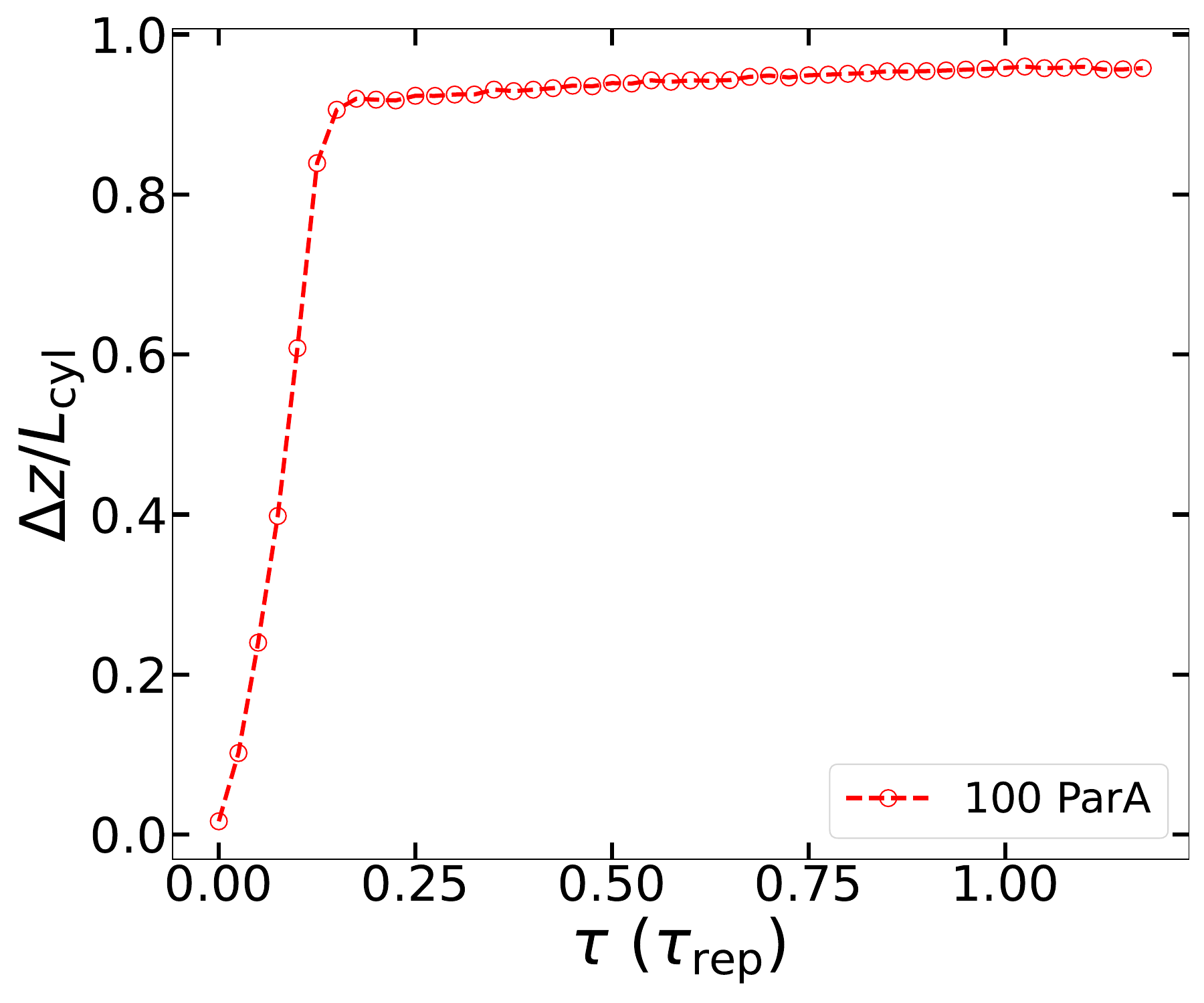}
    \put(-115,80){\large\textbf{(b)}}
    \caption{ \textbf{(a)} We plot the monomer distribution for the case in which daughter DNA-1 monomers only have excluded volume interaction with daughter DNA-1 monomers, and daughter DNA-2 monomers only have excluded volume interaction with daughter DNA-2 monomers. Daughter DNA-2 monomers do not have excluded volume interaction with Mother DNA monomers and daughter DNA-1 monomers. We plot this distribution after completion of whole replication for $120$ independent runs. \textbf{(b)} To characterize the unhindered motion of the two daughter \emph{oriC} loci, we plot the distance between them along the long axis of the cylinder, $\Delta z$, normalized by the instantaneous cylinder length, $L_{\mathrm{cyl}}$. The distance shown in the plot is averaged over $120$ independent runs. }
    \label{fig:SelfEV_mon_distribution}
\end{figure}

\begin{figure}[htbp]
    \centering
    \begin{subfigure}[t]{0.49\linewidth}
        \includegraphics[width=\linewidth]{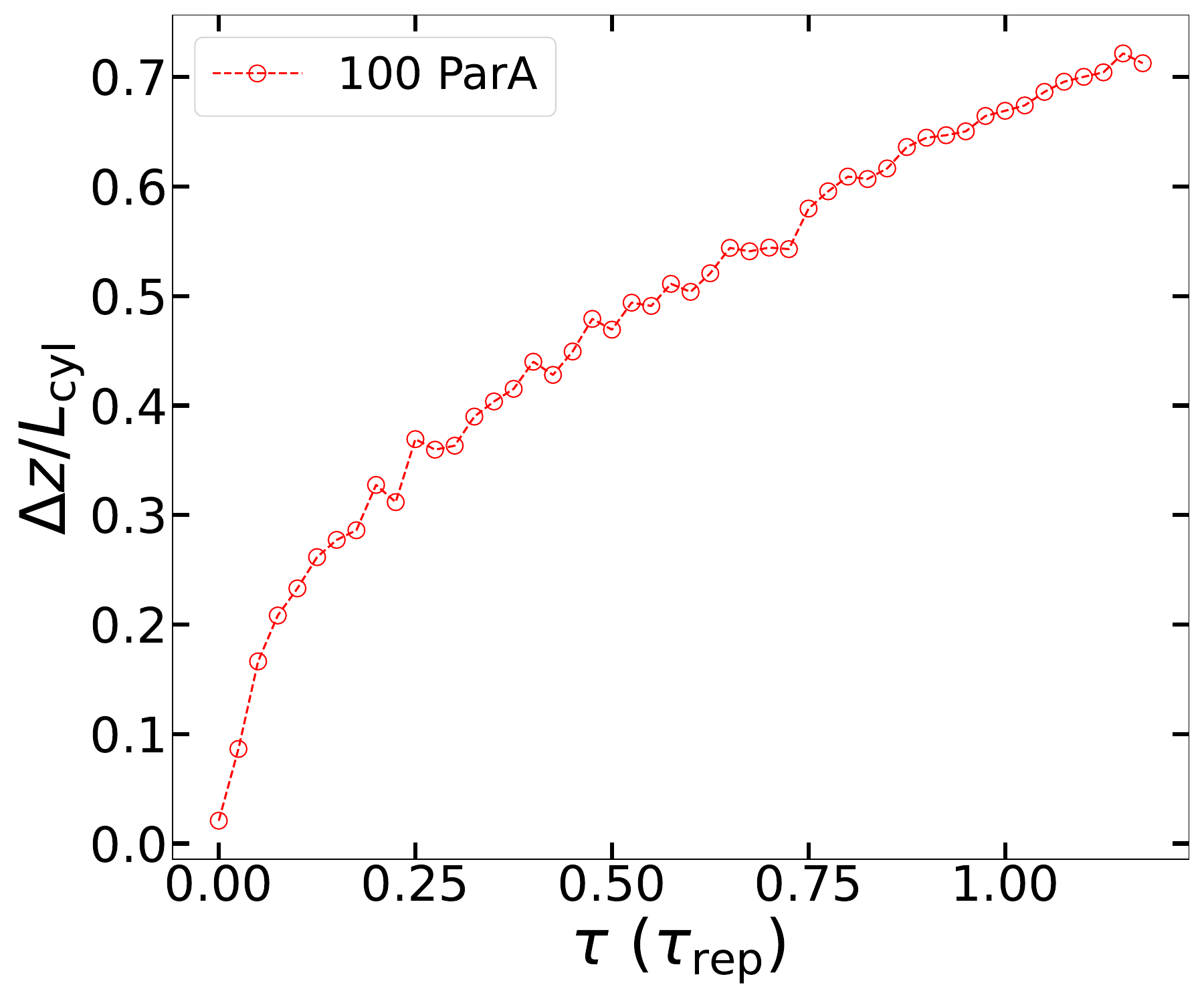}
        \put(-195,168){\large\textbf{(a)}}
    \end{subfigure}
    \begin{subfigure}[t]{0.48\linewidth}
        \includegraphics[width=\linewidth]{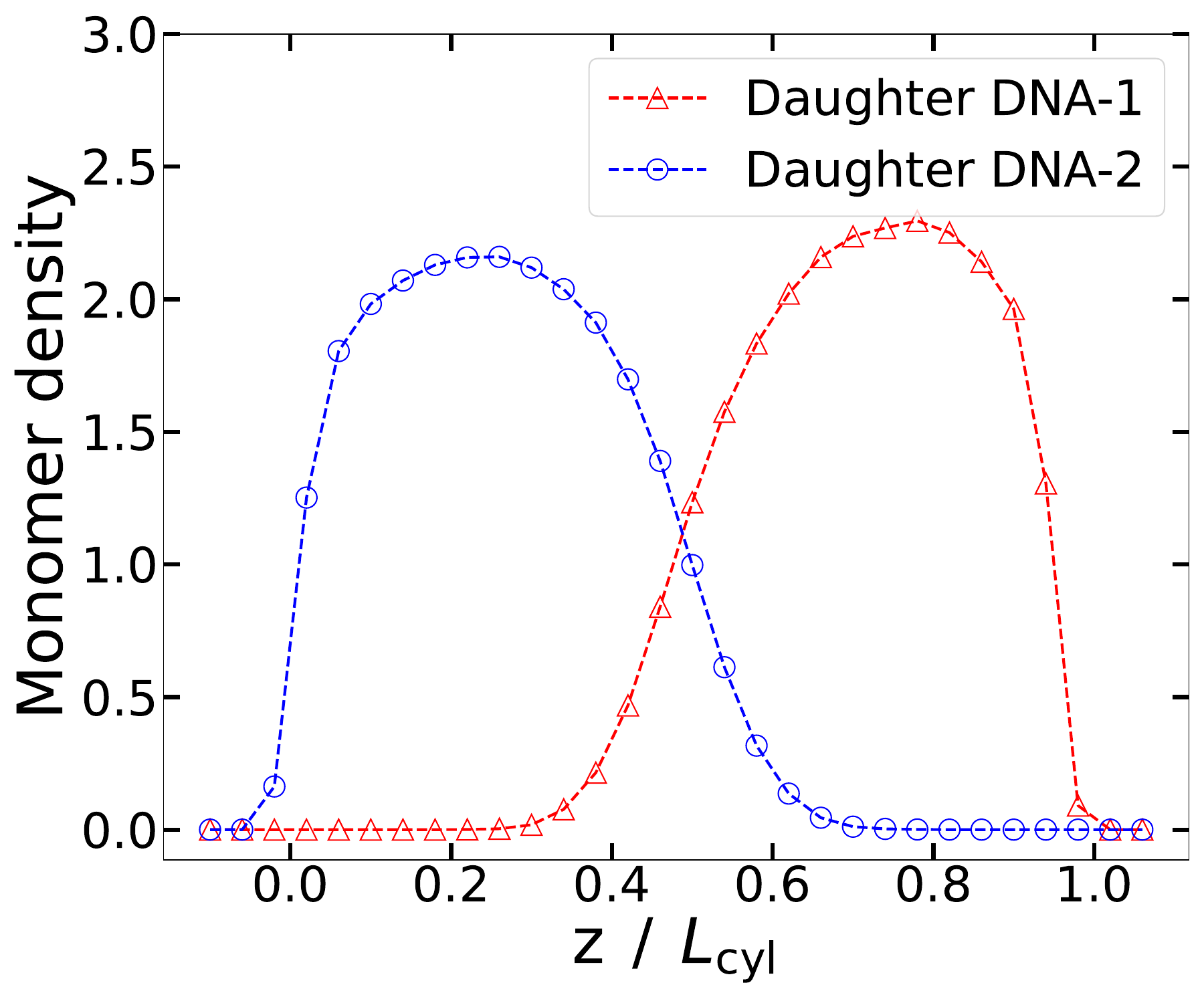}
        \put(-195,180){\large\textbf{(b)}}
    \end{subfigure}

\caption{\textbf{(a)} Plot shows normalized long axis distance $\Delta \mathrm{z} /L_{\mathrm{cyl}}$, between two daughter \emph{oriC} at different instances of replication time. The excluded volume interaction range is set to $0.3a$. The interaction range between ParA and ParB monomers is kept fixed at $r_c^P=1.5a$, while the strength of the attractive interaction to $-5k_BT$ and the repulsive interaction to $5k_BT$; see Fig.~\ref{fig:ParAParBinteractionpotential}(c). \textbf{(b)} The plot shows the distribution of daughter monomers along the long axis, normalized by cell length, after completion of replication of mother DNA. The distribution is calculated from the ``successful" runs among $120$ independent runs.}

\label{fig:SelfEV_testDATA}
\end{figure}

\FloatBarrier

\section{Replisome results in $\mu m$ units}

\begin{figure}[htbp]
    \centering
    \includegraphics[width=0.45\linewidth]{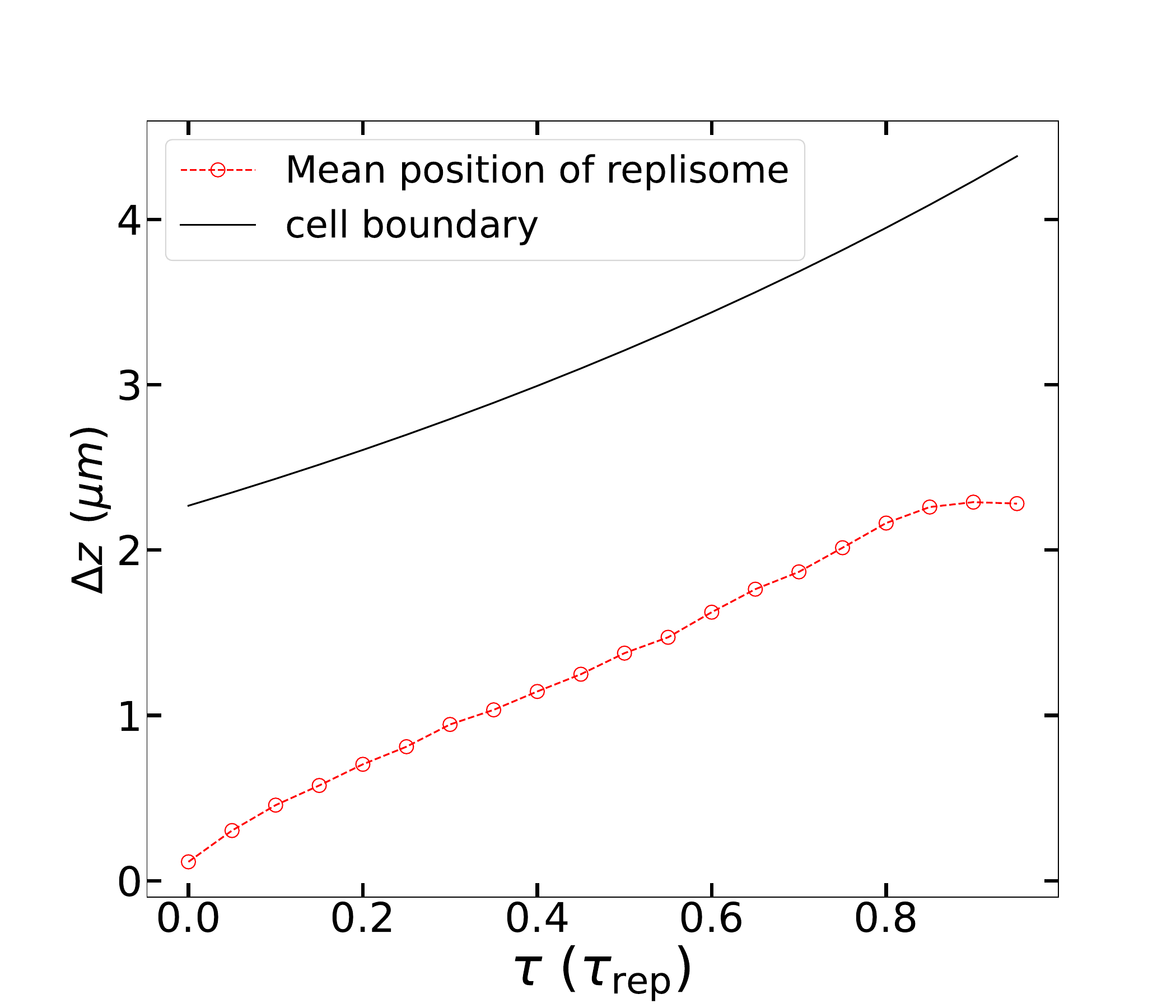}
    \caption{\textbf{Replication fork localization in $\boldsymbol{\mu m}$ for \emph{Model-2}:} The plot shows the
mean longitudinal position $\Delta \mathrm{z}$ of the replisomes (RFs) in $\mu m$ measured from the old pole, averaged over their instantaneous positions, 
as a function of time normalized by the replication time. The distance $\Delta \mathrm{z}$ is measured from the old pole of the cell. The dark black curve represents the position of the new pole/edge of the cell. Data is averaged over the runs that showed successful \emph{oriC} transport to the new pole out of 120 independent runs.}
    \label{fig:placeholder}
\end{figure}
\FloatBarrier

\section{One-to-one mapping of tagged loci and corresponding monomer indices}
A one-to-one mapping between the experimentally tagged loci on the chromosomal arms and the corresponding monomer indices in the polymer model is established.
\begin{figure}[htbp]
    \centering
    \begin{subfigure}{0.38\linewidth}
        \centering
        \includegraphics[width=\linewidth]{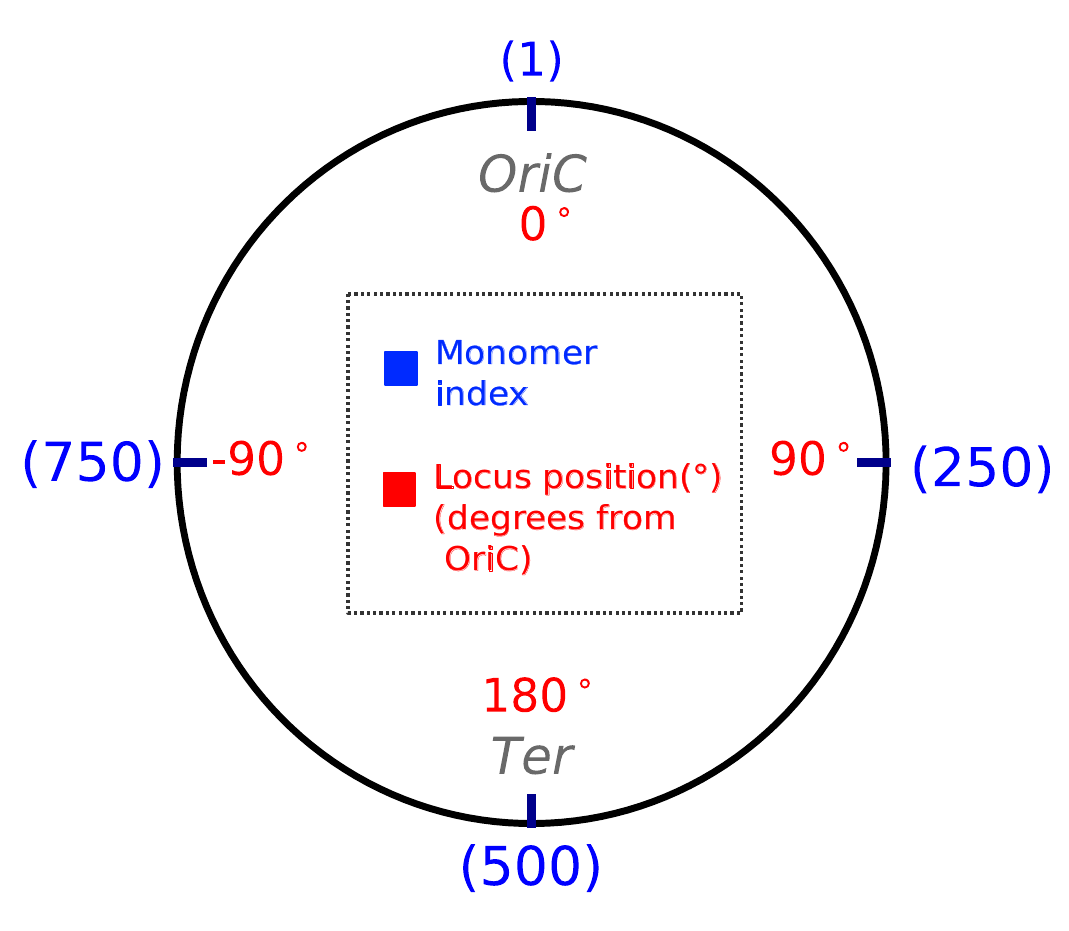}
        \put(-200,150){\large\textbf{(a)}}
        \label{fig:lociindexing}
    \end{subfigure}
    \begin{subfigure}{0.38\linewidth}
        \centering
        \includegraphics[width=\linewidth]{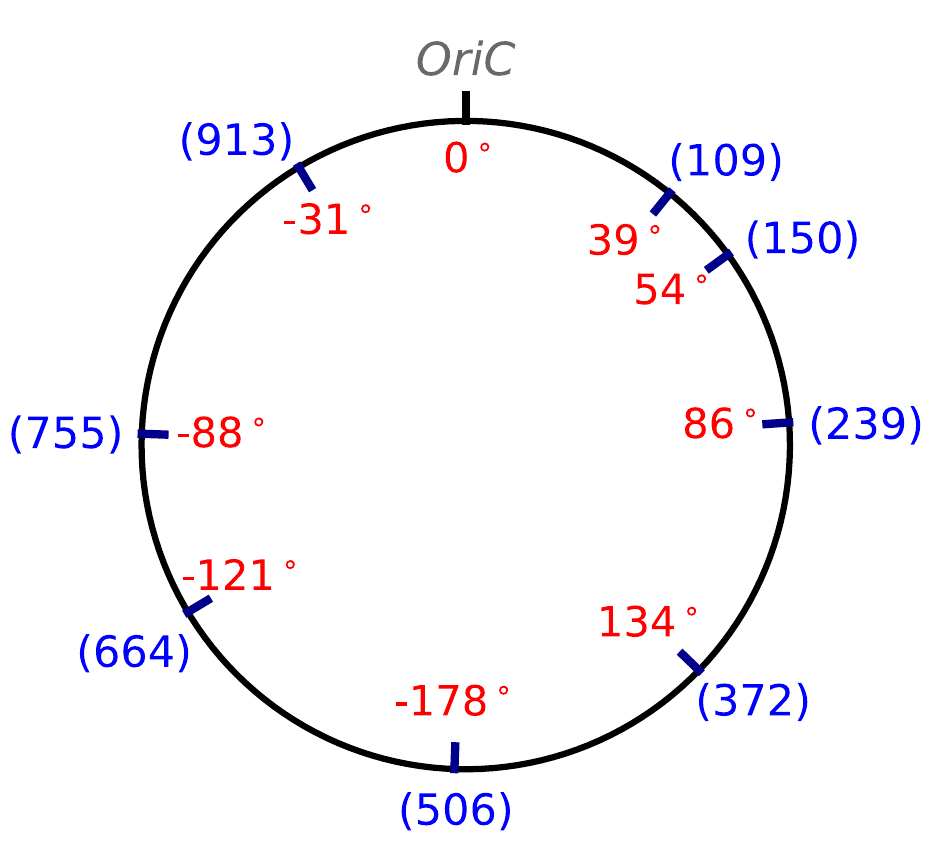}
        \put(-200,150){\large\textbf{(b)}}
        \label{fig:locischeme}
    \end{subfigure}

    \caption{Schematic of the circular chromosome of \emph{Caulobacter crescentus}. \textbf{(a)} The origin of replication (\emph{oriC}) is set to ($0^\circ$), and the angular position of each genomic locus is measured relative to \emph{oriC}, with the left chromosome arm assigned positive angles and the right arm assigned negative angles. The terminus (\emph{ter}) is located at ($180^\circ$) (equivalently,  ($-180^\circ$)). Blue labels indicate monomer indices, while red labels denote the corresponding angular positions. \textbf{(b)} Schematic representation of different genomic loci tagged with fluorescent proteins at specific positions along the chromosome contour \cite{Harju2025.04.03.647011}, together with the corresponding locus indices used in our simulations.}
    \label{fig:loci_schematics}
\end{figure}

\FloatBarrier

\section{Experimental Hi-C maps and Contact maps from 
our simulations:}

\begin{figure*}[htbp]
    \centering
    \includegraphics[width=\linewidth]{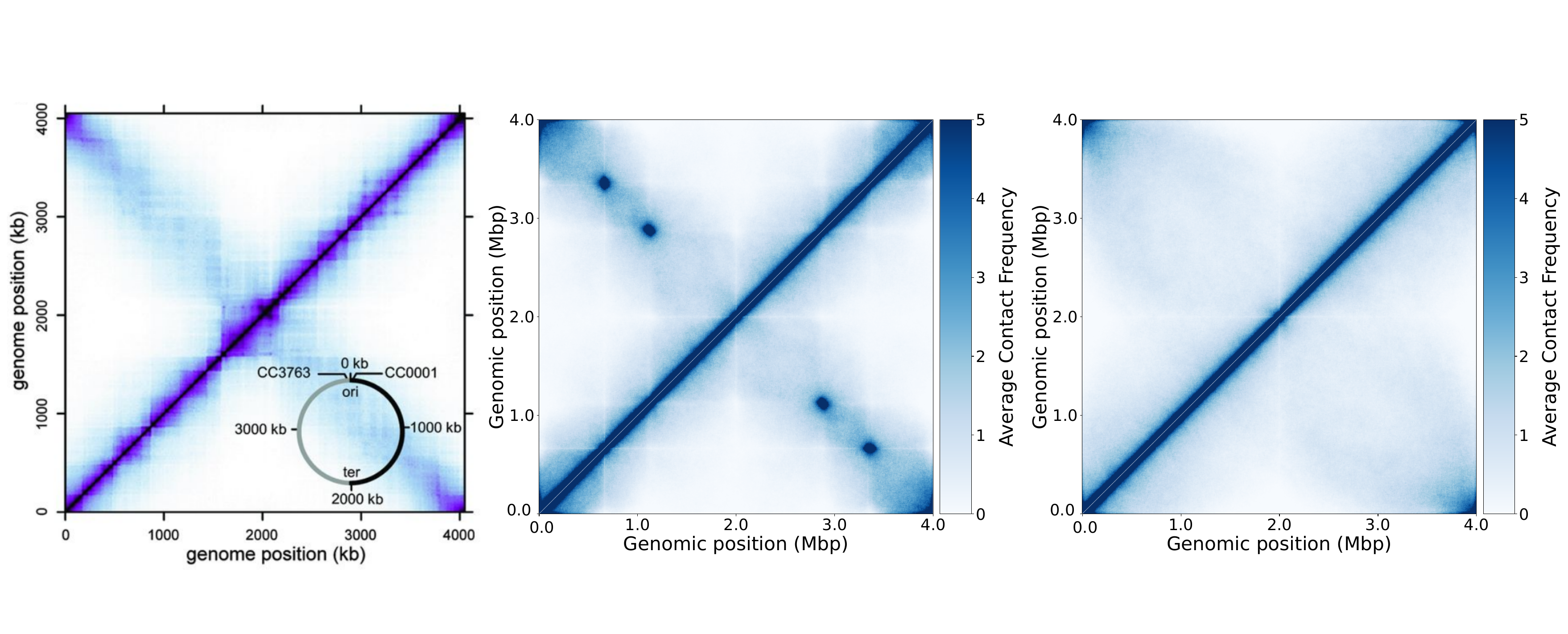}
    \put(-430,155){\large\textbf{(a)}}
    \put(-275,152){\large\textbf{(b)}}
    \put(-100,152){\large\textbf{(c)}}
\caption{\textbf{(a)} Experimentally observed Hi-C Contact map. Reproduced from Le et al., Science, DOI: 10.1126/science.1242059 (2013), AAAS~\cite{le2013high}. \textbf{(b)} Contact maps calculated from Simulations using \emph{Model-2} with modified polymer architecture calculations from configurations after replication is complete. This implies that data for the contact map is collected only from `successful runs' after two Daughter DNA polymers are formed and segregated. We calculate contact maps for the monomers of DNA-1 and DNA-2 separately, and then average the two to obtain the data, which we show in the subfigure. \textbf{(c)} Contact maps from simulations using Model-1, i.e.  without modified polymer architecture. To calculate the contact map, we use the same protocol as described in the figure caption of subfigure (b), using only the `successful' runs. 
}
\end{figure*}

We reproduce the experimental  Hi-C maps from  \cite{le2013high},
the contact map produced from our simulations using \emph{Model-1} (i.e., without extrusion)
and Model-2 ( with extrusion), respectively. Note that the extrusion process creates 
the ring polymer with modified topology with a linear sequence of internal loops,
shown in SI--I(F) of the supplementary information. It is the linear sequence of loops along the chain
contour (created by the extrusion process) which organizes the two arms of the polymer 
along the long axis of the cylinder, and creates the two diagonals of the contact maps.
The linear organization of the chromosome is driven by the entropic repulsion between the 
internal loops, which entropically avoid each other and occupy different segments of the cylinder along the long axis: refer to ~\cite{mitra2022topology} for details.



\section{linear Arrangement}
Established work has suggested that the bacterial chromosome of \emph{C.~crescentus} exhibits a linear organization along the long axis of the cell \cite{viollier2004rapid,le2013high}. 
\begin{figure}[htbp]
    \centering
    \includegraphics[width=0.32\linewidth]{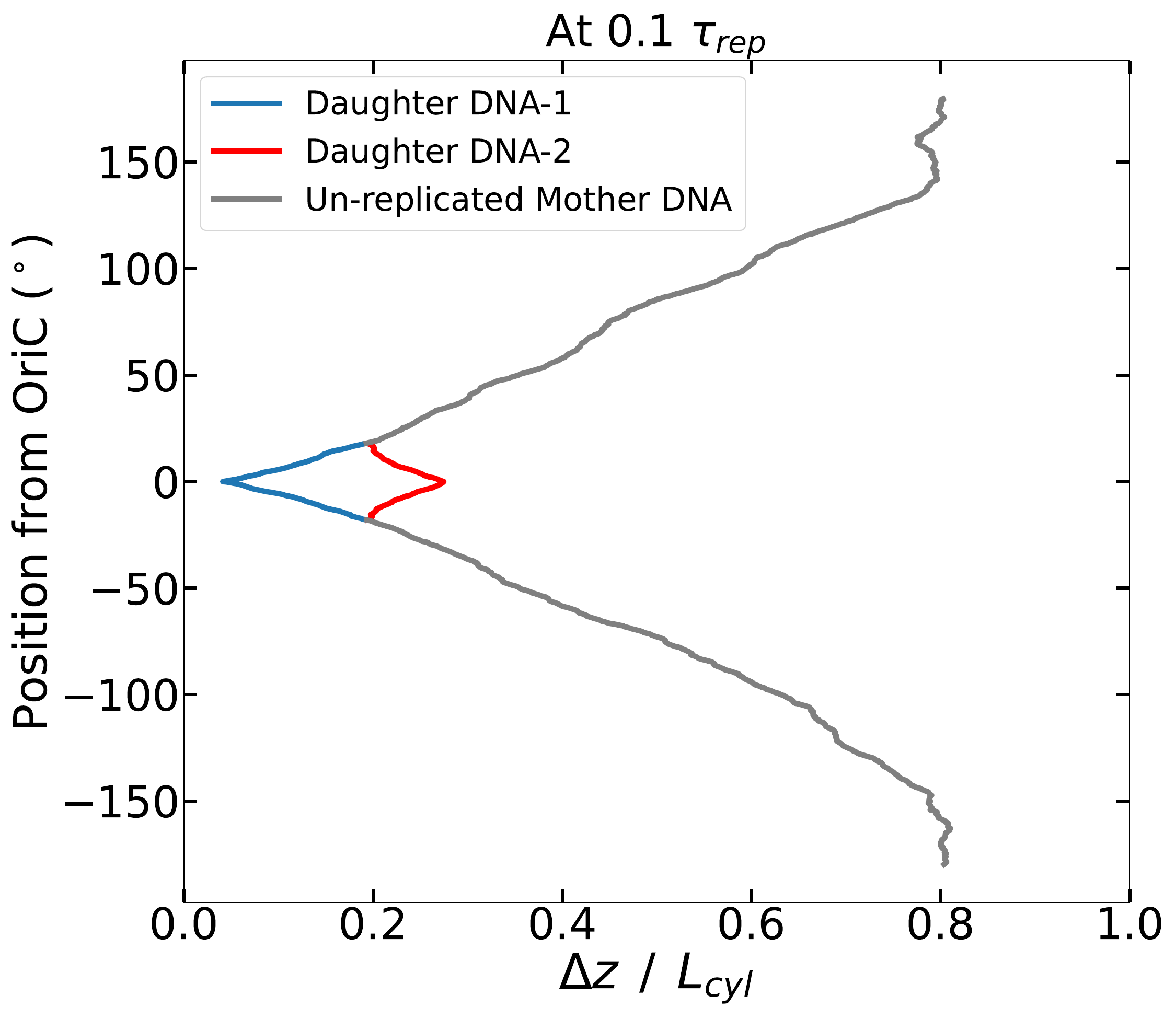}
    \includegraphics[width=0.32\linewidth]{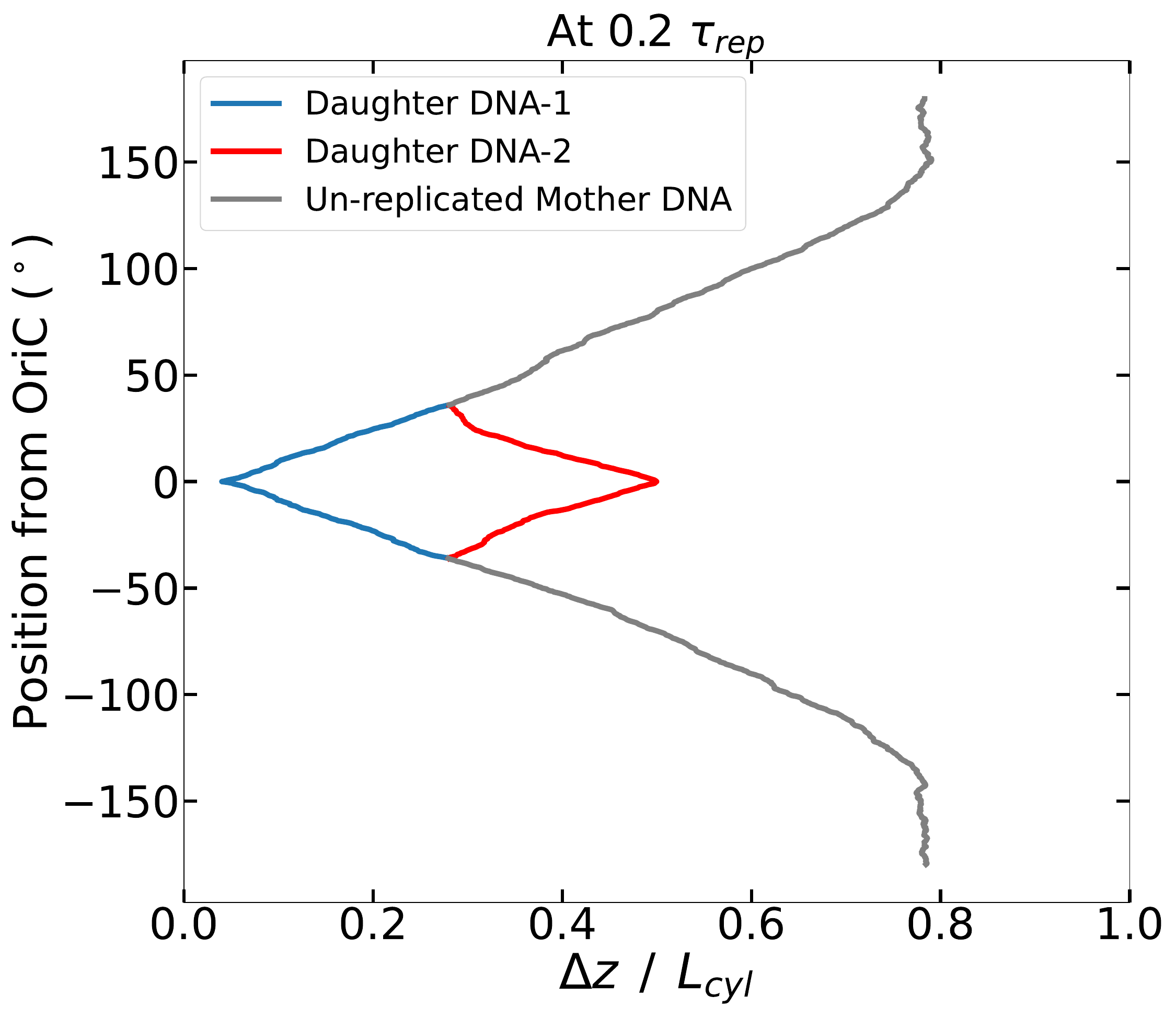}
    \includegraphics[width=0.32\linewidth]{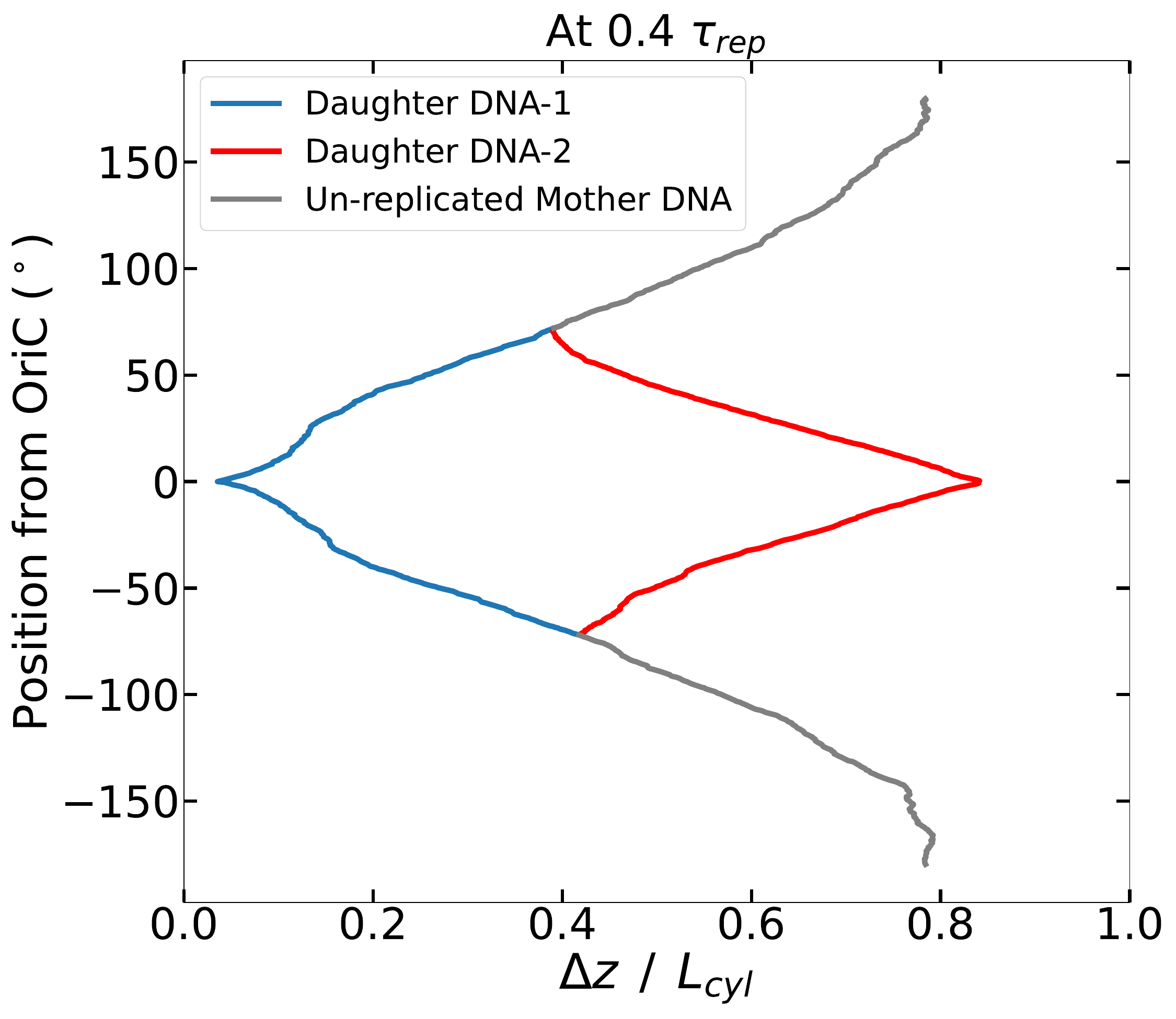}
    \includegraphics[width=0.32\linewidth]{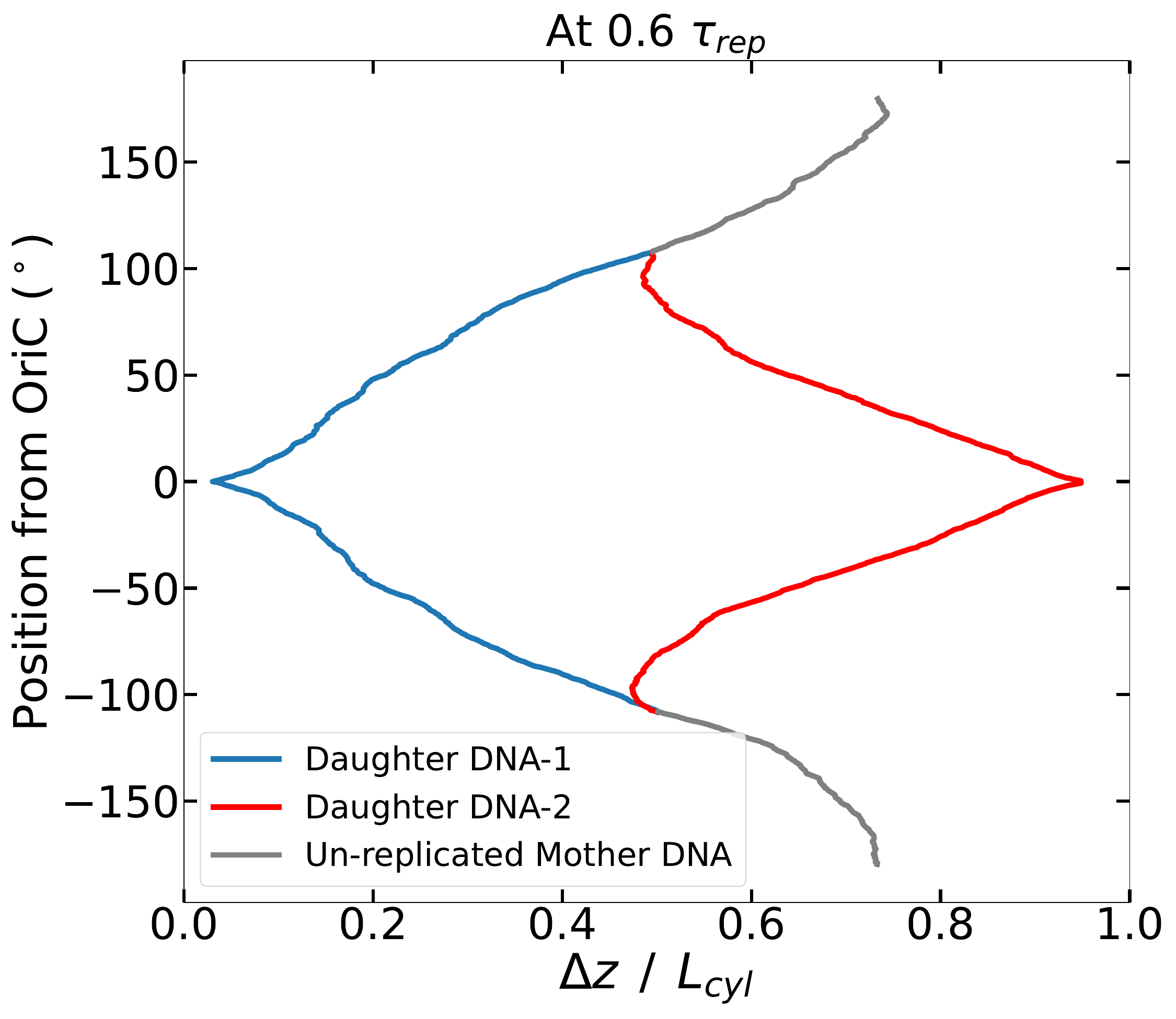}
    \includegraphics[width=0.32\linewidth]{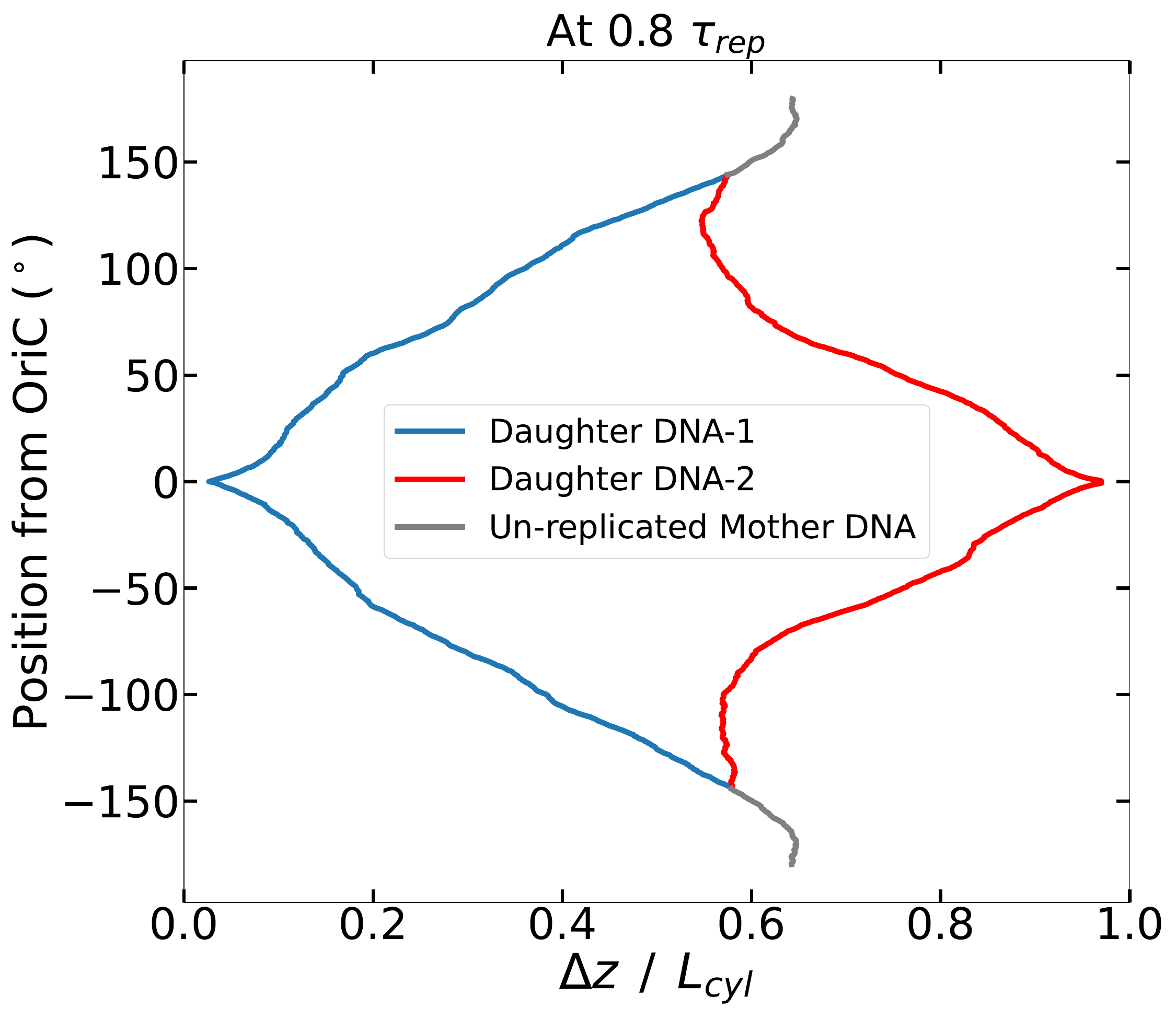}
    \includegraphics[width=0.32\linewidth]{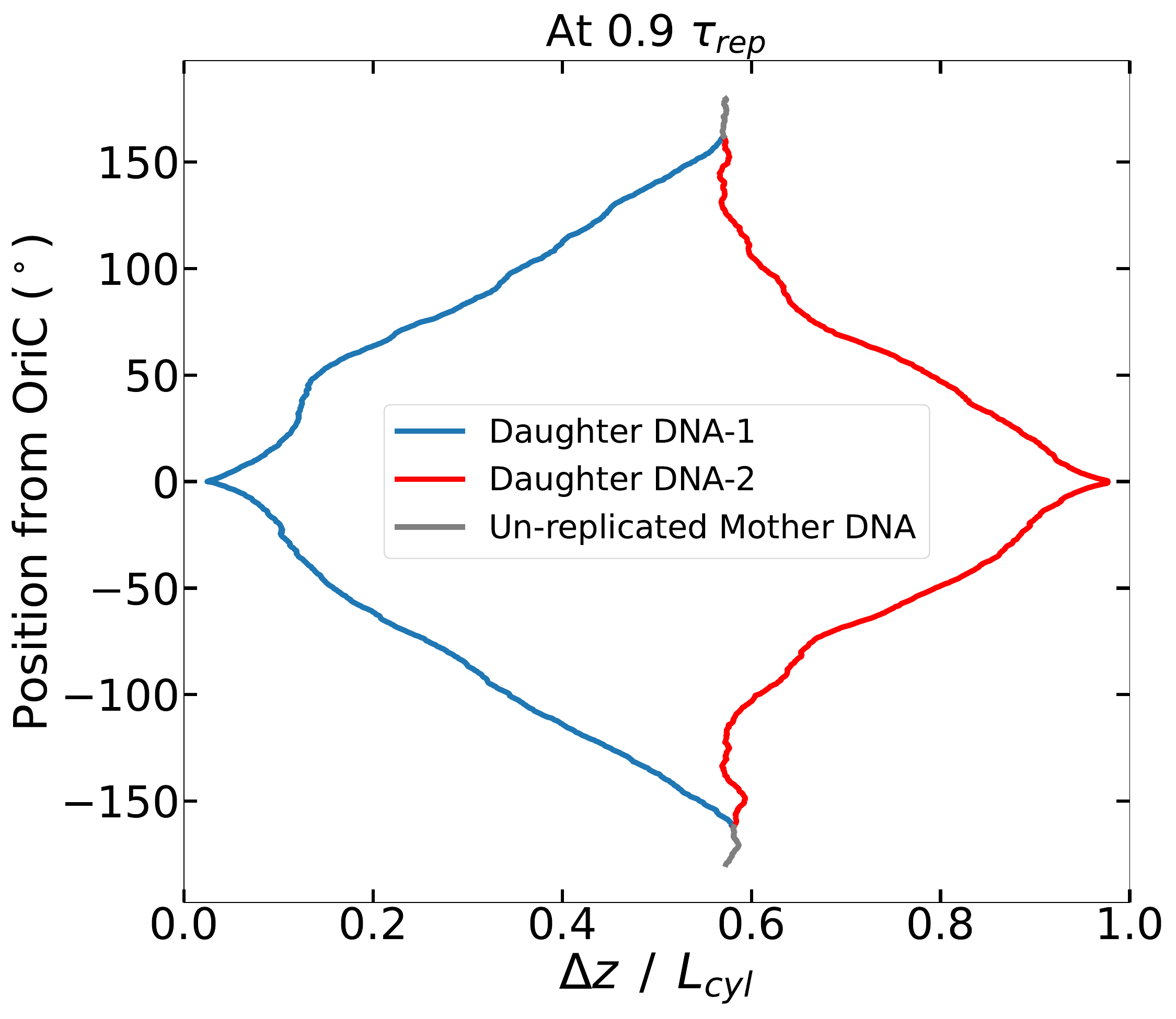}
\caption{\textbf{Linear organization of the chromosome along the long axis of the cell}. The spatial organization of the loci on the unreplicated mother DNA and the two daughter DNA polymers is shown at different stages of replication, corresponding to $10\%$, $20\%$, $40\%$, $60\%$, $80\%$, and $90\%$ of the total replication time $\tau_{rep}$. The replication stage is indicated in the title of each subfigure. The positions of the loci along the chromosome contour, measured from \emph{oriC}, are specified by their angular positions ($^\circ$), while their positions along the cell axis are represented by the normalized coordinate $\Delta z/L_{\mathrm{cyl}}$. Here, $\Delta z$ is the distance along the cell axis measured from the old pole, and $L_{\mathrm{cyl}}$ is the instantaneous cell length. The loci on the two chromosome arms exhibit a predominantly linear organization along the cell axis throughout replication, with the two daughter chromosomes progressively extending toward the new pole as replication proceeds. The blue, red, and gray curves represent Daughter DNA-1, Daughter DNA-2, and the unreplicated mother DNA, respectively.}
\end{figure}

\FloatBarrier

\bibliography{bibliography}